\documentclass[aps,pre,twocolumn,eqsecnum,superscriptaddress]{revtex4-2}

\usepackage{color}
\usepackage{hyperref}
\usepackage{amsmath}
\usepackage{graphicx}
\usepackage{color}
\usepackage{wrapfig}
\usepackage[export]{adjustbox}
\begin{document}


\title{The Spanning Tree Model and the Assembly Kinetics of RNA Viruses.}


\author{Inbal Mizrahi}
\affiliation{Department of Physics and Astronomy, University of California, Los Angeles, CA 90095}
\author{Robijn Bruinsma}
\affiliation{Department of Physics and Astronomy, University of California, Los Angeles, CA 90095}
\affiliation{Department of Chemistry and Biochemistry, University of California, Los Angeles, CA 90095}
\author{Joseph Rudnick}
\affiliation{Department of Physics and Astronomy, University of California, Los Angeles, CA 90095}


\date{\today}

\begin{abstract}
Single-stranded (ss) RNA viruses self-assemble spontaneously in solutions that contain the viral RNA genome molecules and the viral capsid proteins. The self-assembly of \textit{empty} capsids can be understood on the basis of free energy minimization of rather simple models. However, during the self-assembly of complete viral particles in the cytoplasm of an infected cell, the viral genome molecules must be selected from a large pool of very similar host messenger RNA molecules. It is known that the assembly process takes the form of preferential heterogeneous nucleation of capsid proteins on viral RNA molecules (``selective nucleation"). Recently, a simple mathematical model was proposed for the selective nucleation of small ssRNA viruses \cite{PLOS}. 
In this paper we present a statistical physics analysis of the thermal equilibrium and kinetic properties of that model and show that it can account, at least qualitatively, for numerous observations of the self-assembly  of small ssRNA viruses. 
\end{abstract}


\maketitle

\section{Introduction} \label{sec:intro}

In 1956 Crick and Watson \cite{Crick1956} noted that the total length of the genome of a virus should scale with the interior volume of the protein shell (or ``capsid") enclosing the genome. This observation was a central part of their argument that sphere-like viruses should have  icosahedral symmetry. In a similar way, the length of a rod-like capsid should scale with the length of the genome. This approach works well for small, single-stranded (ss) RNA viruses (such as the polio and common cold viruses) as well as for double-stranded DNA viruses (such as the herpes and bacteriophage viruses) that have close-packed genomes \footnote{For double-stranded DNA viruses the genome molecules either are inserted into the interior pre-fabricated spherical capsids by a molecular motor or the genome is pre-condensed prior to assembly.} while for the large ssRNA retroviruses, which do not have close-packed genomes, the genome length scales with the interior surface area of capsid \cite{Ganser-Pornillos2008}. 

Many ssRNA viruses \textit{self-assemble} spontaneously in solutions that contain concentrations of viral capsid proteins and RNA molecules \cite{Fraenkel-Conrat1955, Butler1978}. This is a ``passive" process, driven by free energy minimization. Early work by Aaron Klug \cite{Klug1999} proposed that the RNA genome molecules act as \textit{templates} that direct the viral assembly process. He proposed a physical model in which the repulsive electrostatic interactions between positively charged groups of the capsid proteins are just strong enough to overcome competing attractive hydrophobic interactions, which just prevents the self-assembly of empty capsids under physiological conditions. When viral RNA molecules are added to the solution, the negative charges of the RNA nucleotides neutralize some of the positive charges of the capsid proteins and this tilts the free energy balance towards assembly \footnote{For a quantitative treatment of this model, see ref. \cite{kegel2006}}. 

The genome molecules of ssRNA viruses genome molecules have a tree-like ``secondary structure" produced by Watson-Crick base-pairing between complementary RNA nucleotides of the primary sequence of RNA nucleotides \cite{mathews1999}. The ``co-assembly" picture suggests that there could be connections between the geometrical and topological features of the secondary structure and the capsid geometry. The redundancy of the genetic code allows for the possibility of ``silent'' (or synonymous) mutations that alter the secondary structure of a viral RNA molecule without altering the structure of the proteins encoded by the nucleotide sequence \cite{tubiana}. Viral RNA molecules indeed have specific specific sections, known as \textit{packaging signals}, with enhanced affinity for the capsid proteins \cite{Patel2015}. These packaging signals favor the encapsidation of viral RNA molecules over host mRNA molecules with comparable length. Separately, the topology of viral RNA molecules differs from that of generic single-stranded RNA molecules of the same length: they are significantly more branched and compact than generic RNA molecules of the same length \cite{yoffe2008}. 

Such observations have been understood in terms of the packaging signals reducing the equilibrium free energy of fully assembled viral particles but recent work has focused on the \textit{kinetics} of viral assembly\cite{bruinsma2021}. Kinetic studies of the self-assembly of empty capsids \cite{Prevelige1993, Casini2004, medrano} provided support for a \textit{nucleation-and-growth scenario} where assembly is triggered by the formation of a ``nucleation complex", composed of a small number of capsid proteins. This nucleation complex then extends or elongates by absorbing more proteins until the capsid closes up. When the theory of nucleation-and-growth applied to viral capsids \cite{Zandi2006}, the nucleation complex is identified with the \textit{critical nucleus}, i.e., the state of an assembly of capsid proteins for which the free energy has a maximum as a function of the number of capsid proteins. Formation of the critical nucleus is an energetically ``uphill" process while the subsequent elongation is a ``downhill" process. Under conditions of thermodynamic equilibrium, the critical nucleus would be a half-formed shell. However, experimentally measured values for the interaction energy between capsid proteins are in the range of a few $k_BT$ and this leads to an activation energy barrier for the assembly of capsids under conditions of thermodynamic equilibrium in the range of \textit{hundreds} of $k_BT$. The corresponding assembly rates would be prohibitively low on laboratory time scales. The size of the nucleation complex measured by light scattering \cite{Casini2004} is much smaller than a half-formed shell while assembly rates of empty are generally in the range of $0.01-1.0 Hz$. These observations indicate that capsid assembly takes place under conditions that are well away from thermodynamic equilibrium, more specifically under a relatively high level of \textit{supersaturation}. Finally, a recent study of the assembly kinetics of empty capsids of the Hepatitis B virus indicated that the formation of the critical nucleus takes place along well-defined \textit{assembly pathways} \cite{asor2019}, similar to the distinct assembly pathways that govern the folding of proteins \cite{bryngelson}. 

\begin{figure}[htbp]
\begin{center}
\includegraphics[width=2.5in]{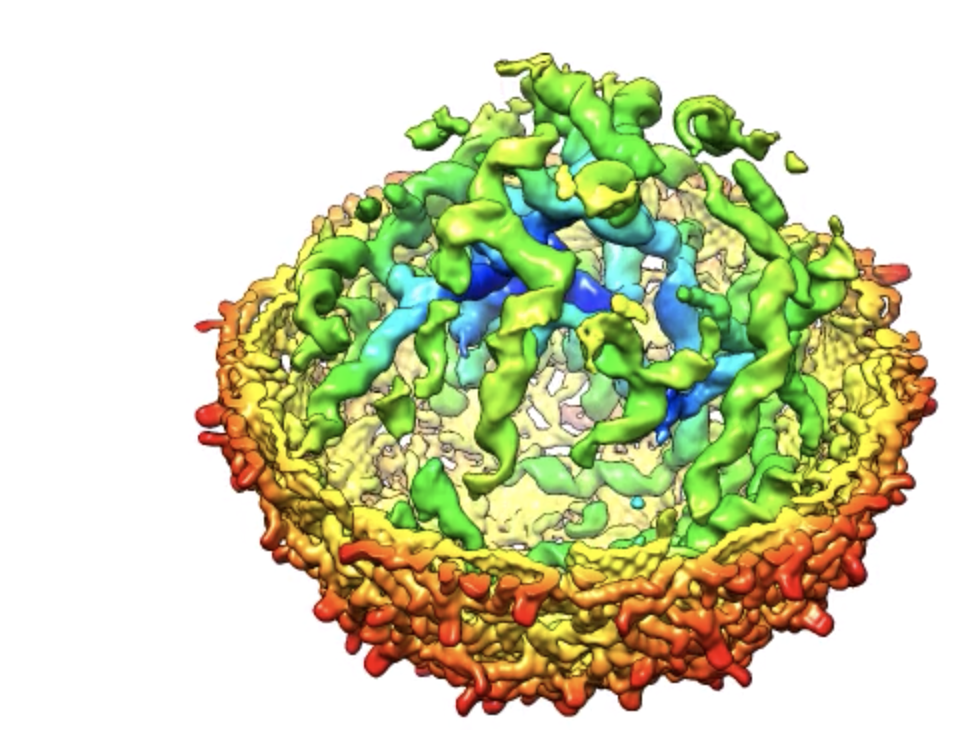}
\caption{CryoEM asymmetric reconstruction of the MS2 virion. The viral genome molecules (in green and blue) associate reproducibly mostly with one half of the capsid shell, which has been removed for clarity. From ref.\cite{dai2017}}
\label{fig:MS2}
\end{center}
\end{figure}

Does the nucleation-and-growth scenario describe the co-assembly of viruses complete with encapsidated RNA genome molecules? Early time-resolved small-angle X-ray scattering experiments on viral assembly in bulk solutions were interpreted in terms of heterogeneous nucleation of capsid proteins on viral RNA molecules with the resulting nucleo-protein complex attracting additional capsid subunits via a ``diffusive antenna" mechanism \cite{hu2007, kler2012}. Recent observations on the assembly kinetics of individual MS2 viruses (a small ssRNA bacteriophage virus) \cite{garmann2019} also provided support for a heterogeneous nucleation-and-growth scenario. An important observation was that the assembly process was characterized by an extremely  broad distribution of time scales. Important information about the kinetics of viral co-assembly has been gleaned as well from static structural studies. Until recently, reconstruction of packaged genome molecules involved ``icosahedral averaging", which resulted in RNA structures with imposed icosahedral symmetry \cite{Baker}. Such studies showed that the interior surface of the icosahedral capsids of certain viruses (e.g., the nodaviruses \cite{Tihova2004, Johnson2004}) is decorated by paired RNA strands lining the edges of the ``capsomers" (i.e., pentameric or hexameric groupings of capsid proteins). Recent progress in cryo-electron tomography has made it possible to image how individual ssRNA genome molecules are packaged inside spherical capsids without icosahedral averaging (``asymmetric reconstruction" \cite{koning, beren}). An important example is again the MS2 virus. It was found that sections of the RNA genome rich in packaging signals reproducibly associated with roughly \textit{half} of the interior surface of the capsid \cite{Dykeman2011, dai2017}, as shown in Fig.\ref{fig:MS2}
This can be interpreted as evidence for a well-defined, reproducible assembly pathway of a critical nucleus that is a compact nucleo-protein complex held together by particular sections of the viral RNA molecule that are rich in packaging sequences. The downhill part involves mostly generic electrostatic RNA-protein attractions and involves non-reproducible RNA configurations. In terms of the theory of nucleation and growth, the packaging signals reduce the height of the assembly energy activation barrier for the packaging of viral RNA molecules. Importantly, this provides a natural mechanism for the selection of viral RNA molecules from a large pool of host messenger RNA (``mRNA") molecules . 

Understanding how the nucleation-and-growth scenario applies to viral assembly is important also for biological reasons. When the molecular components of a single-stranded RNA virus assemble in the cytoplasm of an infected cell and form virions, a limited number of viral RNA molecules compete with a large pool of similar host messenger RNA (``mRNA") molecules for packaging by viral capsid proteins \cite{Dimmock2001}. Reduction of the assembly activation energy barrier by the packaging signals has the effect of selectively increasing the assembly rate of viral particles with respect to the packaging of host mRNA molecules, a mechanism that might be called ``selective nucleation"\footnote{Selective nucleation was proposed by I. Rouzina in the context of the assembly of retroviruses}. In this view, the action of packaging signals is similar to that of enzymes or catalysts that increase the rate of a chemical reaction. 

Nucleation-and-growth is not the only possible assembly scenario. The packaging of linear genome molecules in small capsids has been numerically simulated \cite{Perlmutter2014} and two different packaging scenarios were encountered. This depended on the \textit{affinity ratio} of the protein-genome and protein-protein interactions. For low values of this affinity ratio, a nucleation-and-growth assembly scenario was encountered of the diffusive antenna type. For larger values of the affinity ratio, a so-called ``en masse" scenario was encountered. The first assembly step is, in that case, the formation of a disordered protein/genome condensate that shrinks in size as more proteins are being added. The condensate then undergoes an order-disorder transition in which the spherical symmetry of the condensate is broken and an icosahedral particle forms \cite{rudnick2019}. This form of assembly may have been reproduced in a two-stage in-vitro assembly experiments of CCMV virus-like particles \cite{Garmann}. Micro-mechanical experiments on the encapsidation of linear \textit{double-stranded} genome molecules by capsid proteins \cite{van2020} have also been interpreted according to the en-massed scenario.

In this article we examine the physical properties of a recently proposed mathematical model -- the \textit{``Spanning Tree Model"} -- for the selective nucleation of small RNA viruses. The model is sufficiently simple so that its kinetics can be determined by numerical integration of a Master Equation. The model is used to determine thermodynamic parameters that optimize RNA selectivity, to follow the effect of changing the affinity ratio on the assembly scenario and to explore which topological and geometrical properties of RNA molecules optimize selectivity. The model is a generalization of an early model for the assembly of \textit{empty} dodecahedral capsids in a solution of pentamers that was due to Zlotnick \cite{Zlotnick1994, Endres2002, Zlotnick2007}. The kinetics of the Zlotnick mode obeys a nucleation-and-growth assembly scenario \cite{Morozov2009}. It was used to carry out simulations of the packaging of linear genome molecules \cite{Perlmutter2014}. 

The spanning tree model is introduced in Section II together with the topological and geometrical classification of the model genome molecules followed by a discussion of structural transitions of partial assemblies as a function of the affinity ratio. The thermodynamic properties of the model are discussed in Section III. The Master Equation for the assembly kinetics is introduced in Section IV. Numerical integration of the Master Equation is used to determine the characteristic time scales of the kinetics. In Section V we study the dependence of the packaging efficiency on the topological and geometrical properties of the genome molecules for different values of the affinity ratio. In the concluding Section VI we discuss the possible relevance of these results for our understanding of viral assembly as well as limitations and possible extensions of the model.

\section{The Spanning Tree Model.}

\subsection{Empty capsid assembly}
\begin{figure}[htbp]
\begin{center}
\includegraphics[width=2.5in]{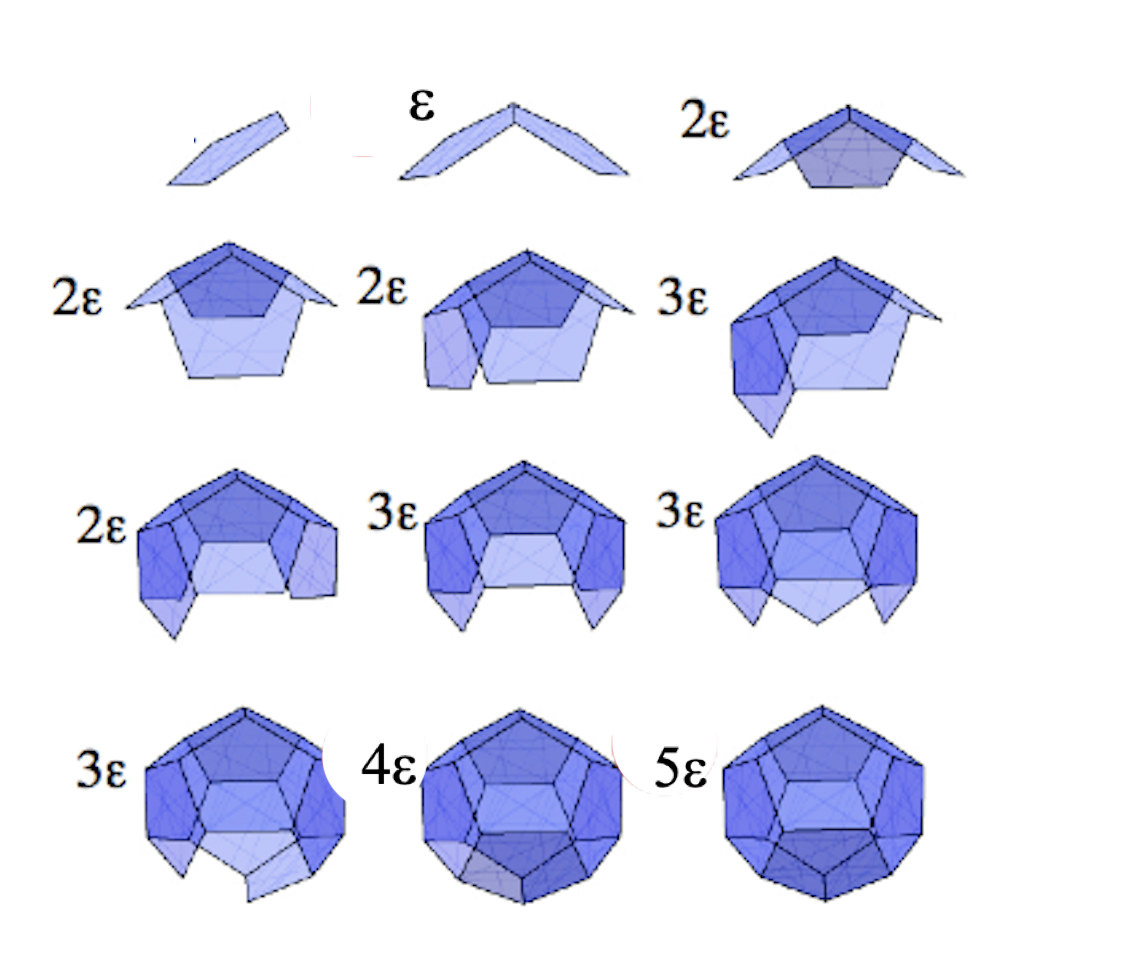}
\caption{Empty capsid assembly pathway. The figure shows a minimum-energy pathway for the assembly of a dodecahedral shell composed of twelve pentamers with adhesive edges. The edge-to-edge binding energy is $\epsilon$. The change in the total energy of the cluster for each added pentamer is indicated. Note that partially assembled capsids are compact.}
\label{fig:Zlotnick model}
\end{center}
\end{figure}

The Zlotnick model treats the capsid as a dodecahedral shell composed of twelve pentamers. The pentamers represent pentameric oligomers of capsid proteins -- known as ``capsomers" -- of a minimal (i.e., T=1) capsid composed of sixty proteins \footnote{The assembly of larger icosahedral capsids can be represented by a version of the model that allows for pentameric and hexameric capsomers.}. Assembly is driven by an attractive edge-edge interaction between pentamers. A \textit{minimum-energy assembly pathway} will be defined as a pentamer-by-pentamer addition sequence where each added pentamer is placed in a location that minimizes the free energy of the partial shell. An example of one of the many (of the order of $10^5$) degenerate minimum-energy assembly pathways is shown in Fig.\ref{fig:Zlotnick model}.   
The Helmholtz free energy $E(n)$ of a partial shell composed of $n$ pentamers is expressed as $E(n)= n_1 \epsilon - n \mu $ with $n_1$ the number of shared pentamer edges of the cluster times and with $\mu$ the chemical potential of a pentamer in solution. Assuming dilute solutions, we will set $\mu=\mu_0 + \ln c_f/\bar{c}$ with $c_f$ the concentration of free pentamers in solution and with $\mu_0$ the chemical potential at the reference pentamer concentration $\bar{c}$. The assembly free energy of a complete capsid then equals $30\epsilon-12 \mu$ for all minimum energy assembly pathways. Assembly equilibrium is defined as the state where the chemical potential of a pentamer in solution is the same as that of a pentamer that is part of a capsid. This is the case if $E(12)=0$ so if $\mu$ equals $\mu^*=(5/2)\epsilon$.  

\begin{figure}[htbp]
\begin{center}
\includegraphics[width=3.5in]{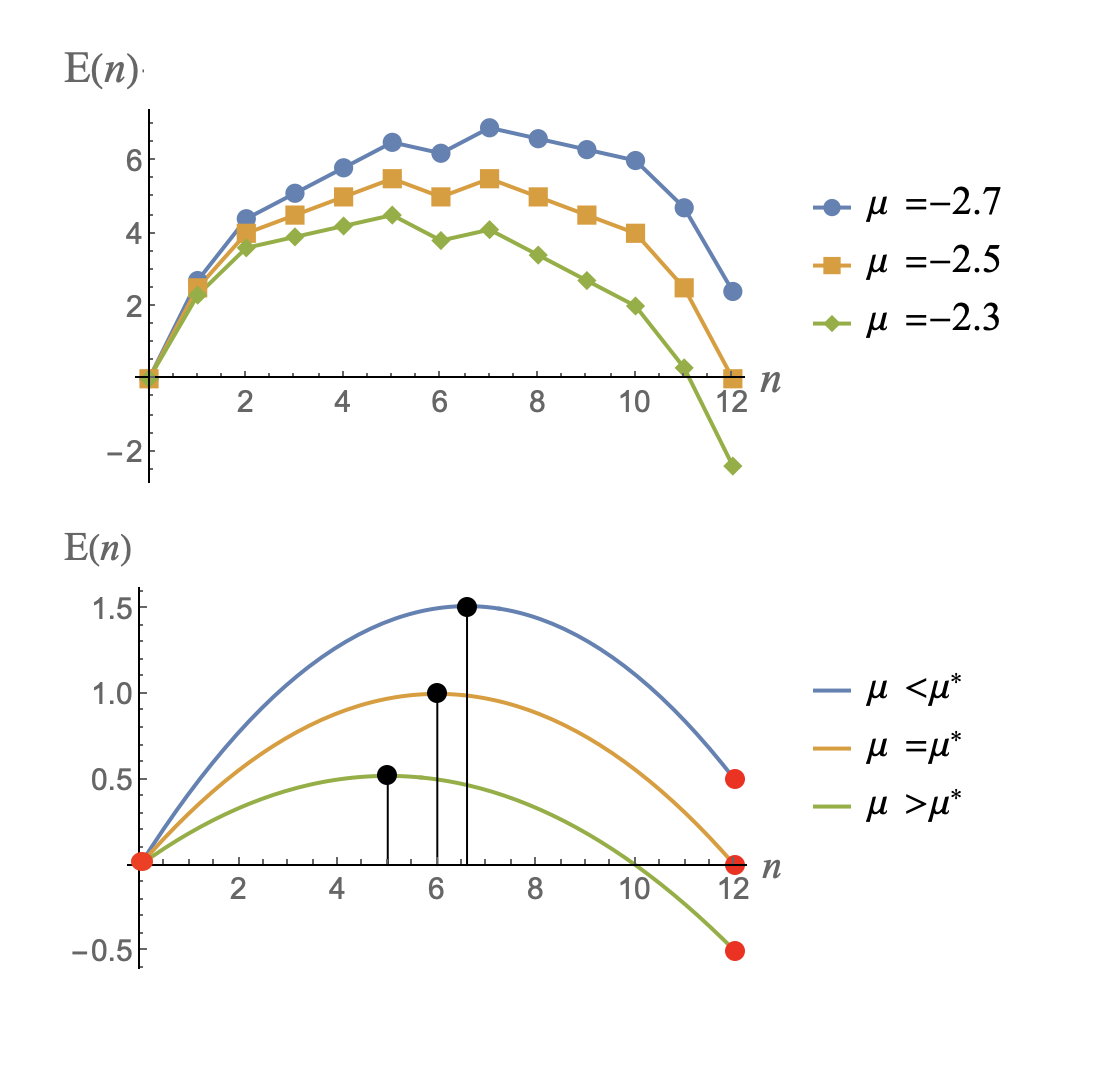}
\caption{Top: Free energy profiles of the minimum assembly energy pathways of the Zlotnick model. Blue dots: The chemical potential $\mu$ is slightly below $\mu^*=-2.5$, the value of the chemical potential  for assembly equilibrium. Orange squares: $\mu$ is equal to $\mu^*$. Green diamonds: $\mu$ is slightly above $\mu^*$. 
Bottom: Assembly free energy profiles according to continuum theory. Solid red dots: energy minima. Solid black dots: energy maxima. For $\mu<\mu^*$, the absolute energy minimum is at $n=0$ while for $\mu>\mu^*$ the absolute minimum is at $n=12$, the assembled capsid. The assembly activation energy barrier of a profile is the height of the maximum. The location $n^*$ of the maximum is for a half-filled shell under equilibrium conditions (around n=6), shifting to lower values as $\mu$ increases. }
\label{fig.2}
\end{center}
\end{figure}

Figure \ref{fig.2} (top) shows the free energy of the partial assemblies shown in Fig.\ref{fig:Zlotnick model} for three different values of the chemical potential $\mu$ close to $\mu^*$ and $\epsilon=-1$ (energy will be expressed in units of $k_BT$).

Figure \ref{fig.2} (bottom) shows the free energy of a spherical cap growing into a spherical shell \cite{Zandi2006}. The initial rise of $E(n)$ with $n$ is due to dominance of the line energy of the perimeter of the cap. The line energy decreases as a function of $n$ for larger $n$. The curve has the standard nucleation-and-growth form. Under assembly equilibrium conditions $\mu=\mu^*$ the critical nucleus is the half-formed shell with the size of the critical nucleus decreasing with increasing levels of supersaturation.

\subsection{Spanning Trees}
\begin{figure}[htbp]
\begin{center}
\includegraphics[width=3in]{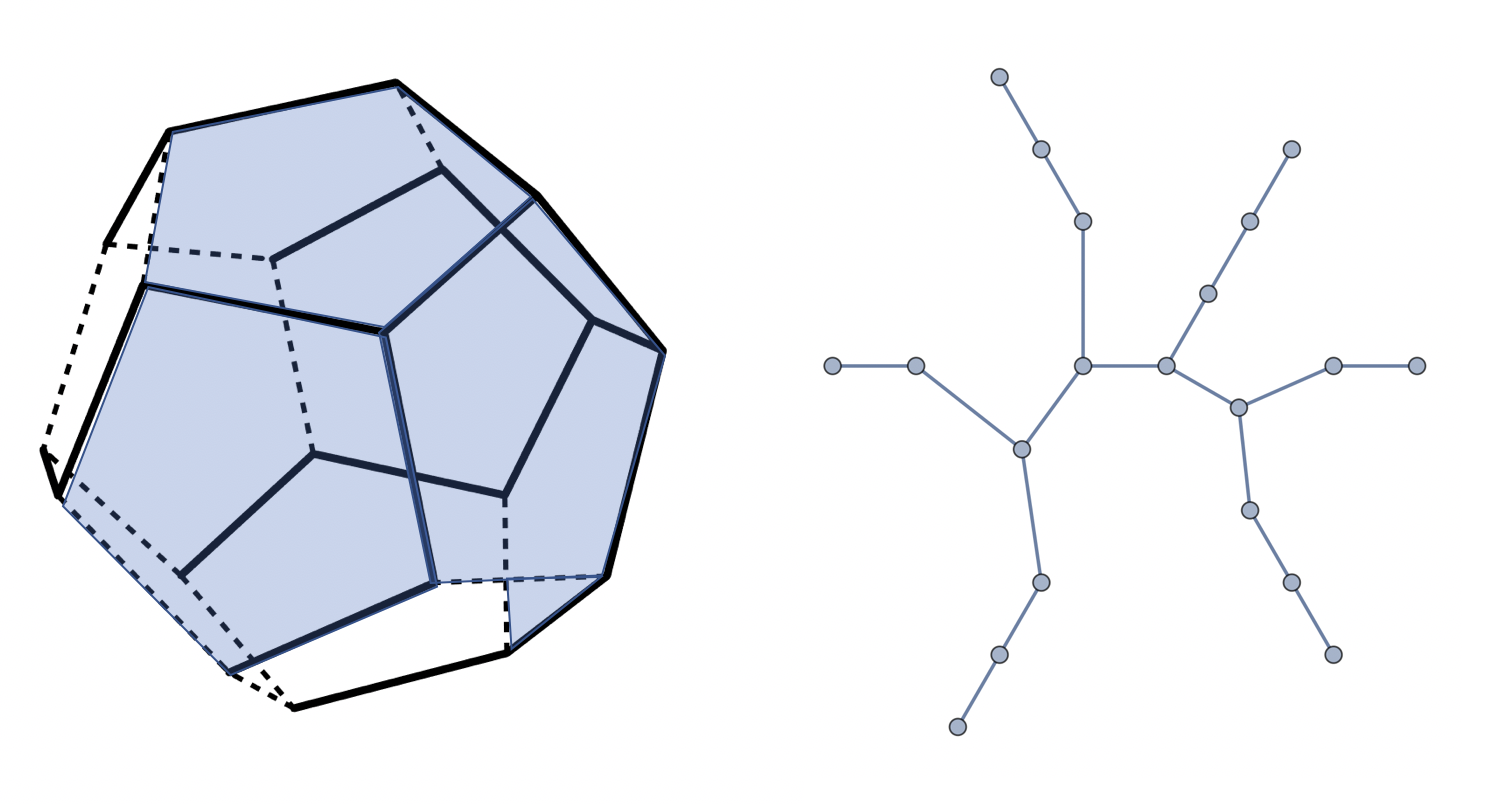}
\caption{Left: Spanning tree ``molecule (1)" connecting the vertices of a dodecahedron (solid lines). The dashed lines indicate edges of the dodecahedron that are not part of the spanning tree. Six pentamers, shown in blue, can be placed on the dodecahedron with each pentamer wrapped by four links of the tree. Right: Planar graph of the same spanning tree.}
\label{fig:dodtrees}
\end{center}
\end{figure}

The second part of the definition of the model is the specification of the section of the encapsidated RNA molecule that interacts with the capsid proteins. The secondary structure of this outer section is represented by a tree graph. Tree graphs are collections of nodes connected by links such that there is one and only one path of links connecting any pair of nodes \cite{Bollobas} (see Fig.\ref{fig:dodtrees}, right). The nodes represent in our case specific interaction sites between an encapsidated RNA molecule and the interior of the dodecahedral capsid.  A \textit{spanning tree graph} of a polyhedron is defined as a tree graph whose nodes are located on the vertices of the polyhedron with just enough links to connect the nodes together in a tree structure \cite{Graham}. The outer section of the packaged RNA molecule is, in the model, assumed to be a spanning tree of the dodecahedron. Each spanning tree of the dodecahedron has the same number of vertices (twenty) and the same number of links (nineteen). A spanning tree thus covers only nineteen of the thirty edges of the dodecahedron. The remaining edges of the dodecahedron represent parts of the capsid that associate with the genome only through non-specific interactions. It should however be emphasized that the spanning tree model is not supposed to be a realistic representation of any particular virus.

Figure \ref{fig:dodtrees} (left) shows an example spanning tree graph of the dodecahedron. The projection of this spanning tree on the plane is shown on the right (we will refer to this example as ``molecule (1)"). 
Now place pentamers on the spanning tree. Each pentamer interacts with five edges of the dodecahedron. From the geometry of the spanning tree it is easy to see that a pentamer can only interact with four links of a spanning tree. If interaction of pentamer edges with the links of the spanning tree is energetically favorable -- as we will assume -- then a cluster of pentamers minimizes the interaction free energy between pentamers and spanning tree by maximizing the number of pentamers in contact with four links of the spanning tree. The figure shows (in blue) that a maximum of six pentamers can be positioned in this fashion on the spanning tree.

\subsection{Classification of Spanning Tree Molecules.}
 
The set of spanning trees that leave the dodecahedron invariant (modulo symmetry operations of the dodecahedron) has about $10^5$ members. A topological characteristic that has been previously applied to viral RNA in terms of packaging efficiency is the \textit{Maximum Ladder Distance} (or MLD) \cite{yoffe2008, fang}. This is the maximum number of paired RNA nucleotides separating any two nucleotides. The MLD of a spanning tree graph is defined as the maximum number of links separating any pair of nodes. In graph theory, the ladder distance between two nodes of a tree graph is called the \textit{distance} while the MLD is known as the \textit{diameter} of a tree graph \cite{Bollobas}. The MLD of the tree molecule shown in Fig.\ref{fig:dodtrees} is nine. It can be demonstrated that the smallest possible MLD for a spanning tree of the dodecahedron is nine (as shown in Appendix \ref{app:A}), while the largest possible MLD of a spanning tree is nineteen. Minimum MLD spanning trees resemble Cayley trees while maximum MLD spanning trees are \textit{Hamiltonian Paths}, i.e., walks without self-intersection that visit all vertices of a polyhedron \cite{Rudnick2005, Dykeman2013b}. It has been shown that, in the absence of interactions, the solution radius of gyration of a branched polymeric molecule increases with the MLD as a power law \cite{gutin}. A systematic comparison between the genomic RNA molecules of RNA viruses confirms that they have significantly lower MLDs than randomized versions of the same molecules \cite{yoffe2008, fang}. 

\begin{figure}[htbp]
\begin{center}
\includegraphics[width=3in]{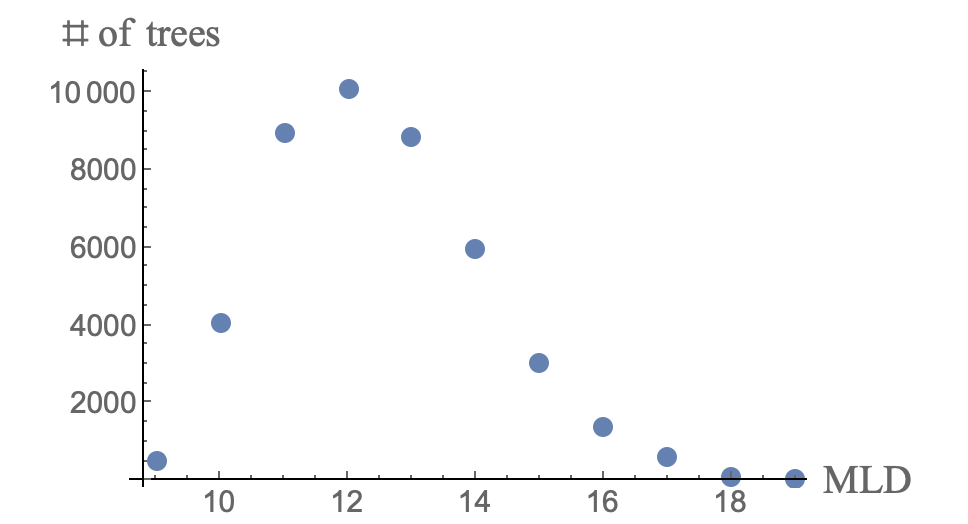}
\caption{The number of spanning trees on the dodecahedron as function of the maximum ladder distance (MLD). Two spanning trees of the dodecahedron that are related by a symmetry operation of the dodecahedron are treated as the same. }
\label{fig:dodlogplot}
\end{center}
\end{figure}

Figure \ref {fig:dodlogplot} is a plot of the number of spanning trees of the dodecahedron as a function of the MLD.
The plot has a pronounced maximum around MLD twelve. By comparison, the configurational entropy of an annealed branched polymer composed of nineteen monomers that is not constrained to be a spanning tree depends on the MLD as $19-MLD^2/19$ \cite{gutin} and has a maximum at the smallest possible MLD. It follows that the demand that a tree molecule with a certain number of links is also a spanning tree of a dodecahedron greatly constrains the branching statistics.

Could the MLD fully characterize a spanning tree? Figure \ref{fig:HP} shows a linear molecule with nineteen links, so the MLD is equal to nineteen.  In the form of a nineteen step Hamiltonian walk, it also is a spanning tree: 
\begin{figure}[htbp]
\begin{center}
\includegraphics[width=2.0in]{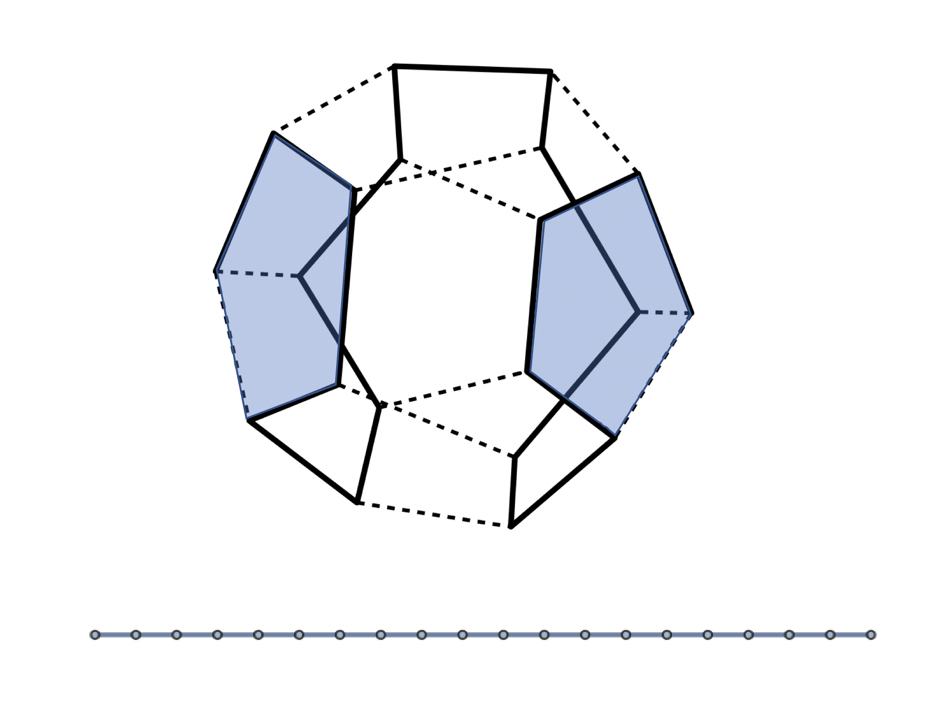}
\caption{A possible Hamiltonian path for a linear genome molecule (``molecule (2)''). Only two of the pentamers can be maximally wrapped.}
\label{fig:HP}
\end{center}
\end{figure}
For this ``molecule (2)", only two pentamers can be placed on the dodecahedron. After enumerating all possible Hamiltonian paths, one finds that the wrapping numbers of a Hamiltonian path can be two, three, or four (see Fig.\ref{fig:NP_MLD}) yet all these spanning trees have the same MLD. This indicates that the MLD does not suffice as a useful characteristic of a spanning tree.  

A characteristic that, unlike the MLD, is sensitive not only to the topology but also to the folding geometry of a spanning tree is the \textit{wrapping number} (denoted by $N_P$). The wrapping number of a spanning tree of the dodecahedron is the maximum number of pentamers that can be placed on the spanning tree with four of its edges wrapped by a link of the spanning tree. For the spanning tree shown in Fig.\ref{fig:dodtrees}, $N_P$ equals six. The maximum $N_P$ for a spanning tree of the dodecahedron is eight while the minimum is two. The distribution of wrapping numbers is shown in Fig.\ref{fig:NP}:
\begin{figure}[htbp]
\begin{center}
\includegraphics[width=3in]{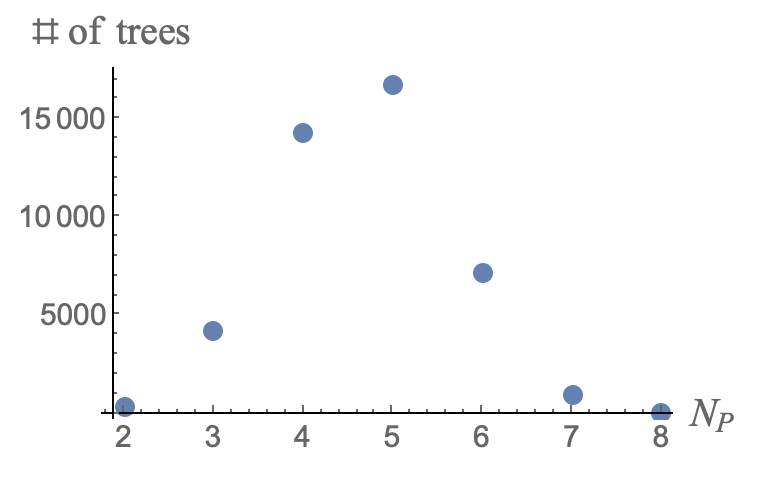}
\caption{The number of spanning trees on the dodecahedron as a function of the wrapping number. Two spanning trees of the dodecahedron that are related by a symmetry operation of the dodecahedron are treated as the same. }
\label{fig:NP}
\end{center}
\end{figure}
The wrapping number distribution has a maximum at $N_P=5$. A \textit{compact} cluster of pentamers that are maximally wrapped by the spanning tree would be a candidate of a nucleation complex with a low energy activation barrier. However, the wrapping number does not take compactness into account. In particular, it does not distinguish whether maximally wrapped pentamers have shared links (Fig.\ref{fig:HP} is an example of maximally wrapped pentamers with no shared edges). The wrapping number also does not seem to be a sufficient characteristic to predict whether a spanning tree has a low assembly energy activation barrier.

Are the wrapping number and MLD characteristics independent? Figure \ref{fig:NP_MLD} is a plot of the range of allowed wrapping numbers for given MLD: 
\begin{figure}[htbp]
\begin{center}
\includegraphics[width=3in]{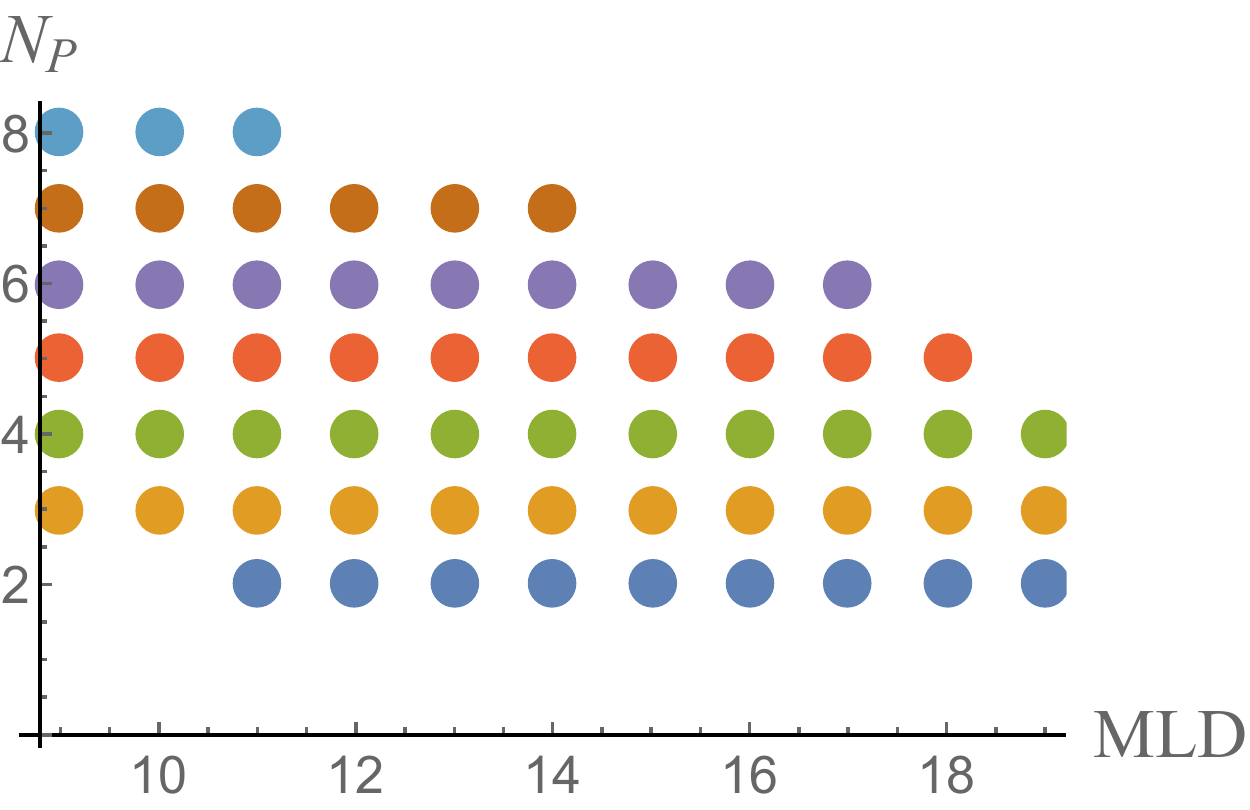}
\caption{Plot of the range of wrapping numbers ($N_P$) of the spanning tree molecules of the dodecahedron as a function of the maximum ladder distance (MLD).}
\label{fig:NP_MLD}
\end{center}
\end{figure}
There is a correlation between the two characteristics. For the largest MLDs where the wrapping number is restricted to be between two and four this correlation is strong while it is weaker for the smallest MLDs when the wrapping number can adopt nearly its full range. 


\subsection{Assembly Energy Profiles.}

Next we construct the minimum free energy assembly pathways and energy profiles. The initial state will be a spanning tree molecule folded over the edges of a mathematical dodecahedron with no pentamers. Physically, this starting state can be viewed as representing a folded or pre-condensed form of the viral ssRNA genome molecule(s) \footnote{In actuality, condensation of the RNA genome molecules takes place \textit{during} encapsidation.}. It will be assumed that different spanning trees have the same folding energy prior to assembly so the spanning trees have the same \textit{a-priori} probability. Next, pentamers are placed on the dodecahedron, one after the other. The free energy of a cluster of $n$ pentamers associated with spanning tree $i$ is defined as: 

\begin{equation}
E^i(n) = n_1\epsilon_1 + n_2\epsilon_2+n_3\epsilon_3 + n_4\epsilon_4-n\mu
\end{equation}

Here, $n_1$ is the number of links of the spanning tree that lie along a pentamer edge that is not shared with another pentamer; $n_2$ is the number of spanning tree links that lie along a pentamer edge that is shared with another pentamer; $n_3$ is the number of edges shared between two pentamers that are not covered by a spanning tree link and finally $n_4$ is the number of pentamer edges that is not shared with another pentamer and that also is not associated with a spanning tree link. The associated energies are given as $\epsilon_i$ with $i =1,2,3,4$. An energy scale will be used in which the energy $\epsilon_4$ of a bare edge is equal to zero. Energies are expressed in units of an overall energy scale $\epsilon_0$, itself expressed in units of the thermal energy $k_BT$. In these units, the interaction energy $\epsilon_3$ between two pentamer edges is equal to $-1$ while the physical meaning of $-\epsilon_1$ is that of the ratio of the affinity of a spanning tree link with a pentamer edge over the pentamer-pentamer affinity. Next, interactions between edges and spanning tree links are assumed to be additive with $\epsilon_2=-1+2\epsilon_1$. This leaves $\epsilon_1$, $\mu$, and $\epsilon_0$ as remaining free parameters. The total assembly energy of a complete particle is equal to $\Delta E = 19\epsilon_2 -11-12 \mu$ for all spanning trees. The model is thus constructed so there is no thermodynamic bias for the different spanning tree so packaging selectivity is necessarily kinetic in origin. 

\subsubsection{Small $|\epsilon_1|$}

\begin{figure}[htbp]
\begin{center}
\includegraphics[width=3.5in]{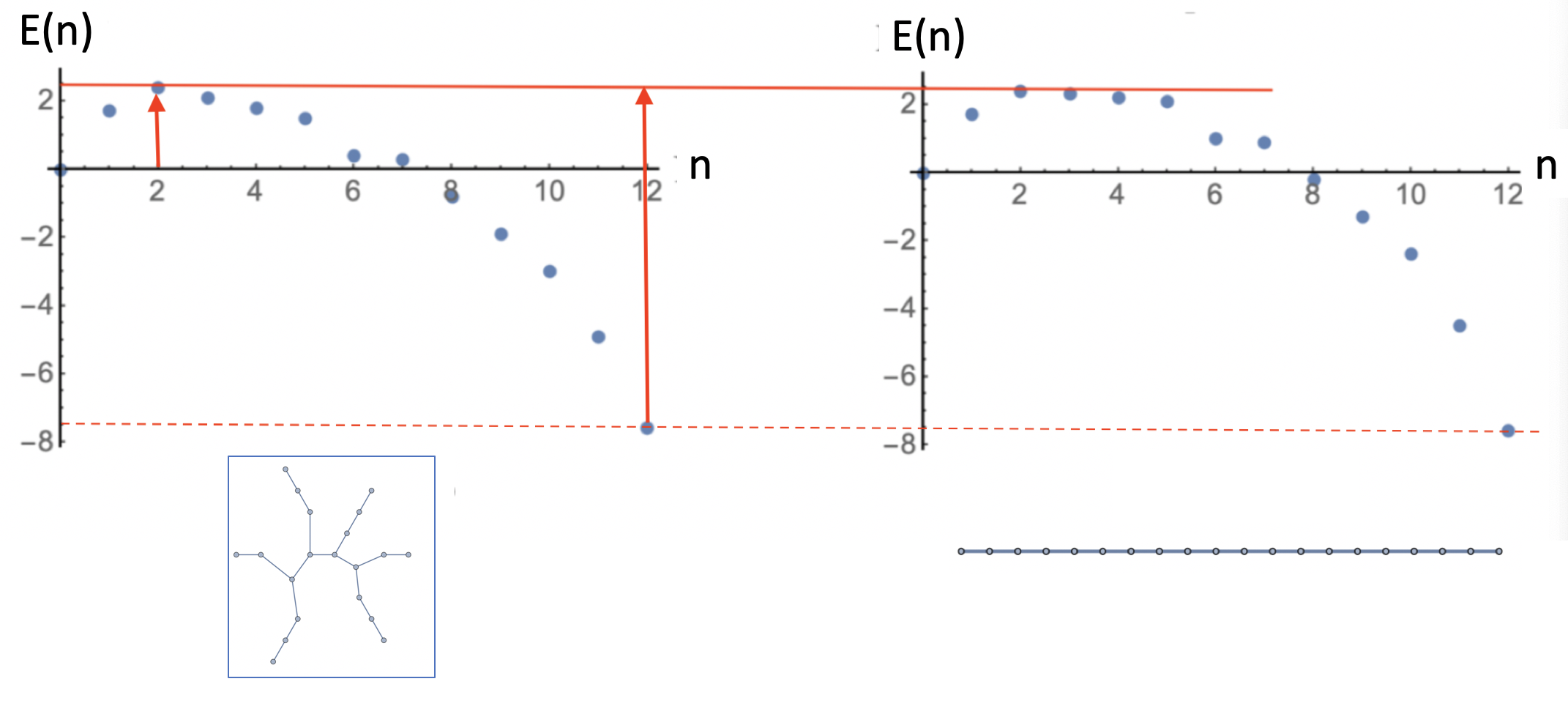}
\caption{Minimum free energy assembly pathways for the $MLD=9$, $N_P=6$ spanning tree of Fig.\ref{fig:dodtrees} (left, molecule(1)) and the $MLD=19$, $N_P=2$ spanning tree of Fig.\ref{fig:HP} (right, molecule (2)) at the reference concentration. Energy parameters are $\epsilon_1=-0.2$, $\epsilon_0=1$, and $\mu=-2.5$. The total assembly energy is indicated by a dashed red line while the assembly activation barrier by a solid red line. The red arrows indicate the activation energy for assembly (left) and disassembly (right).} 
\label{example2}
\end{center}
\end{figure}

Minimum free energy assembly pathways are constructed in the same way as before. Figure \ref{example2} compares the minimum free energy assembly profiles for the $MLD=9$, $N_P=6$ tree of Fig.\ref{fig:dodtrees} (from here on referred to as ``molecule (1)") and for the $MLD=19$, $N_P=2$ Hamiltonian walk of Fig.\ref{fig:HP} (from here on referred to as ``molecule (2)") for $\epsilon_1=-0.2$, so in the regime of relatively low affinity of RNA for capsid proteins 
Both energy profiles are similar to that of an empty capsid (i.e., the case $\epsilon_1=0.0$) and also resemble each other, though the two molecules have very different structures. The total assembly energies (solid red lines) are the same by construction but the heights of the assembly energy activation barriers also are practically the same. The same is true for the energy barrier for \textit{disassembly} separating the n=12 and n=2 states (right red arrow, about 10.0). The main difference is that the width of the energy barrier for molecule (2) is somewhat larger than that of molecule (1). If the assembly kinetics is viewed as a random walk over the energy landscapes of Fig.\ref{example2} then molecule (1) is expected to have somewhat faster assembly kinetics than molecule (2).  

\subsubsection{Large $|\epsilon_1|$.}

\begin{figure}[htbp]
\begin{center}
\includegraphics[width=3.5in]{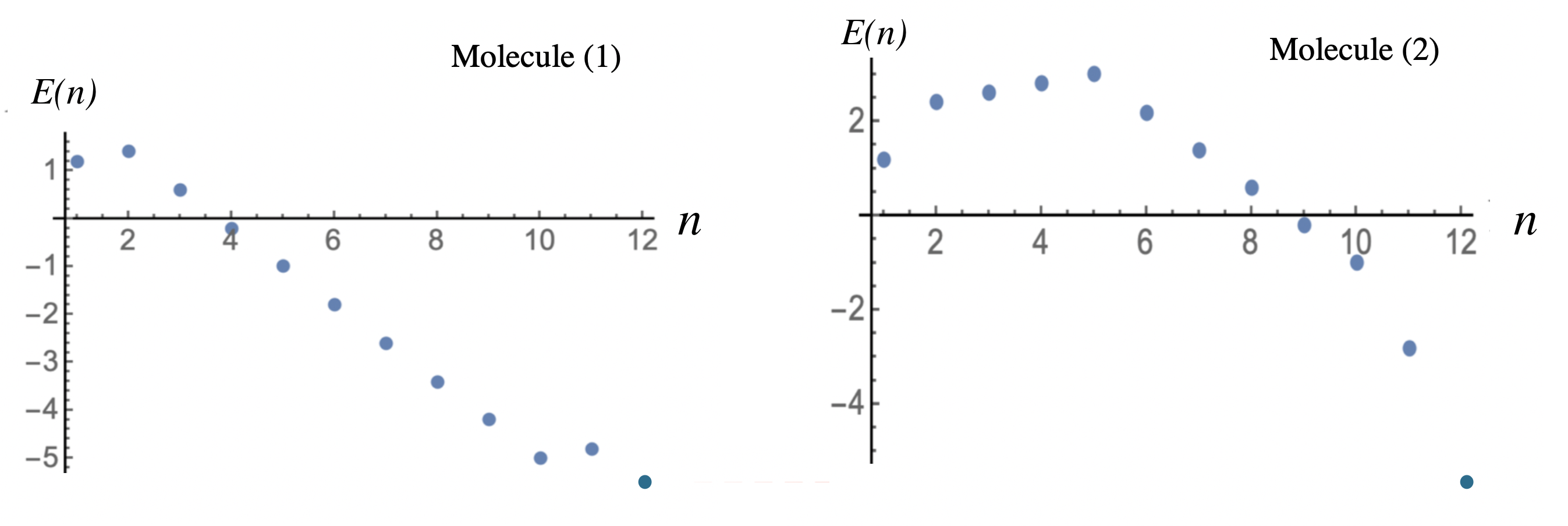}
\caption{Assembly energy profiles for molecule (1) (left) and molecule (2) (right) for energy parameters $\epsilon_1=-1.0$, $\epsilon=-1$, and $\mu=-5.2$}
\label{RSN}
\end{center}
\end{figure}

\begin{figure}[htbp]
\begin{center}
\includegraphics[width=2.2in]{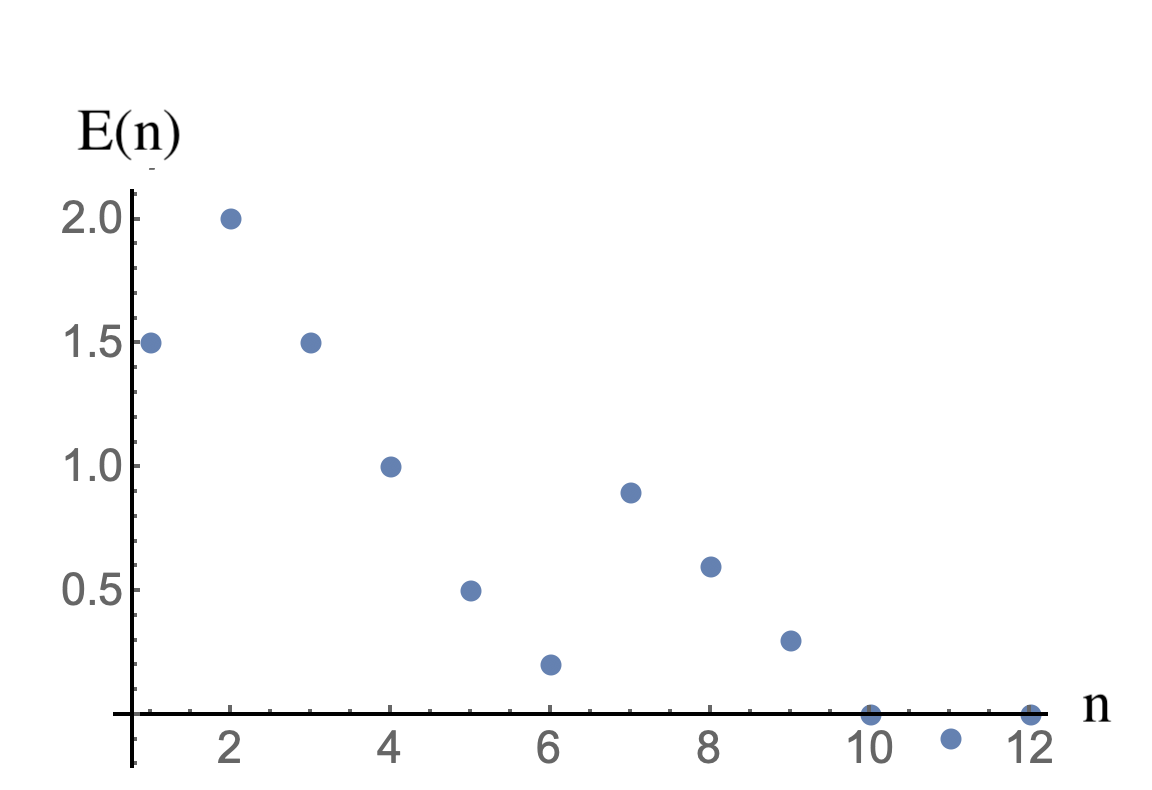}
\caption{Assembly energy profile of molecule (1) for $\epsilon_1=-1.2$ (bottom) under conditions of assembly equilibrium.}
\label{fig:dis3}
\end{center}
\end{figure}

Figure \ref{RSN} shows the energy profiles of molecules (1) and (2) for $\epsilon_1=-1.0$, corresponding to a higher RNA/protein affinity. The value of $\mu$ was decreased from $-2.5$ to $-5.2$ in order to keep the activation energy barrier in the same range as for $\epsilon=-0.2$. 
The assembly activation energy of molecule (1) is more than twice as small as that of molecule (2) while the energy profile has developed a second minimum at $n=10$. Metastable intermediate states are familiar from experimental studies of viral assembly \cite{Parent2006, Tuma2008, Basnak2010} as well as from numerical simulations \cite{Johnston2010, Hagan2011, Baschek2012, Perlmutter2015b}. Known as ``kinetic traps", they may retard or block assembly. Intermediate metastable states become progressively more pronounced as the chemical potential $\mu_0$ is reduced towards assembly equilibrium as illustrated in Fig.\ref{fig:dis3} for the case of $\epsilon_1=-1.2$ of molecule (1)  
The energy profile has acquired a deep metastable minimum at $n=6$, which is the exact opposite of what is expected from a nucleation-and-growth scenario. Moreover, the minimum energy state is no longer the $n=12$ assembled viral particle but rather the defected $n=11$ state. In short, for larger ratios of the pentamer-tree molecule and pentamer-pentamer affinity, the nucleation-and-growth scenario breaks down. 

\subsubsection{Partial Assemblies and Structural Transitions.}

\begin{figure}[htbp]
\begin{center}
\includegraphics[width=2.0in]{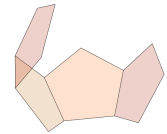}
\caption{The minimum energy $n=4$ assembly state of molecule (2) for $\epsilon_1=-1.2$.}
\label{fig:F}
\end{center}
\end{figure}

For small values of $|\epsilon_1|$, the structure of a partial assembly for given $n$ is the same as that of the empty capsid, as shown shown in Fig.\ref{fig:Zlotnick model}. However, for larger $|\epsilon_1|$ the first $N_p$ pentamers will be placed on maximum wrapping sites. It follows that partial assemblies are expected to undergo a \textit{structural transition} as a function of $|\epsilon_1|$. For molecule (1) this structural transition has secondary effects. For small $|\epsilon_1|$, a six-pentamer clusters has five-fold symmetry with one central pentamer sharing its five edges with five other pentamers that share three edges with their neighbors (see Fig.2). For large $|\epsilon_1|$ the minimum energy $n=6$ cluster has its six pentamers placed on the six available maximum wrapping sites of an $N_P=6$ spanning tree shown in Fig.4. By moving only one pentamer, this structure can be transformed into the small $|\epsilon_1|$ $n=6$ structure with five-fold symmetry. The transition takes place at $\epsilon_1=-1$. However, for molecule (2) with $N_p=2$, the transition is more dramatic. The $n=4$ pentamer cluster of an empty caspid has a two-fold symmetry axis (see Fig.2). Only two pentamers can be maximally wrapped (see Fig.6). The full $n=4$ minimum energy structure is shown in Fig.\ref{fig:F}

This structural transition has an interesting feature. If one would relax the constraints that keep the links of the genome molecule associated with the edges of a mathematical dodecahedron, allow tree links to rotate freely around the nodes of the tree, and finally allow pentamers to swivel around shared edges then the pentamers of the empty-capsid partial assemblies of Fig.2 would -- for $n>2$ -- not be able to move with respect to each other without breaking pentamer-pentamer bonds. The empty-capsid partial assemblies can be said to be \textit{mechanically rigid}. The same is true as well for the $n=6$ structure of Fig.4. However, this is not the case for the four-pentamer structure shown in Fig.\ref{fig:F}. If this structure were released from the dodecahedron, and allowed to fluctuate freely, then the four pentamers could freely swivel along the three shared edges. We will call such an assembly \textit{flaccid}. Transitions from rigid to flaccid as a function of $\epsilon_1$ become are common for larger values of the MLD. We would expect that actual flaccid RNA/capsid protein assembly intermediates free in solution would be characterized by strong conformational thermal fluctuations.  

\section{Boltzmann Distribution.}
In the configuration space of minimum energy partial assemblies, the Boltzmann distribution for the pentamer occupation probabilities $P^i_n$ of spanning tree $i$ is
\begin{equation}
P^i_{n}\propto\exp-E^i(n)
\label{BD}
\end{equation}
By substitution, it can be checked that
\begin{equation}
\frac{c_f^{12} r_0^i}{c_{12}^i}=K
\label{LMA}
\end{equation}
with $c_f$ the concentration of free pentamers, $r_0^i$ the concentration of pentamer-free spanning trees of type $i$ and $c_{12}^i=r^i P_{12}$ the concentration of assembled particles. Here, $r^i$ is the total concentration of this type of tree molecules while $r_0^i=r^i P_0$. Concentrations are all expressed in units of the reference concentration $\bar{c}$. Finally, $K=\exp E_0(12)$ where $E_0(12)$ is the particle assembly free energy for $c_f=1$. The superscript $i$ was dropped here because $E_0(12)$ is independent of $i$. This relation has the form of the \textit{Law of Mass Action} (LMA) of physical chemistry where $K$ is known as the dissociation constant. 

\subsection{Conservation Laws}
The probability distribution $P^i_{n}$ is related to conservation laws for the number of pentamers and of tree molecules. Conservation of tree molecules requires that $r_0^i+\sum_{n=1}^{12} c_n^i = r^i$ where $c_n^i=r^i P_{n}$ is the concentration of partial assemblies with $n$ pentamers. This conservation law is assured if the probabilities sum to one for each type of tree molecules, i.e., if  $\sum_{n=0}^{12} P^i_{n} = 1$. Next, conservation of pentamer molecules is assured if $c_f=c_0 \Gamma\thinspace([P])$ with $c_0$ the total pentamer concentration and with
 \begin{equation}
\Gamma\thinspace([P]) \equiv 1 - (D/12)\sum_i\sum_{n=1}^{12} n P^i_{n} 
\label{eq:gammadef}
\end{equation}
the probability that a pentamer is free in solution. The quantity $D=12 r_0/c_0$ plays the role of the \textit{mixing ratio}, i.e., the ratio of the total number of RNA molecules over the total number of capsid proteins. Here, $r_0=\sum_i r^i$ is the total concentration of tree molecules. If $D=1$ then there are exactly enough pentamers to encapsidate all tree molecules. At first sight it would seem that $D=1$ should awlays be the best mixing ratio if one wanted to minimize the number of unpackaged RNA molecules and free pentamers at the end of the assembly process. D=1 is similar to the \textit{stoichiometric ratio} of physical chemistry. Because the concentration of free pentamers $c_f=c_0 \Gamma\thinspace([P])$ depends on the Boltzmann distributions of all types of tree molecules, the different distribution are coupled.

\subsection{Self-Assembly Diagram.}

\begin{figure}[htbp]
\begin{center}
\includegraphics[width=3in]{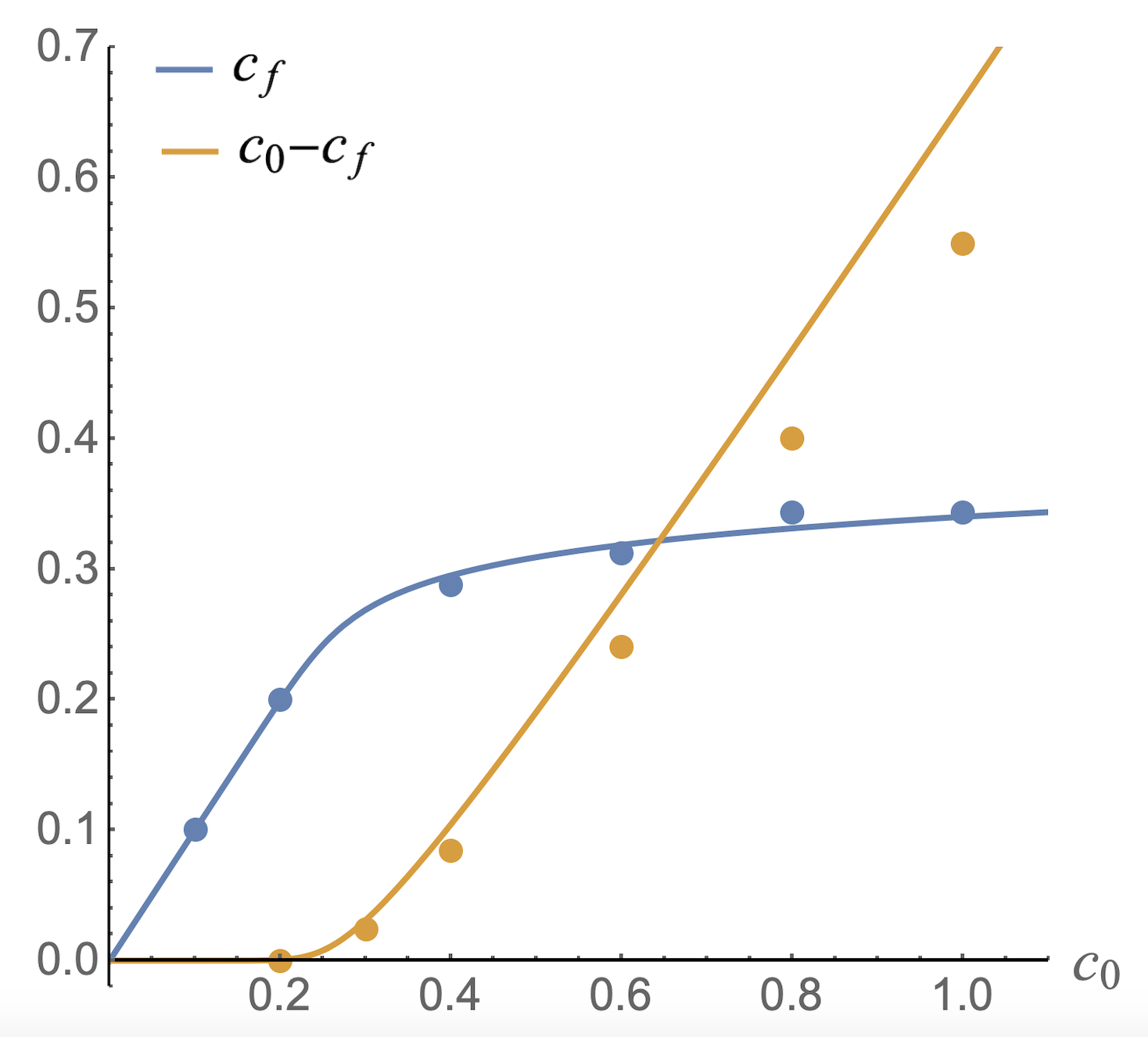}
\caption{Equilibrium self-assembly diagram for molecule (1) for $\epsilon_1=-0.2$, $\epsilon_0=1$, $\mu=-2.5$, and $D=1$. Horizontal axis: total pentamer concentration $c_0$. Vertical axis: either the free pentamer concentration $c_f$ (blue) or the concentration $c_0-c_f$ of pentamers that are associated with a tree molecule (ochre). Solid lines: solution of Eq.\ref{C1}}
\label{fig:EA}
\end{center}
\end{figure}

A standard diagnostic plot for self-assembly processes are plots of the concentration of free building blocks and of assembled particles as a function of the total concentration of building blocks \cite{safran}. Such a plot was computed from the Boltzmann distribution Eq.\ref{BD}, assuming that only one type of tree molecule as in solution, (molecule (1), with the energy parameters of Fig.\ref{example2}. The result is shown in Fig.\ref{fig:EA}. The dots show the concentrations of free pentamers in solution and of pentamers that are part of an assembled particle, both as a function of the total pentamer concentration $c_0$.
For low pentamer concentrations, nearly all pentamers are free in solution so the concentration of free pentamers is close to $c_0$. As $c_0$ increases, the concentration of free pentamers stops increasing and then saturates. The concentration of pentamers that are part of an assembled particle now starts to increase proportional to $c_0$. The transition point, which is around $c_0=0.2$ between these two regimes is known in the soft-matter physics literature as the \textit{critical aggregation concentration} (or CAC) \cite{safran}.

The intermediate occupation probabilities $P_n$ with $2\leq n\leq11$ had low values for this parameter setting. If one entirely neglects these intermediates then the conservation law for genome molecules reduces to $P_{0} \simeq (1-P_{12})$ and that of pentamers to $c_f \simeq c_0(1-DP_{12})$. Inserting these two relations into the LMA equation Eq.\ref{LMA} produces an closed form expression for the concentration $c_f$ of free pentamers:
\begin{equation}
\left(\frac{c_f}{c_0}\right)^{12}\left(\frac{D-1+\frac{c_f}{c_0}}{1-\frac{c_f}{c_0}}\right)=\left(\frac{K}{c_0^{12}}\right)
\label{C1}
\end{equation}
Importantly, like the LMA, this relation is independent of the type of tree molecules (Explicit forms in certain limiting regimes are given in Appendix \ref{app:B}). Figure \ref{fig:EA} compares Eq.\ref{C1} (solid lines) with the numerical solution of the thirteen coupled Boltzmann equations (dots) that does allow for assembly intermediates.

\subsection{Phase-Diagram.}

\begin{figure}[htbp]
\begin{center}
\includegraphics[width=3.5in]{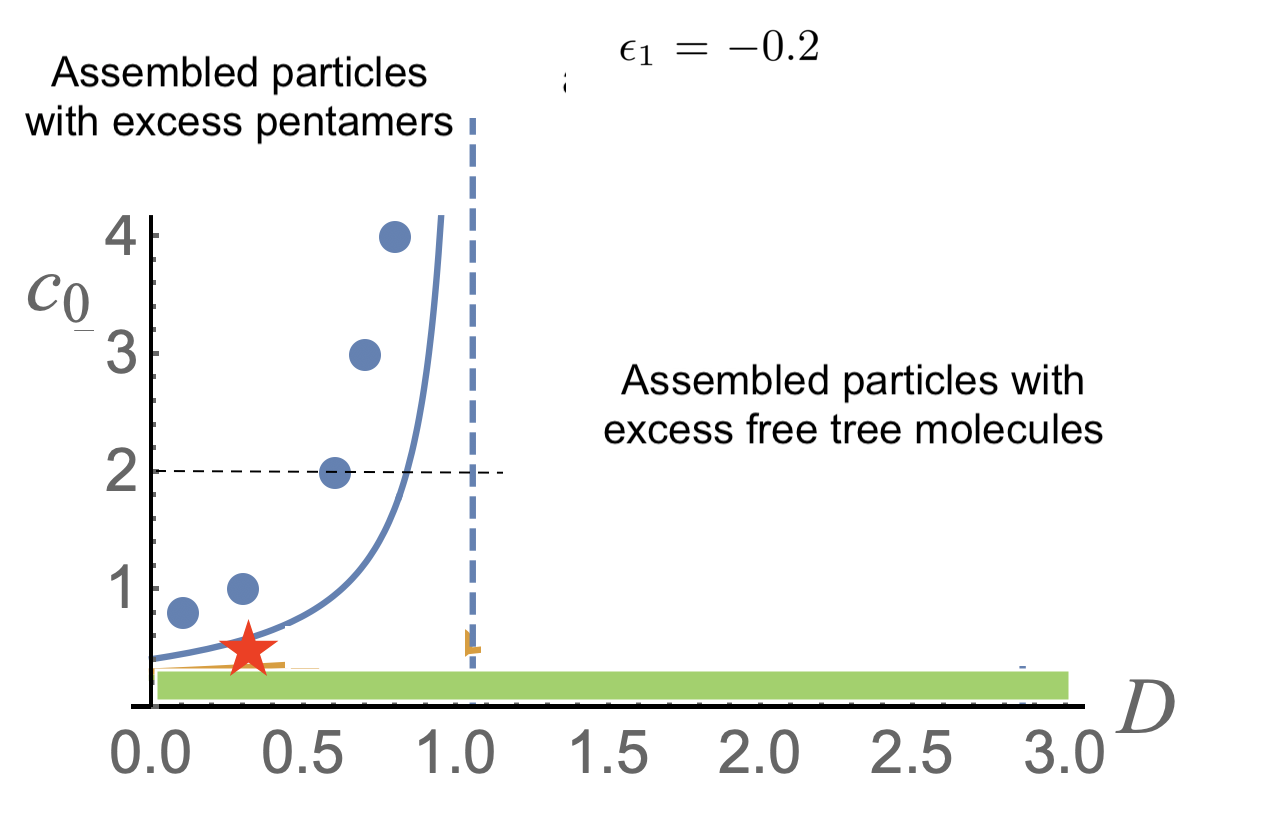}
\caption{Quasi phase-diagram for $\epsilon_1=-0.2$, $\epsilon_3=-1$, and $\mu_0=-2$. Horizontal axis: Depletion factor $D$. Vertical axis: Pentamer concentration $c_0$. Blue dots: points where 95 percent of the genome molecules have been packaged according to the Boltzmann distribution. Solid blue line: same as computed from Eq.\ref{contour}. In the green sector there is practically no capsid assembly. The red star marks a possible operating point for viral assembly inside infected cells.}
\label{fig:PD}
\end{center}
\end{figure}

A second way to display self-assembly measurements under equilibrium conditions is in the form of a quasi phase-diagram that shows the dominant type of assembly as a function of thermodynamic parameters \footnote{Since a virus is a system of limited size true phase transitions are not possible.} For viral assembly, the protein and RNA concentrations are a natural choice for such a phase-diagram. For the case of the spanning-tree model, we will use the pentamer concentration $c_0$ and the mixing ratio $D$ as thermodynamic parameters. The blue dots in Fig.\ref{fig:PD} show points in a $c_0$ vs $D$ diagram where 95 percent of the spanning trees of molecule (1) are fully encapsidated. 
To the right of the blue dots, most tree molecules are encapsidated and coexist with excess free pentamers. To the left of the blue dots most pentamers are part of assembled particles and coexist with excess free tree molecules. The blue dots can be viewed as optimal mixing states that leave minimal excess free pentamers of tree molecules. For high pentamer concentrations, the line of blue dots approaches $D=1$, which was the expected optimal mixing state. This is however definitely not the case for smaller values of $c_0$. 

If one neglects assembly intermediates then it can be shown from Eq.\ref{C1} that the relation $c_0(D)$ for 95 percent occupancy is a hyperbola in the $c_0-D$ plane:
\begin{equation}
c_0(D)=\frac{1}{(1-D\thinspace P_{12})}\left(\frac{K P_{12}}{1-P_{12}}\right)^{1/12}
\label{contour}
\end{equation}
with $P_(12)=0.95$. The hyperbola diverges at $D=1/P_{12}$, which is close to one for a 95 percent packaging fraction. It shifts to smaller values of $D$ as the pentamer concentration $c_0$ is reduced with $c_0(D)$ always larger than $K^{1/12}$, which is of the order of the CAC. The entropy-dominated green sector in Fig.\ref{fig:PD} is below the CAC. 

\begin{figure}[htbp]
\begin{center}
\includegraphics[width=2.5in]{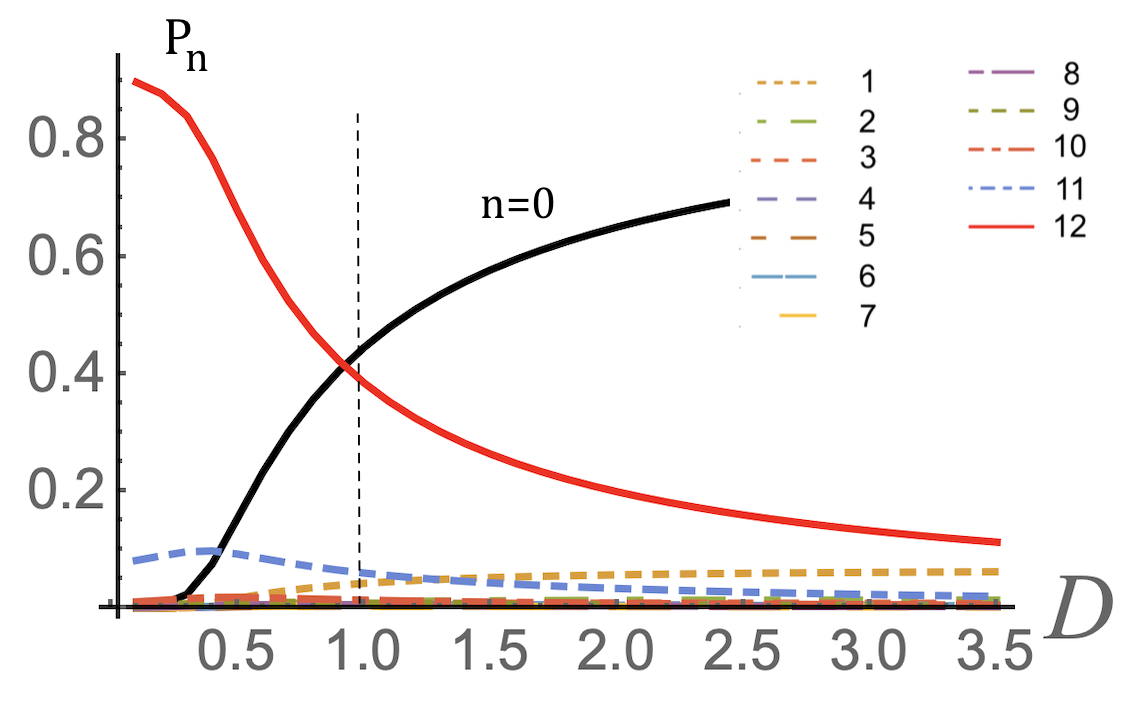}
\includegraphics[width=2.5in]{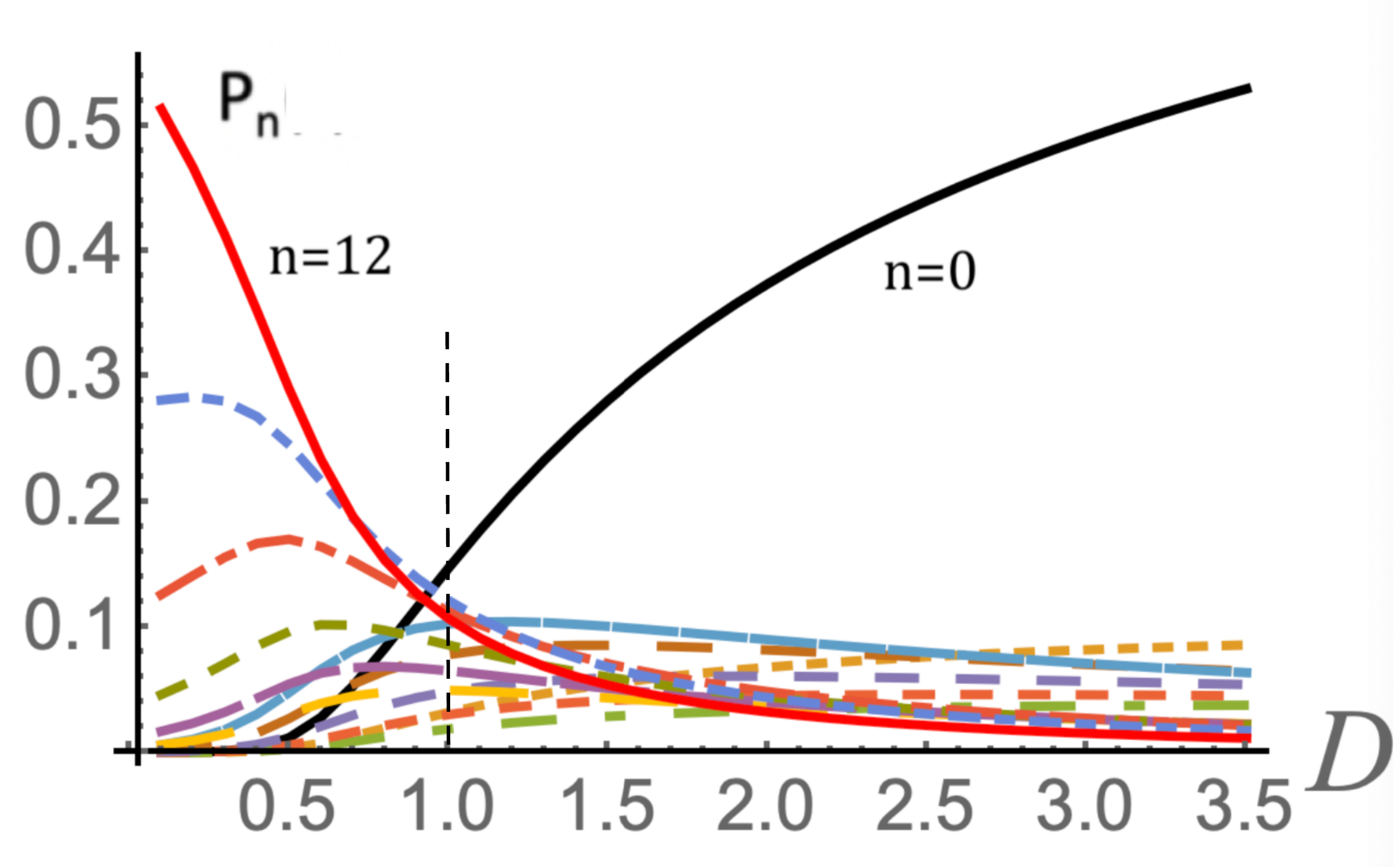}
\caption{Equilibrium occupation numbers as a function of the depletion factor $D$ for fixed $c_0=1.0$. Top: same parameters as Fig.\ref{fig:PD}. Bottom: $-\epsilon_1$ is increased from $0.2$ to $1.2$ and the reference chemical potential $\mu_0$ is reduced from $-2.5$ to $-5$ so the total assembly energy remains approximately the same. }
\label{fig:D}
\end{center}
\end{figure}

Figure \ref{fig:D}(Top) shows how the occupation probabilities change as one crosses the boundary between the pentamer-rich and genome-rich states for fixed $c_0$  
The transition region between the two states is quite broad. Note that for large $D$ the occupation probabilities of assembly intermediates start to become comparable to the occupation probability of assembled capsids. This is due to the fact that as $D$ increases, more and more pentamer binding sites become available. Breaking up assembled particles and distributing the resulting pentamers over the additional binding sites increases the system entropy. When $-\epsilon_1$ is increased to $1.2$ (see Fig.\ref{fig:D}(Bottom)), this entropic effect becomes dominant and causes partially assembled particles to overwhelm fully assembled particles for larger $D$. The structural transition around $\epsilon_1=-1$ of assembly intermediates discussed in the previous section here shows up as an \textit{order-disorder transition} for mixing ratios larger than one.

\section{Assembly and Disassembly Kinetics.}

\subsection{Master Equation.}

The kinetics of the model is defined in terms of the thirteen time-dependent occupation probabilities $P_n(t)$. We will again limit ourselves to the case of a single spanning tree and drop the subscript $i$ \footnote{The case of competing spanning trees is discussed in the paper, submitted to Plos Biocomputation.}. The assembly kinetics is assumed to obey Markov chain statistics \cite{Perkett2014}, in which case the occupation probabilities evolve in time according to the Master Equation \cite{vanKampen}:
\begin{equation}
\frac{dP_n(t)}{dt}=\sum_{m= n\pm1}\left(W_{m,n}P_m(t)-W_{n,m}P_n(t)\right)
\label{eq:ME}
\end{equation}
Here, $W_{m,n}$ is a thirteen-by-thirteen matrix of transition rates from state $m$ to state $n$. We only include transitions with $m=n \pm1$ so only with one pentamer gained or lost at a time. The matrix $W_{m,n}$ is defined in terms of a simplified form of the diffusion-limited chemical kinetics of Eq. 8.35 of ref.\cite{schulten}. The addition of a pentamer to a genome molecule associated with a pentamer cluster of size $n$ is treated as a bi-molecular reaction with a reaction rate $k_{n,n+1} {c_{n}}{c_f}$. The on-rate $k_{n,n+1}$ is given by
\begin{equation}
 k_{n,n+1}=\lambda
 \begin{cases}
 e^{-(E(n+1)-E(n))}\quad\quad\thickspace\thickspace \text{if}\thinspace \quad E(n+1) > E(n)\\
1\qquad\qquad\qquad\quad\qquad \text{if} \quad E(n+1) < E(n)
 \end{cases}
 \end{equation}
Here, $\lambda$ is a base rate that depends on molecular quantities such as diffusion coefficients and reaction radii but not on the various concentrations. If the addition of a pentamer reduces the free energy of a partial assembly then the on-rate is equal to this base rate. If there is an energy cost, then the base rate is reduced by an Arrhenius factor. Importantly, the assembly energies $E(n)$ the on-rates should be computed here with $ c_f = c_0$.

In terms of the on-rates, the rate matrix is given by $W_{n,n+1}=k_{n,n+1}{c_f}$ after canceling a common factor $r_t$. The entries of the rate matrix that correspond to adding a pentamer are then
 \begin{equation}
 W_{n,n+1}=c_0 \Gamma\thinspace([P])
 \begin{cases}
e^{-\Delta E_{n,n+1}}\qquad\thickspace\thinspace\text{ if   } \Delta E_{n,n+1}>0\\
1 \qquad\qquad\qquad\thickspace\thickspace\text{ if   }  \Delta E_{n,n+1}<0
 \end{cases}
 \end{equation}
where $\Delta E_{n,n+1} \equiv E(n+1)-E(n)$ again with $c_f = c_0$. The prefactor $\lambda$ has been absorbed in a redefinition of the time scale. As for the equilibrium case
 \begin{equation}
\Gamma\thinspace([P]) = 1 - (D/12)\sum_{n=1}^{12} n P_{n}(t) 
\label{eq:gammadef}
\end{equation}
is the probability for a pentamer to be free in solution. The off-rate entries $W_{n+1,n}$ are determined by the on-rates through the condition of detailed balance:
\begin{equation}
\frac{W_{n+1,n}}{W_{n,n+1}}=\frac{P_{n}}{P_{n+1}}
\end{equation}
where the right hand side the time-independent equilibrium Boltzmann Distribution Eq. (\ref{BD}) must be inserted. 

\begin{figure}[htbp]
\begin{center}
\includegraphics[width=3.3in]{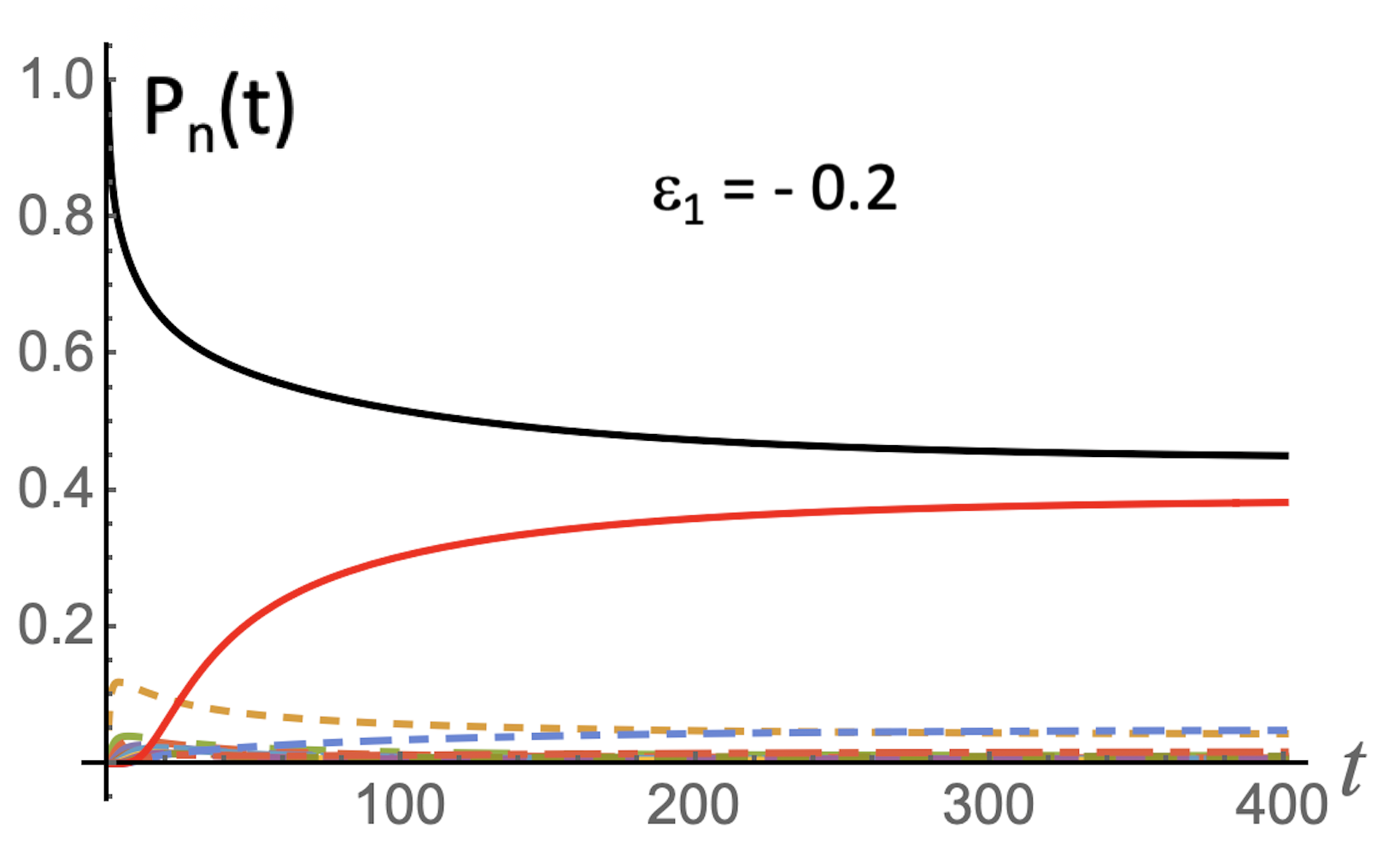}
\includegraphics[width=3.3in]{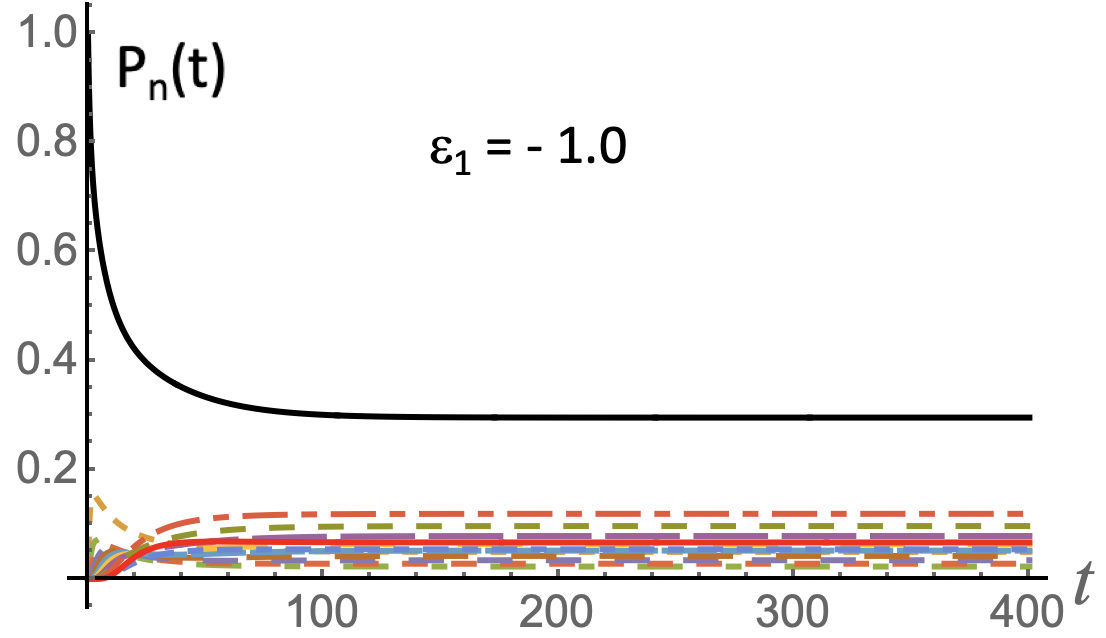}
\caption{Numerical solution of the master equation for molecule (1). Shown are the occupation probabilities $P_n(t)$. The color code is the same as that of Fig.\ref{fig:D}. Time is in dimensionless units of $1/\lambda$. Other parameter values are those of Fig.\ref{example2} with $\epsilon_1=-0.2$. Bottom: same except that $\epsilon_1=-1.0$. }
\label{ME}
\end{center}
\end{figure}

Typical examples of numerical solutions of the Master Equation are shown in Fig.\ref{ME} for molecule (1) for reference conditions $c_0=D=\beta=1$. 
The top figure shows the case where the energy parameters are those of Fig.\ref{fig:EA}), so $\epsilon_1=-0.2$. The red curve is the probability $P_{12}(t)$ for a tree molecule to be encapsidated while the black curve is the probability $P_0(t)$ for a tree molecules to be free of pentamers. The probabilities for partial assemblies are small compared to $P_{12}(t)$ and $P_{0}(t)$. Eventually, about forty percent of the tree molecules are encapsidated in assembled particles while about five percent are associated with partial assemblies. The dominant assembly pathway is the same as the one shown in Fig.\ref{fig:Zlotnick model}. The occupation probabilities exponentially approach constant values at late times that agree with the Boltzmann distribution. The bottom figure shows what happens when the strength of the interaction between pentamers and tree molecules is increased, with $\epsilon_1=-1.0$. The reference chemical potential is reduced to $\mu_0=-5.2$ in order to keep the activation energy roughly the same. The role of partial assemblies becomes significantly larger. 

\subsection{Time Scales for Assembly and Disassembly.}
\begin{figure}[htbp]
\begin{center}
\includegraphics[width=3.5in]{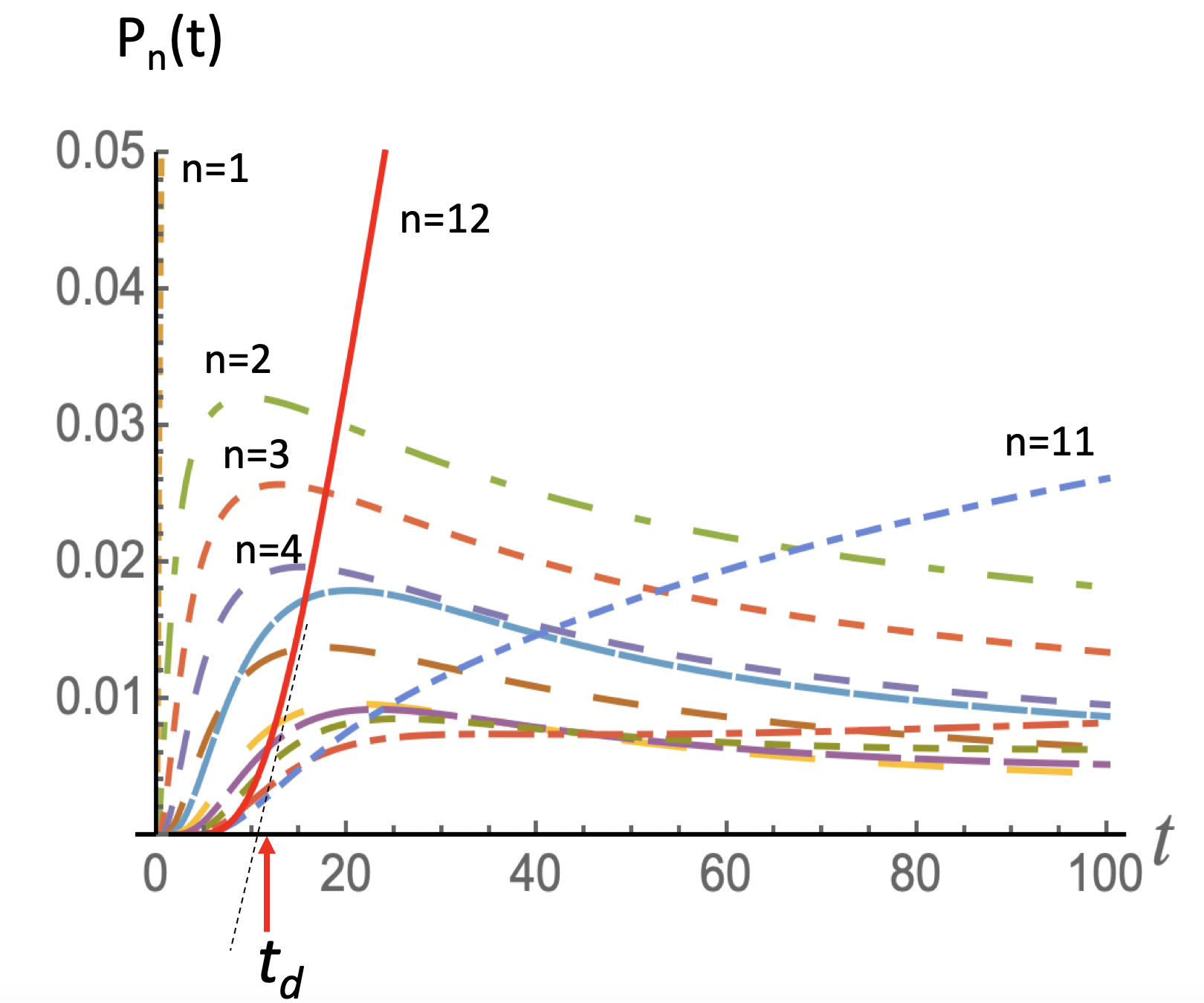}
\caption{Assembly shock-wave for molecule (1) for $\epsilon=-0.2$ in terms of the occupation probabilities $P_n(t)$. The intersection of the maximum tangent of the $P_{12}(t)$ curve with the time axis (solid black line) defines the assembly lag time $t_d$. The color coding is the same as that of Fig.17.}
\label{fig:master3}
\end{center}
\end{figure}

\begin{figure}[htbp]
\begin{center}
\includegraphics[width=3.5in]{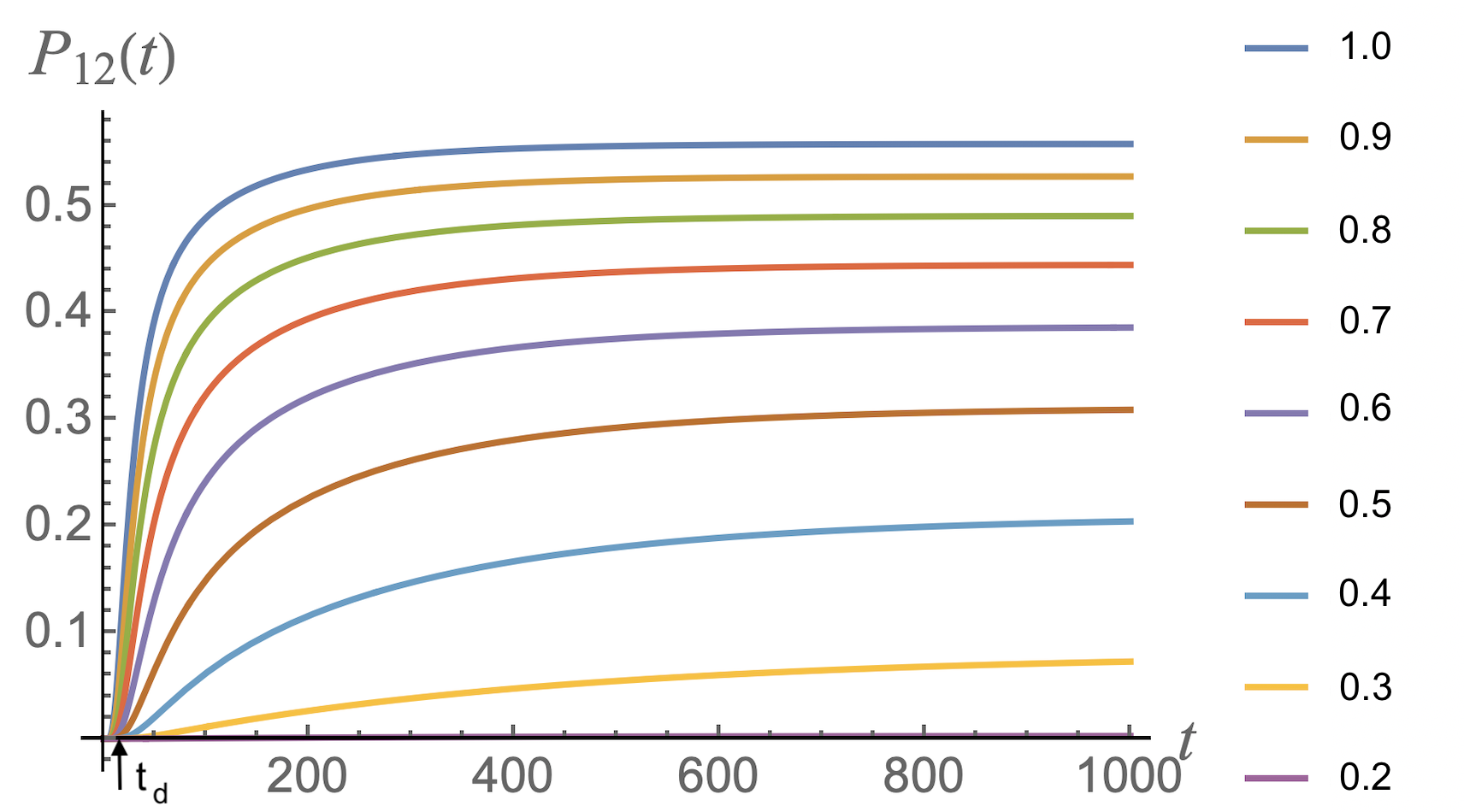}
\caption{Plots of the fraction $P_{12}(t)$ of genome molecules that are fully encapsidated for increasing total concentration for different total pentamer concentration $c_0$ ranging from 1.0 to 0.2. Particles do not form for $c_0$ less than 0.2. The arrow shows the assembly delay time. }
\label{LS}
\end{center}
\end{figure}

The next step is to determine the characteristic time scales. The occupation probabilities display a rather complex kinetics, as can seen by amplifying the scale, which is done in Fig.\ref{fig:master3}.
The functions $P_n(t)$ in Fig.\ref{fig:master3} have, for $n$ less than 12, a maximum as a function of $n$.  The peak times of the different distributions increase with $n$. This describes a type of \textit{assembly shock-wave} propagating in configuration space from small to large $n$. Similar assembly shock waves have been reported for the dynamical versions of the Zlotnick model discussed in Section II \cite{Endres2002, Morozov2009}. A characteristic \textit{assembly delay time} $t_d$ can be defined from the early time behavior of $P_{12}(t)$ as the intercept of the maximum tangent of $P_{12}(t)$ with the time axis. For the case of molecule (1), this gives $t_d\simeq11$. The delay time is roughly an average of the various peak times. Assembly delay times are a familiar feature of the assembly kinetics of empty capsids \cite{Casini2004} and of aggregation phenomena in general \cite{Wu1992}. Figure \ref{LS} shows that the assembly delay time is not strongly dependent on the pentamer concentration. 

 Figure \ref{LS} shows the effect of varying the total pentamer concentration $c_0$ on the assembly delay time. 
The assembly delay time depends only weakly on the pentamer concentration. Note that assembly stops when the pentamer concentration drops below a threshold of about $0.2$, consistent with the CAC of equilibrium thermodynamics (see Figs \ref{fig:EA} and \ref{fig:PD}). The values of $P_{12}(t)$ at late times also are  consistent with the Boltzmann distribution \footnote{Similar plots are obtained by time-dependent light scattering studies of the assembly of empty capsids with varying protein concentrations. \cite{Casini2004}.}.

It is clear from Fig.\ref{LS} that final thermal equilibration takes places on time scales that are much longer than $t_d$ and that do depend strongly on $c_0$. To compute the equilibration time, complete the definition of the transition matrix by introducing diagonal entries $W'_{n,n}=-\sum_{m\neq n}W(m,n)$. The column elements of the resulting matrix $W'_{m,n}$ add to zero. The Master Equation can be rewritten in terms of  $W'_{m,n}$ as $\frac{d\bf{P}}{dt}=\bf{W' P}$. The eigenvalues of the completed transition matrix $W'_{m,n}$ are then (minus) the decay rates of the different modes of relaxation. These modes correspond to the eigenvectors of $W'_{m,n}$. The smallest eigenvalue of $W'_{m,n}$ must correspond to the Boltzmann distribution and has to be zero. The rate of approach to final thermal equilibrium is the negative of the next smallest eigenvalue of $W'_{m,n}$. 

The largest eleven eigenvalues of $W'_{m,n}$ are all of the order of $1/t_d$. The twelfth eigenvalue $1/t_r$ is however much smaller, with $t_r\simeq644$ while the thirteenth eigenvalue is indeed consistent with zero within our numerical precision. We will refer to $t_r$ as ``the" relaxation time. The dependence of these two key time scales on the overall energy scale $\epsilon_0$ is quite different. For $\epsilon_0=3$, $t_d$ barely changes ($t_d\simeq11.7$) but the relaxation time increases exponentially to $t_r\simeq1.7\times 10^5$. Figure \ref{LS} shows that the relaxation time strongly increases as the pentamer concentration decreases.

\begin{figure}[htbp]
\begin{center}
\includegraphics[width=2.7in]{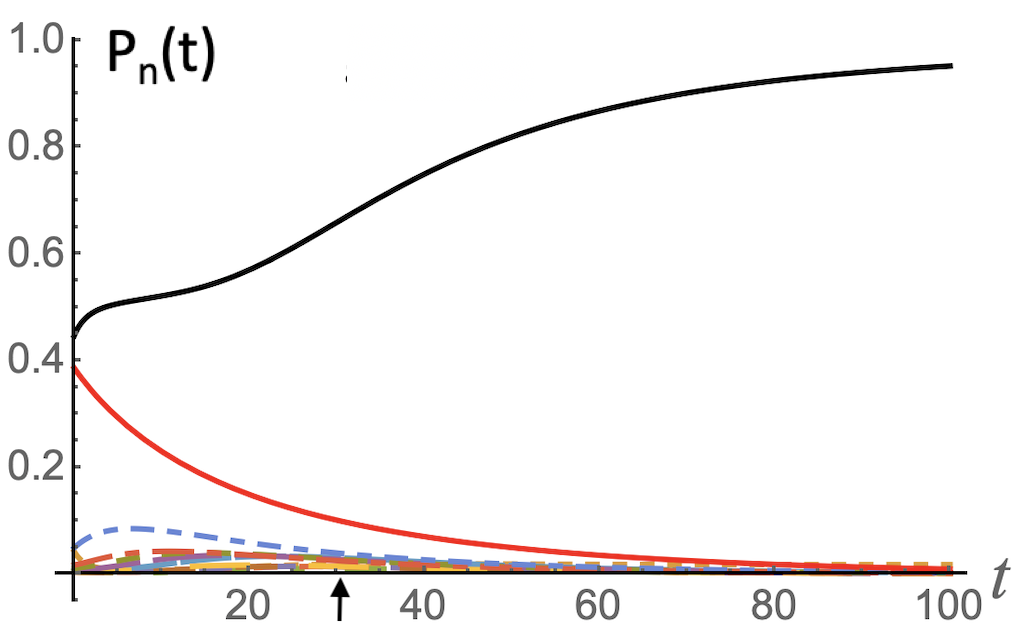}
\includegraphics[width=2.7in]{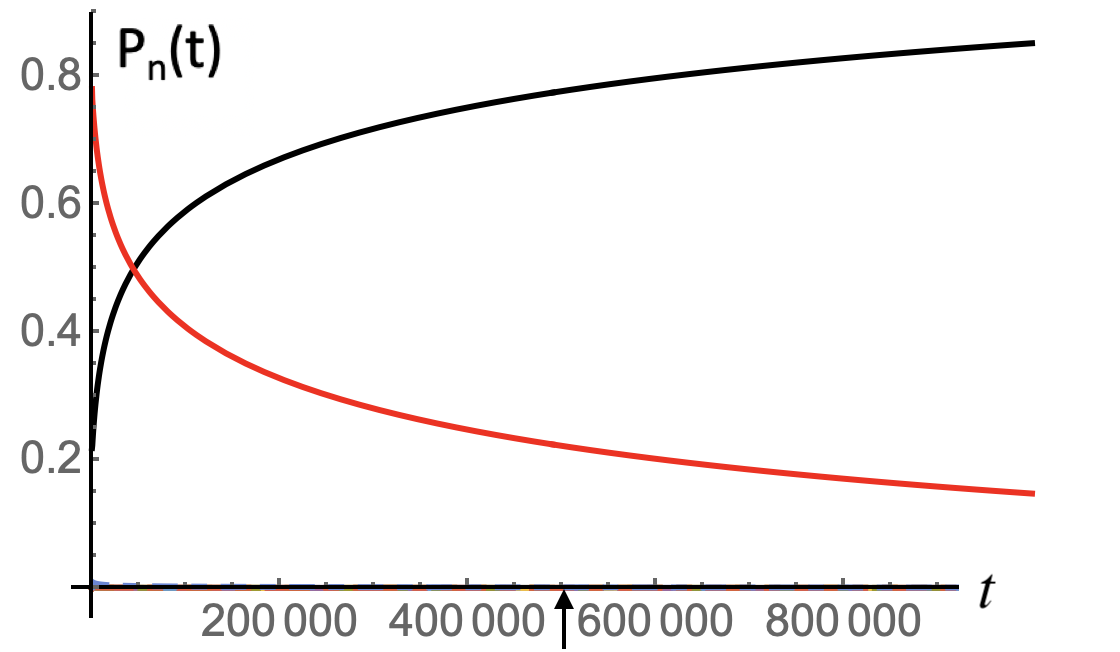}
\caption{Top: The assembly calculation of Fig. \ref{ME} at $c_0=1$ is followed by a disassembly calculation at $c_0=0.1$. The color coding is that of Fig.\ref{fig:D}. Bottom: Same except that the overall energy scale is increased to $\epsilon_0=3$. The arrows indicate the relaxation time $t_r$.}
\label{fig:dis}
\end{center}
\end{figure}

In order to investigate how assembled particles ``fall apart" when the pentamer concentration is lowered below the CAC, we used the outcome of an assembly calculation with $c_0$ above the CAC as the initial condition for a second calculation with $c_0$ reduced below the CAC. The results are shown in Fig. \ref{fig:dis}, still for molecule (1).
The assembled particles disintegrate on time scales of the order of the relaxation time $t_r\simeq 31.6$. Repeating the calculation for $\epsilon_0=3$ (bottom figure), disassembly is now much slower. The characteristic disassembly time scale remains of the order of the relaxation time, now with $t_r\simeq 5\times 10^5$.  Assembled viral particles must be at least metastable in solutions where the concentration of capsid proteins is well below the CAC where they are subject to an entropic thermodynamic drive towards disassembly. In the following, we will focus on time scales large compared to the assembly delay time but short compared to the relaxation time, which governs the disassembly process. This is possible only if there is a large separation of time scales, so for values of $\epsilon_0$ that significantly larger than one. We will investigate how the topology and geometry of the spanning trees affects assembly under these conditions.

\section{Single-Stage and Two-Stage Assembly.}

 We first examined normal pentamer-by-pentamer (``single-stage") assembly.  
\subsection{Single-Stage Assembly.}
\begin{figure}[htbp]
\begin{center}
\includegraphics[width=3.7in]{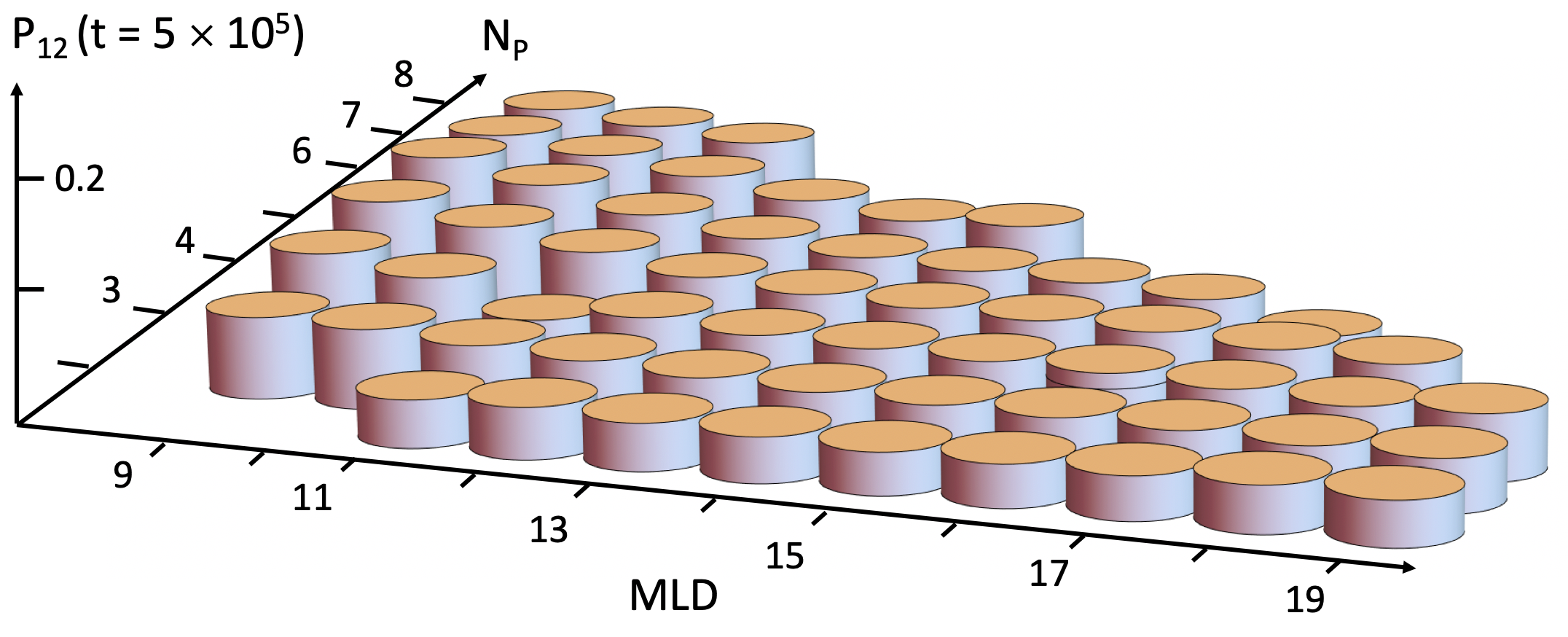}
\caption{Plot of the probability $P_{12}(t = 5\times 10^5)$ (vertical axis) that a genome molecule is fully packaged. The parameters are $\epsilon_1 = 0.2$, $\epsilon_3 =-1.0$, $\mu=-2.5$, $\epsilon_0=3$, $c_0=0.22$ and $D=0.3$.}
\label{fig:outcome1}
\end{center}
\end{figure}

A full survey was performed for every spanning tree of the probability $P_{12}(\tau)$ of a tree molecule to be fully encapsidate at a time $\tau = 5\times 10^5$. This was done for the parameter setting $\epsilon_0=3$,  $\epsilon_1 = 0.2$, $\epsilon_3 =-1.0$, $\mu=-2.5$, $\epsilon_0=3$, $c_0=0.22$ and $D=0.3$ since for this setting, (i) partial assemblies are compact and rigid, (ii) the nucleation-and-growth scenario applies, (iii) the system is close to the CAC (as is probably the biologically relevant regime, see Conclusion) with few assembly intermediates, and (iv) there is large separation between the assembly time scale $t_d$ and the disassembly time scale $t_r$. For these values, $\tau$ is about $0.25\thinspace t_r$. Finally, for every pair of MLD and $N_p$ numbers, the average of $P_{12}(\tau)$ and the standard deviation were computed over all spanning trees that belonged to that category. These averages were then plotted in the form of a three-dimensional bar graph with the MLD and $N_P$ numbers as the horizontal axes, as shown in Fig.\ref{fig:outcome1}. 

The largest value of $P_{12}(\tau)$ that we encountered was for MLD=9 and $N_p=8$ with $P_{12}(\tau)=0.155\pm0.016$. There were practically no assembly intermediates so the remaining tree molecules were not associated with pentamers, as was the case for other MLD and $N_p$ pairs. The lowest value of $P_{12}(\tau)$ was $0.033$ for MLD=16 and $N_p=2$. In general, the value of $P_{12}(\tau)$ systematically decreased when $N_p$ was decreased for given MLD. For given $N_p$ larger than two, the packaging probability decreased with increasing MLD, at least for MLD less than about thirteen. For MLD larger than thirteen the packaging probability was not much dependent on the MLD. This was also the case for $N_p=2$. The outliers that did not follow these trends, such as $MLD=11$ with $N_p=4$, were consistent with the statistical error bars. Single-stage assembly selects reasonably well for spanning-trees with large Wrapping Number and small MLD.

\subsection{Two-Stage Assembly.}
In the introduction it was pointed out that a second mode of assembly -- ``en-masse" -- is encountered for higher values of the affinity between the RNA molecules and the capsid proteins. In the first stage, a disordered protein-RNA condensate forms. This condensate undergoes an ordering transition, either after a certain period of time has elapsed or after an increase in the protein-protein affinity induced by a change in the thermodynamic parameters of the system. In order to mimic the en-masse assembly scenario, we carried out assembly calculations in two steps. During the first step, pentamers had a significant affinity for the tree molecule while the interaction between pentamers was made weakly repulsive. The outcome of this first calculation was then used as the starting state of a second calculation now with standard parameter settings. 

\begin{figure}[htbp]
\begin{center}
\includegraphics[width=2.5in]{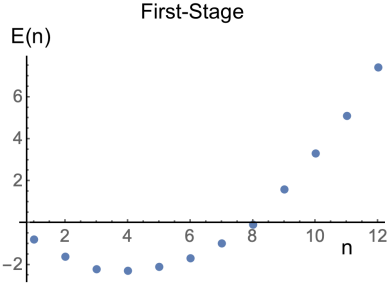}
\includegraphics[width=2.in]{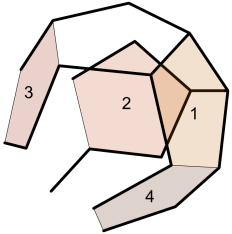}
\caption{Top: Assembly energy profile for the first assembly stage of a molecule (1) for $\epsilon_1=-0.2$, $\epsilon_3=0.5$, $\epsilon_0=3$, $D=1$, $c_0=1$ and $\mu=0$. Bottom: Minimum energy assembly path.}
\label{fig:stage1_EP}
\end{center}
\end{figure}

We again used molecule (1). During the first assembly stage $\epsilon_3$ was equal to $+0.5$, which means that the pentamer-pentamer interaction was repulsive. The reference chemical potential $\mu_0$, which includes a contribution associated with conformational changes required to allow for bonding between proteins, was set to zero. During the second assembly stage, the pentamer-pentamer affinity and the reference chemical potential were reset to the values of Fig.\ref{fig:outcome1}. The assembly energy profile and occupation probabilities of the first stage are shown in Fig.\ref{fig:stage1_EP}.

The top figure shows the minimum energy assembly profile at the reference concentration. The minimum at $n=4$ can be understood from the bottom figure, which shows the placing sequence of the first four pentamers. The first two pentamers can be placed on maximum wrapping sites without being forced to share any edges, which would be energetically unfavorable. Even though additional maximum wrapping sites remain available, the third pentamer is placed on a site where it makes contact with only three links of the tree molecules in order to avoid pentamer-pentamer repulsion. Even that option is no longer possible for the fourth pentamer, which is forced to share an edge with the first pentamer. The assembly energy no longer decreases significantly when the fourth pentamer is added. Note that the $n=4$ structure is flaccid, as is typical for first-stage structures. 

\begin{figure}[htbp]
\begin{center}
\includegraphics[width=3.5in]{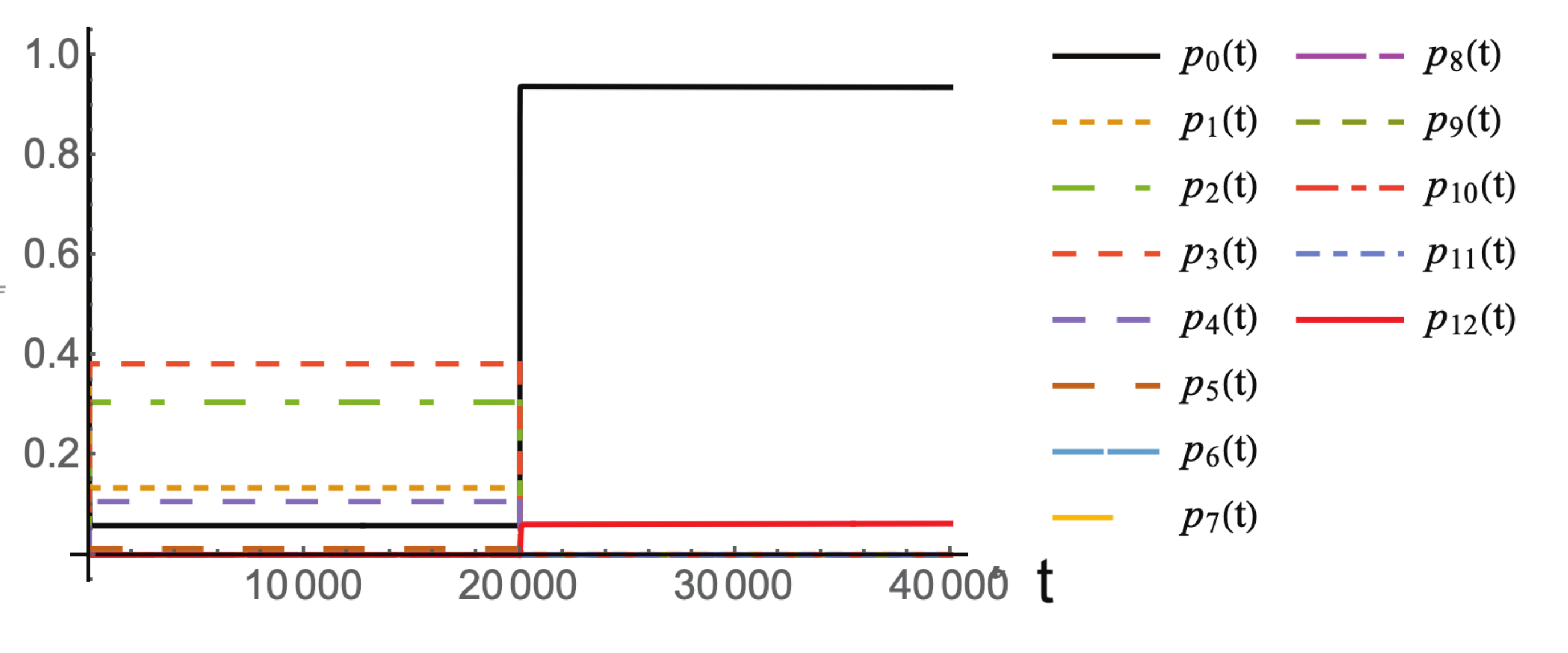}
\caption{Two-stage assembly of molecule (1). The parameter settings of the first stage for times less than $t=20.000$ are those of Fig.\ref{fig:stage1_EP} except that $c_0=0.22$.  At $t=20,000$, the parameters of the second stage were set to those of Fig.\ref{fig:outcome1}.}
\label{fig:two_stage2}
\end{center}
\end{figure}

The occupation probabilities for the first stage, computed from this energy profile, are shown in Fig.\ref{fig:two_stage2} for times less than $t=20.000$. The parameter settings are the same as Fig.\ref{fig:stage1_EP}, except that the concentration was reduced from the reference concentration to $c_0=0.22$, which is at the border of the CAC. This increases the free energy of the $n=4$ structure with respect to the $n=3$ structure by an amount $-\ln 0.22$ and by $-2\ln 0.22$ for the $n=2$ structure. As a result, the occupation probabilities of the $n=2$ and $n=3$ structures are increased with respect to that of the $n=4$ structures, as can be seen in Fig.\ref{fig:stage1_EP}. The occupation probabilities are time-independent, indicating that the system has equilibrated, which is expected since the delay time $t_d$ and the relaxation time $t_r$ are only of the order of $10$.

Next, at $t=20.000$, the parameters were returned to same values as for single-stage assembly (so with $\epsilon_3=-1$ and with $\mu=-2,5$). The parameter reset changes all minimum-energy assembly and disassembly pathways so for every one of the thirteen starting states produced by the first assembly stage at $t=20.000$, a different minimum-energy assembly/disassembly pathway must be generated for the second pathway. The outcome of such a two-stage assembly calculation is shown in Fig.\ref{fig:two_stage2}

Fully assembled particles appear practically immediately at $t=20.000$. The occupation probabilities remain constant for times larger than $t=20.000$. This is rather surprising because the relaxation time now is longer than $20.000$ (about $10^5$). The kinetics of two-stage assembly somehow does not seem to ``see" the relaxation time. While the solution was poly-disperse for times less than $t=20.000$, for times greater than $t=20.000$ the tree molecules are either fully encapsidated (red line, about 5 percent) or pentamer free (black line, about 95 percent). There were no assembly intermediates. Next, the calculation was repeated except that the reference chemical potential was reset from $\mu_0=-2.5$ to the assembly equilibrium condition with $\mu_0=-3.14$. Curiously, this did not change the results for the second stage in Fig.\ref{fig:two_stage2}. 

These results can be understood by comparing the second-stage minimum energy assembly profiles.

\begin{figure}[htbp]
\begin{center}
\includegraphics[width=3.5in]{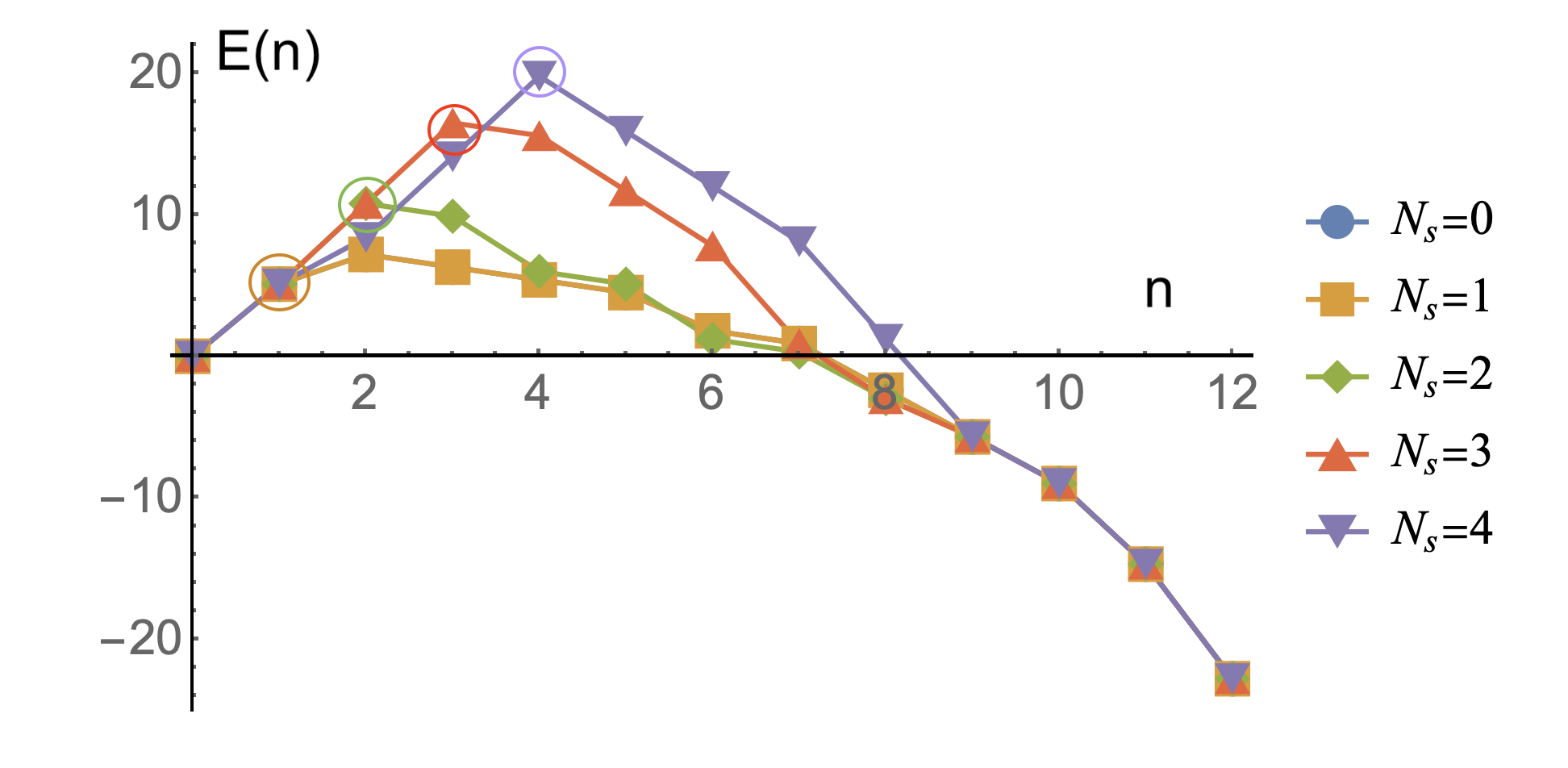}
\includegraphics[width=3.5in]{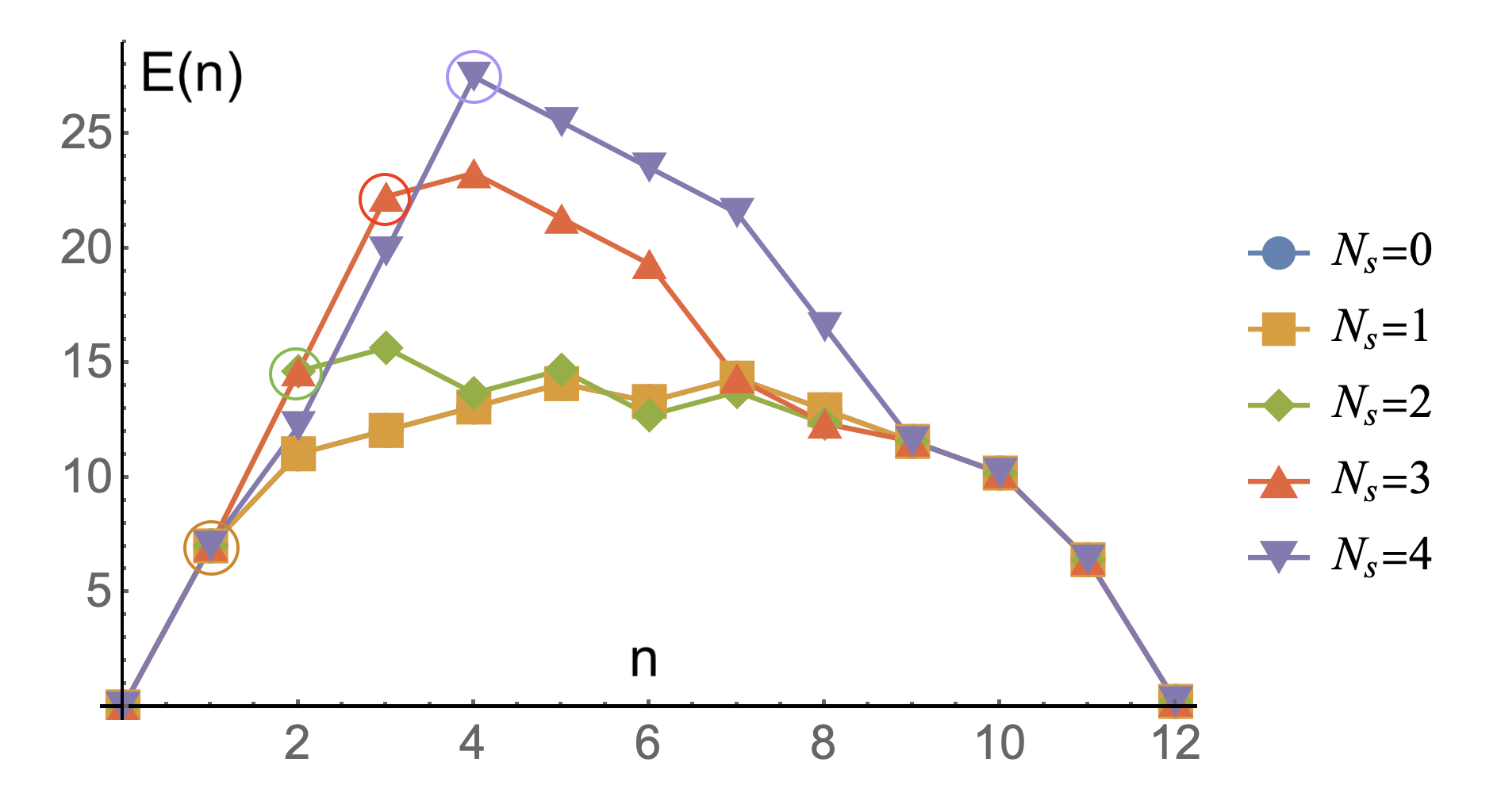}
\caption{Second-stage assembly energy profiles of the thirteen different states produced by the first stage. The index $N_s$ gives the occupancy at the start of the second stage. Only the profiles for $N_s=0,1,2,3,4$ are shown. The energy parameters of the second stage are $\epsilon_1=-0.2$, $\epsilon_3=-1.0$, $\epsilon_0=3$, with $\mu_0=-2.5$ (top) and $\mu_0=-3.1$ (bottom). The starting points for the second stage are indicated by circles. The vertical energy scale is in units of $\epsilon_0$.}
\label{fig:two_stage3}
\end{center}
\end{figure}
These plots show the minimum energy assembly profiles for different values of $N_s$ (color indexed), which is the number of pentamers that are initially present on the tree molecule as a result of the first assembly stage. The probabilities for the different $N_s$ states are simply the occupation probabilities of the first stage prior to the parameter switch. Second stage kinetics thus always starts from the state $n=N_s$ (circled). For example, in the top figure the initial state $n=4$ for the $N_s=4$ energy profile, shown in purple, is circled. Note that $n=4$ is the maximum of the energy profile (``transition state"). Imagine the state of the system as a particle performing a random walk along this energy profile, starting at $n=4$. At the first time step, the particle can hop either to $n=3$ or to $n=5$. Both are energetically downhill. For our kinetic scheme, the partition ratio $\frac{W_{4,5}}{W_4,3}$ between the rates to hop forwards -- from $n=4$ to $n=5$ -- or backwards -- from $n=4$ to $n=3$ -- is fixed by detailed balance to equal the ratio $(P_4/P_3) \exp\Delta E_{3,4}$, which in turn is equal to $c_f$. The left/right partition ratio for the $n=2$ and $n=3$ initial state, both of which are transition states, is also $c_f$. The energy cost of uphill steps is about $5 \epsilon_3 \simeq 15 k_BT$. If one neglects such up-hill steps, then the particle will end up either at $n=0$ (i.e., free tree molecules) or at $n=12$ (i.e., encapsidated tree molecules) with a partition ratio that is equal to the concentration $c_f\simeq c_0$ of free pentamers in the equilibrium state (in units of the reference concentration).

The bottom plot shows the case where the chemical potential is tuned to the state of assembly equilibrium. For single-stage assembly, starting at $n=0$, the energy activation barrier would be a prohibitive $30 \epsilon_3\simeq 100 k_BT$. However, because assembly now starts (mostly) at $N_s=3$ or $N_s=2$, there is a reasonable probability that the particle will reach the $n=12$ state. The partition ratio is not exactly the same as for the previous case because the $n=2$ state and the $n=3$ state no longer are transition states. Adding a pentamer is now in both cases an energetically uphill step. However, the energy cost of the uphill step is rather modest in this case (about $2\epsilon_3$). This should lead to only a modest reduction in the fraction of assembled viral particles, provided $c_f$ remains the same. This last condition is consistent with the observation that the partition ratio did not change significantly with respect to the previous case.The physical mechanism that allows for this amplification effect is that when a few pentamers are already present on the tree molecule prior to the parameter switch, because of the first stage ``priming" process, then these pentamers will act as nucleation sites for the completion of the capsid. The production yield of assembled particles can be much larger in the two-stage process.

\subsection{Two-Stage Selectivity.}
Finally, we investigated the selectivity of two-stage assembly in terms of the Wrapping Number and the MLD. A full survey of tens of thousands of spanning trees was not practical because two-stage assembly is computationally quite intensive. Instead, single representatives were picked from each pair of Wrapping Number and MLD. The results are shown in Fig.\ref{fig:outcome1}
\begin{figure}[htbp]
\begin{center}
\includegraphics[width=3.5in]{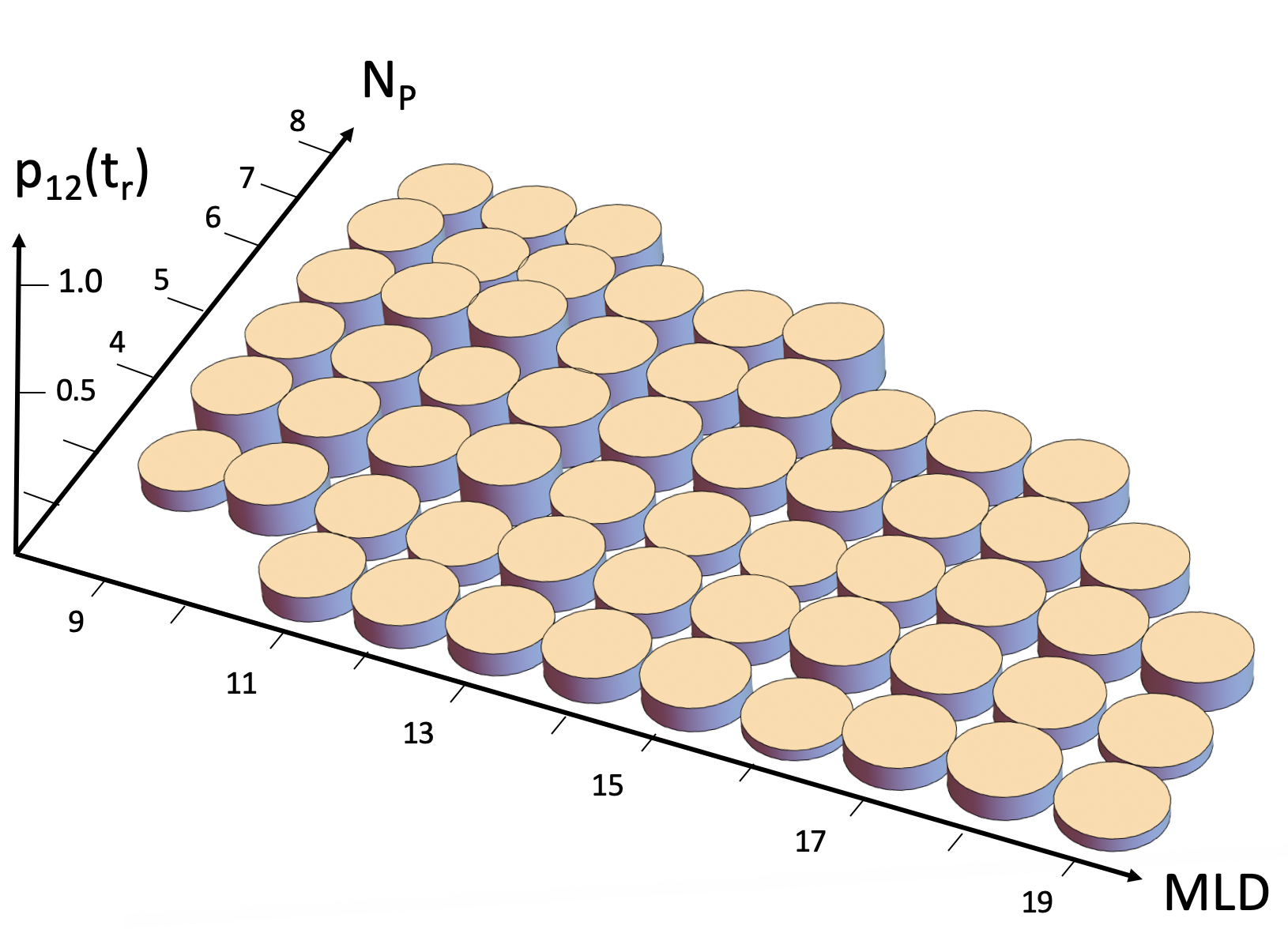}
\caption{Plot of the probability $P_{12}(t_r)$ (vertical axis) that a tree molecule is fully packaged for two-stage assembly. The parameters are the same as for Fig.\ref{fig:outcome1}}
\label{fig:outcome2}
\end{center}
\end{figure}
For low values of $N_p$, there appears to be no selectivity with respect to the MLD. Similarly, for low values of the MLD, there appears to be no selectivity with respect to $N_p$ except that $N_p=3$ and $N_p=8$ have lower packaging probability. While there is no clear selectivity, the packaging probabilities are on average higher for $N_p$ between four and seven and MLD less than twelve.

\section{Conclusion}
We analyzed the thermodynamics and kinetic properties of the spanning-tree model for dodecahedral particles using numerical solutions of a set of coupled master equations. 

The dependence on the time-dependent occupation probabilities of partial and complete assemblies on the topology and geometry of the spanning-tree molecules -- which represent the outer part of condensed viral RNA -- was shown to be largely determined by two correlated criteria: the maximum ladder distance (MLD), which is a topological measure of the degree of branching of the spanning tree, and the Wrapping Number ($N_P$), which is geometrical measure of the spanning tree that counts the number of sites that can accommodate the pentamers --  which represent capsid protein oligomers -- with maximum binding energy to the spanning tree. 

The assembly kinetics is characterized by two important time scales: the delay time $t_d$ for the onset of production of fully assembled particles and $t_r$, the relaxation time for thermal equilibration. While delay time increases modestly as a function of the energy scale $\epsilon_0$ of the interactions, the relaxation increases exponentially. When the solution concentration $c_0$ of pentamers is lowered below the critical aggregation concentration of the equilibrium phase diagram, assembled particles remain stable against spontaneous disassembly on time scales shorter than $t_r$. The dependence of the probability for a spanning tree to be completely packaged on the $N_P$ and MLD numbers is a \textit{purely kinetic effect} that disappears on time-scales large compared to the relaxation time. This selectivity is due to small differences between the activation energy barriers spanning trees with different $N_P$ and MLD (``selective nucleation"). Selective nucleation only ``works" if the assembly activation energy barrier is large compared to the thermal energy.

As a function of the ratio $|\epsilon_1|$ between protein/RNA and protein/protein affinity, partial assemblies undergo a structural transition around $|\epsilon_1|=1$ from predominantly rigid to predominantly flaccid structures, which has the character of an order-disorder transition in the equilibrium phase diagram. For $|\epsilon_1|$ small compared to one, assembly kinetics has the character of the classical nucleation-and-growth scenario under conditions of supersaturation. This scenario breaks down for $|\epsilon_1|$ of the order of one or larger. At the same point, assembled particles start to readily disintegrate when the pentamer concentration is reduced after assembly. The minimum-energy assembly pathway is then characterized by pronounced kinetic traps. 

Finally, we carried out an exhaustive survey of the assembly kinetics of all spanning trees and found that, for direct assembly, the probability for the appearance of fully assembled particles progressively increases with increasing Wrapping Number and decreasing MLD. When direct single-stage assembly was replaced by two-stage assembly -- as a model for the en-masse scenario -- then the yield of fully assembled particles sharply increased, even under conditions of assembly equilibrium where single-stage assembly is blocked. However, the dependence of two-stage assembly on the $N_P$ and MLD numbers is weak and non-systematic as compared to single-stage assembly.

We mentioned in the introduction a number of experimental observations that motivated the construction of the model. The first was an in-vitro study of the co-assembly of CCMV with non-CCMV RNA molecules \cite{Comas-Garcia}, which reported that when the RNA-to-protein mixing ratio was low, virus-like particles (VLPs) formed with excess proteins, in agreement with simple arguments based on phase-coexistence. When the mixing ratio exceeded a threshold, the virus-like particles were replaced by disordered RNA-protein aggregates instead of the expected coexistence of VLPs with excess RNA while the threshold point separating the two regimes was well below the stoichiometric ratio. In the spanning tree model, the mixing ratio is represented by $D$. If $D$ is increased at fixed total protein concentration $c_0$ in the quasi phase-diagram Fig.\ref{fig:PD} then for smaller $D$ practically all tree molecules are encapsidated (to the left of the blue line in Fig.\ref{fig:PD}) while for larger values of $D$ an increasing fraction of the tree molecules are not encapsidated. As $|\epsilon_1|$ increases, a significant fraction are partially encapsidated. The transition point separating the two regimes drops below the stoichiometric ratio when $c_0$ approached the critical aggregation concentration. These results are consistent with the experimental observations.  

Another important motivation was the asymmetric reconstruction of the MS2 virus, which provided evidence that a subsection of the RNA genome that is rich in packaging signals, reproducibly associates with a compact cluster of capsid proteins \cite{Dykeman2011}. Our model reproduces this observation for genome molecules with small MLD and large $N_P$, in which case the intermediate states that regulate the assembly rate are compact and rigid for smaller values of $\epsilon_1$. In the spanning-tree model, the role of packaging signals is played by the $N_p$ maximum wrapping sites. This reinforces the notion that packaging selection is carried not by individual RNA stem-loops but rather by spatially correlated patterns of stem-loops, as proposed in ref.\cite{Dykeman2011}.

An important aspect of the model that should be improved is the fact that it does not account for conformational fluctuations of the genome molecules, prior to assembly, as well as thermal fluctuations of flaccid intermediates. For the same reason, the model also does not include the ``antenna" effect for the diffusive influx of capsid proteins to partial assemblies \cite{Hu2006}. Another aspect of the model that should be improved in the context of asymmetric reconstructions is a more realistic treatment of the the condensed RNA molecule. Such a generalization would start from a determination of the key interaction sites between capsid proteins and the surface of the enclosed RNA. These sites would span a polyhedron with many more sites than the vertices of a dodecahedron. The spanning tree should reproduce the outer density of the RNA molecule, such as Fig.1. Classifying the set of all spanning trees would be significantly more challenging and require more extensive numerical work.

\begin{acknowledgements}
We would like to thank Alexander Grosberg for drawing our attention to spanning trees and Ioulia Rouzina for introducing us to the concept of selective nucleation. We would like to thank Reidun Twarock, Charles Knobler and William Gelbart for reading a first draft and commenting on it. We also benefitted from discussions with Chen Lin, Zach Gvildys and William Vong. RB would like to thank the NSF-DMR for continued support under CMMT Grant No.1836404.
\end{acknowledgements}

\clearpage

\begin{appendix}
\section{Demonstration hat the smallest MLD for spanning trees on the dodecahedron is nine} \label{app:A}
\begin{figure}[htbp]
\begin{center}
\includegraphics[width=2.5in]{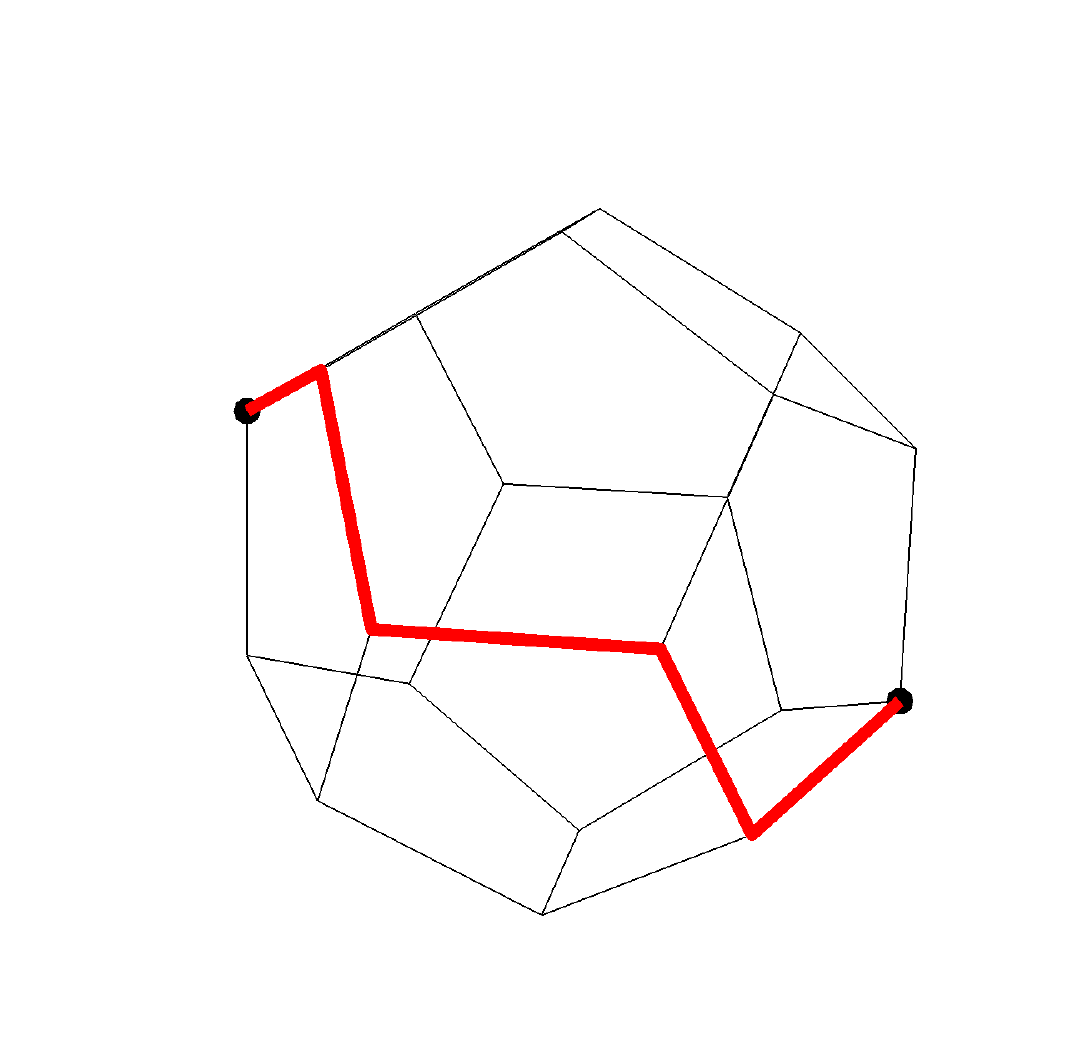}
\caption{Two maximally separated vertices on the dodecahedron and one of the 12 shortest paths consisting of five edges that join them.}
\label{fig:app1}
\end{center}
\end{figure}

We begin by noting that for every vertex on the dodecahedron there is a vertex on the opposite side of the polyhedron that is a ladder distance five away. That is, getting from one of the two vertices to the other requires traversing at least five edges. Figure \ref{fig:app1} shows such a path.
For each such pairs of vertices there are 12 minimal paths. 

\begin{figure}[htbp]
\begin{center}
\includegraphics[width=2.5in]{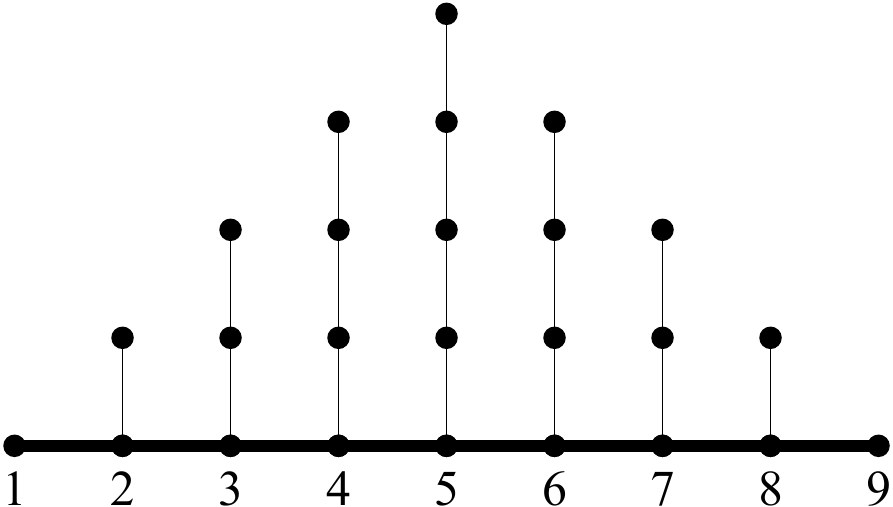}
\caption{A ladder distance 8 path in the hypothetical MLD 8 spanning tree on the dodecahedron. The path is shown as a thick line, and the nine vertices are labeled for easy reference. The thinner vertical lines represent longest allowed paths branching off the ladder distance 8 path. }
\label{fig:app2}
\end{center}
\end{figure}

Now, assume that there is a spanning tree with MLD 8. In such a case, we can pick out a path of ladder distance eight in that tree. All other elements of the tree will consist of trees that branch out from that path. Figure \ref{fig:app2} is a figurative depiction of the path along with the longest allowed branch sprouting off each vertex on that path. The likelihood of branching off those ``side branch'' paths is ignored; such branching does not alter the argument below. 

Consider  first the central vertex on the ladder distance eight path, labeled 5 in Fig.\ref{fig:app2}. The side path with ladder distance four is the longest that can attach to it. A longer path increases the MLD of the tree. Clearly, there is no possibility of reaching a point a ladder distance five from vertex 5 along any path with ladder distance four, so the path shown cannot connect the central vertex to the vertex a distance five away from it.  Next, consider the two sites flanking the central vertices, labeled 4 and 6. Attached to each is the longest possible path branching out from them, Such a path has ladder distance three. If either of these paths reached to the vertex a ladder distance five away from the central vertex, then there would be a ladder distance four (or less) path from that vertex through one of the flanking vertices to the maximally separated vertex, and we know that no such path exists. We can continue this argument to encompass all allowed paths sprouting from vertices on the chosen path. Thus, there is a vertex on the dodecahedron that cannot be a part of the MLD  8 tree containing this path. Consequently no tree with MLD 8 can be a spanning tree on the dodecahedron. The argument above can clearly be applied to the possibility of a spanning tree with MLD less than eight. That there is a spanning tree with MLD 9 is readily established by construction.

\newpage

\section{Limiting solutions of Eq.\ref{C1}} \label{app:B}
For $\left(\frac{c_0^{12}}{K}\right)$ small compared to one, the equation has a solution with $c_f$ close to $c_0$ given by:
\begin{equation}
\frac{c_f}{c_0}\simeq 1 - \frac{D c_0^{12}}{K}
\end{equation}
For $\left(\frac{c_0^{12}}{K}\right)$ large compared to one and mixing ratio $D$ larger than one, the equation has a solution with $c_f$ independent of $c_0$:
\begin{equation}
{c_f}\simeq \left(\frac{K}{D-1}\right)^{1/12}
\end{equation}
Finally, for $\left(\frac{c_0^{12}}{K}\right)$ large compared to one but the mixing ratio $D$ less than one, the equation has a different solution with $c_f$ independent of $c_0$:
\begin{equation}
\frac{c_f}{c_0}\simeq 1 - D +\frac{K}{c_0^{12}}\frac{D}{(1-D)^{1/2}}
\end{equation}
There is thus a change in regimes near the point where the mixing ratio is equal to one. For the special case that $D=1$, the LMA equation reduces to
\begin{equation}
\left(\frac{c_f}{c_0}\right)^{13}\left(\frac{1}{1-\frac{c_f}{c_0}}\right)\simeq\left(\frac{K}{c_0^{12}}\right)
\label{B}
\end{equation}
\end{appendix}


\begin{thebibliography}{64}%
\makeatletter
\providecommand \@ifxundefined [1]{%
 \@ifx{#1\undefined}
}%
\providecommand \@ifnum [1]{%
 \ifnum #1\expandafter \@firstoftwo
 \else \expandafter \@secondoftwo
 \fi
}%
\providecommand \@ifx [1]{%
 \ifx #1\expandafter \@firstoftwo
 \else \expandafter \@secondoftwo
 \fi
}%
\providecommand \natexlab [1]{#1}%
\providecommand \enquote  [1]{``#1''}%
\providecommand \bibnamefont  [1]{#1}%
\providecommand \bibfnamefont [1]{#1}%
\providecommand \citenamefont [1]{#1}%
\providecommand \href@noop [0]{\@secondoftwo}%
\providecommand \href [0]{\begingroup \@sanitize@url \@href}%
\providecommand \@href[1]{\@@startlink{#1}\@@href}%
\providecommand \@@href[1]{\endgroup#1\@@endlink}%
\providecommand \@sanitize@url [0]{\catcode `\\12\catcode `\$12\catcode
  `\&12\catcode `\#12\catcode `\^12\catcode `\_12\catcode `\%12\relax}%
\providecommand \@@startlink[1]{}%
\providecommand \@@endlink[0]{}%
\providecommand \url  [0]{\begingroup\@sanitize@url \@url }%
\providecommand \@url [1]{\endgroup\@href {#1}{\urlprefix }}%
\providecommand \urlprefix  [0]{URL }%
\providecommand \Eprint [0]{\href }%
\providecommand \doibase [0]{https://doi.org/}%
\providecommand \selectlanguage [0]{\@gobble}%
\providecommand \bibinfo  [0]{\@secondoftwo}%
\providecommand \bibfield  [0]{\@secondoftwo}%
\providecommand \translation [1]{[#1]}%
\providecommand \BibitemOpen [0]{}%
\providecommand \bibitemStop [0]{}%
\providecommand \bibitemNoStop [0]{.\EOS\space}%
\providecommand \EOS [0]{\spacefactor3000\relax}%
\providecommand \BibitemShut  [1]{\csname bibitem#1\endcsname}%
\let\auto@bib@innerbib\@empty
\bibitem [{PLO()}]{PLOS}%
  \BibitemOpen
  \href@noop {} {}\bibinfo {note} {I. Mizrahi, R. Bruinsma and J. Rudnick,
  submitted to Plos Biocomputation.}\BibitemShut {Stop}%
\bibitem [{\citenamefont {Crick}\ and\ \citenamefont
  {Watson}(1956)}]{Crick1956}%
  \BibitemOpen
  \bibfield  {author} {\bibinfo {author} {\bibfnamefont {F.~H.~C.}\
  \bibnamefont {Crick}}\ and\ \bibinfo {author} {\bibfnamefont {J.~D.}\
  \bibnamefont {Watson}},\ }\href@noop {} {\bibfield  {journal} {\bibinfo
  {journal} {Nature}\ }\textbf {\bibinfo {volume} {177}},\ \bibinfo {pages}
  {473} (\bibinfo {year} {1956})}\BibitemShut {NoStop}%
\bibitem [{Note1()}]{Note1}%
  \BibitemOpen
  \bibinfo {note} {For double-stranded DNA viruses the genome molecules either
  are inserted into the interior pre-fabricated spherical capsids by a
  molecular motor or the genome is pre-condensed prior to
  assembly.}\BibitemShut {Stop}%
\bibitem [{\citenamefont {Ganser-Pornillos}\ \emph {et~al.}(2008)\citenamefont
  {Ganser-Pornillos}, \citenamefont {Yeager},\ and\ \citenamefont
  {Sundquist}}]{Ganser-Pornillos2008}%
  \BibitemOpen
  \bibfield  {author} {\bibinfo {author} {\bibfnamefont {B.~K.}\ \bibnamefont
  {Ganser-Pornillos}}, \bibinfo {author} {\bibfnamefont {M.}~\bibnamefont
  {Yeager}},\ and\ \bibinfo {author} {\bibfnamefont {W.~I.}\ \bibnamefont
  {Sundquist}},\ }\href@noop {} {\bibfield  {journal} {\bibinfo  {journal}
  {Curr. Opin. Struct. Biol.}\ }\textbf {\bibinfo {volume} {18}},\ \bibinfo
  {pages} {203} (\bibinfo {year} {2008})}\BibitemShut {NoStop}%
\bibitem [{\citenamefont {Fraenkel-Conrat}\ and\ \citenamefont
  {Williams}(1955)}]{Fraenkel-Conrat1955}%
  \BibitemOpen
  \bibfield  {author} {\bibinfo {author} {\bibfnamefont {H.}~\bibnamefont
  {Fraenkel-Conrat}}\ and\ \bibinfo {author} {\bibfnamefont {R.~C.}\
  \bibnamefont {Williams}},\ }\href@noop {} {\bibfield  {journal} {\bibinfo
  {journal} {Proc. Natl. Acad. Sci. U. S. A.}\ }\textbf {\bibinfo {volume}
  {41}},\ \bibinfo {pages} {690} (\bibinfo {year} {1955})}\BibitemShut
  {NoStop}%
\bibitem [{\citenamefont {Butler}\ and\ \citenamefont
  {Klug}(1978)}]{Butler1978}%
  \BibitemOpen
  \bibfield  {author} {\bibinfo {author} {\bibfnamefont {P.~J.~G.}\
  \bibnamefont {Butler}}\ and\ \bibinfo {author} {\bibfnamefont
  {A.}~\bibnamefont {Klug}},\ }\href@noop {} {\bibfield  {journal} {\bibinfo
  {journal} {Sci. Am.}\ }\textbf {\bibinfo {volume} {239}},\ \bibinfo {pages}
  {62} (\bibinfo {year} {1978})}\BibitemShut {NoStop}%
\bibitem [{\citenamefont {Klug}(1999)}]{Klug1999}%
  \BibitemOpen
  \bibfield  {author} {\bibinfo {author} {\bibfnamefont {A.}~\bibnamefont
  {Klug}},\ }\href@noop {} {\bibfield  {journal} {\bibinfo  {journal} {Philos.
  Trans. R. Soc. Lond. B. Biol. Sci.}\ }\textbf {\bibinfo {volume} {354}},\
  \bibinfo {pages} {531} (\bibinfo {year} {1999})}\BibitemShut {NoStop}%
\bibitem [{Note2()}]{Note2}%
  \BibitemOpen
  \bibinfo {note} {For a quantitative treatment of this model, see ref. \cite
  {kegel2006}}\BibitemShut {NoStop}%
\bibitem [{\citenamefont {Mathews}\ \emph {et~al.}(1999)\citenamefont
  {Mathews}, \citenamefont {Sabina}, \citenamefont {Zuker},\ and\ \citenamefont
  {Turner}}]{mathews1999}%
  \BibitemOpen
  \bibfield  {author} {\bibinfo {author} {\bibfnamefont {D.~H.}\ \bibnamefont
  {Mathews}}, \bibinfo {author} {\bibfnamefont {J.}~\bibnamefont {Sabina}},
  \bibinfo {author} {\bibfnamefont {M.}~\bibnamefont {Zuker}},\ and\ \bibinfo
  {author} {\bibfnamefont {D.~H.}\ \bibnamefont {Turner}},\ }\href@noop {}
  {\bibfield  {journal} {\bibinfo  {journal} {Journal of molecular biology}\
  }\textbf {\bibinfo {volume} {288}},\ \bibinfo {pages} {911} (\bibinfo {year}
  {1999})}\BibitemShut {NoStop}%
\bibitem [{\citenamefont {Tubiana}\ \emph {et~al.}(2015)\citenamefont
  {Tubiana}, \citenamefont {Bo{\v{z}}i{\v{c}}}, \citenamefont {Micheletti},\
  and\ \citenamefont {Podgornik}}]{tubiana}%
  \BibitemOpen
  \bibfield  {author} {\bibinfo {author} {\bibfnamefont {L.}~\bibnamefont
  {Tubiana}}, \bibinfo {author} {\bibfnamefont {A.~L.}\ \bibnamefont
  {Bo{\v{z}}i{\v{c}}}}, \bibinfo {author} {\bibfnamefont {C.}~\bibnamefont
  {Micheletti}},\ and\ \bibinfo {author} {\bibfnamefont {R.}~\bibnamefont
  {Podgornik}},\ }\href@noop {} {\bibfield  {journal} {\bibinfo  {journal}
  {Biophysical journal}\ }\textbf {\bibinfo {volume} {108}},\ \bibinfo {pages}
  {194} (\bibinfo {year} {2015})}\BibitemShut {NoStop}%
\bibitem [{\citenamefont {Patel}\ \emph {et~al.}(2015)\citenamefont {Patel},
  \citenamefont {Dykeman}, \citenamefont {Coutts}, \citenamefont {Lomonossoff},
  \citenamefont {Rowlands}, \citenamefont {Phillips}, \citenamefont {Ranson},
  \citenamefont {Twarock}, \citenamefont {Tuma},\ and\ \citenamefont
  {Stockley}}]{Patel2015}%
  \BibitemOpen
  \bibfield  {author} {\bibinfo {author} {\bibfnamefont {N.}~\bibnamefont
  {Patel}}, \bibinfo {author} {\bibfnamefont {E.~C.}\ \bibnamefont {Dykeman}},
  \bibinfo {author} {\bibfnamefont {R.~H.~A.}\ \bibnamefont {Coutts}}, \bibinfo
  {author} {\bibfnamefont {G.~P.}\ \bibnamefont {Lomonossoff}}, \bibinfo
  {author} {\bibfnamefont {D.~J.}\ \bibnamefont {Rowlands}}, \bibinfo {author}
  {\bibfnamefont {S.~E.~V.}\ \bibnamefont {Phillips}}, \bibinfo {author}
  {\bibfnamefont {N.}~\bibnamefont {Ranson}}, \bibinfo {author} {\bibfnamefont
  {R.}~\bibnamefont {Twarock}}, \bibinfo {author} {\bibfnamefont
  {R.}~\bibnamefont {Tuma}},\ and\ \bibinfo {author} {\bibfnamefont {P.~G.}\
  \bibnamefont {Stockley}},\ }\bibfield  {journal} {\bibinfo  {journal} {Proc.
  Natl. Acad. Sci. U.S.A.}\ }\href {https://doi.org/10.1073/pnas.1420812112}
  {10.1073/pnas.1420812112} (\bibinfo {year} {2015})\BibitemShut {NoStop}%
\bibitem [{\citenamefont {Yoffe}\ \emph {et~al.}(2008)\citenamefont {Yoffe},
  \citenamefont {Prinsen}, \citenamefont {Gopal}, \citenamefont {Knobler},
  \citenamefont {Gelbart},\ and\ \citenamefont {Ben-Shaul}}]{yoffe2008}%
  \BibitemOpen
  \bibfield  {author} {\bibinfo {author} {\bibfnamefont {A.~M.}\ \bibnamefont
  {Yoffe}}, \bibinfo {author} {\bibfnamefont {P.}~\bibnamefont {Prinsen}},
  \bibinfo {author} {\bibfnamefont {A.}~\bibnamefont {Gopal}}, \bibinfo
  {author} {\bibfnamefont {C.~M.}\ \bibnamefont {Knobler}}, \bibinfo {author}
  {\bibfnamefont {W.~M.}\ \bibnamefont {Gelbart}},\ and\ \bibinfo {author}
  {\bibfnamefont {A.}~\bibnamefont {Ben-Shaul}},\ }\href@noop {} {\bibfield
  {journal} {\bibinfo  {journal} {Proceedings of the National Academy of
  Sciences}\ }\textbf {\bibinfo {volume} {105}},\ \bibinfo {pages} {16153}
  (\bibinfo {year} {2008})}\BibitemShut {NoStop}%
\bibitem [{\citenamefont {Bruinsma}\ \emph {et~al.}(2021)\citenamefont
  {Bruinsma}, \citenamefont {Wuite},\ and\ \citenamefont
  {Roos}}]{bruinsma2021}%
  \BibitemOpen
  \bibfield  {author} {\bibinfo {author} {\bibfnamefont {R.~F.}\ \bibnamefont
  {Bruinsma}}, \bibinfo {author} {\bibfnamefont {G.~J.}\ \bibnamefont
  {Wuite}},\ and\ \bibinfo {author} {\bibfnamefont {W.~H.}\ \bibnamefont
  {Roos}},\ }\href@noop {} {\bibfield  {journal} {\bibinfo  {journal} {Nature
  Reviews Physics}\ ,\ \bibinfo {pages} {1}} (\bibinfo {year}
  {2021})}\BibitemShut {NoStop}%
\bibitem [{\citenamefont {Prevelige}\ \emph {et~al.}(1993)\citenamefont
  {Prevelige}, \citenamefont {Thomas},\ and\ \citenamefont
  {King}}]{Prevelige1993}%
  \BibitemOpen
  \bibfield  {author} {\bibinfo {author} {\bibfnamefont {P.~E.}\ \bibnamefont
  {Prevelige}}, \bibinfo {author} {\bibfnamefont {D.}~\bibnamefont {Thomas}},\
  and\ \bibinfo {author} {\bibfnamefont {J.}~\bibnamefont {King}},\ }\href@noop
  {} {\bibfield  {journal} {\bibinfo  {journal} {Biophys. J.}\ }\textbf
  {\bibinfo {volume} {64}},\ \bibinfo {pages} {824} (\bibinfo {year}
  {1993})}\BibitemShut {NoStop}%
\bibitem [{\citenamefont {Casini}\ \emph {et~al.}(2004)\citenamefont {Casini},
  \citenamefont {Graham}, \citenamefont {Heine}, \citenamefont {Garcea},\ and\
  \citenamefont {Wu}}]{Casini2004}%
  \BibitemOpen
  \bibfield  {author} {\bibinfo {author} {\bibfnamefont {G.~L.}\ \bibnamefont
  {Casini}}, \bibinfo {author} {\bibfnamefont {D.}~\bibnamefont {Graham}},
  \bibinfo {author} {\bibfnamefont {D.}~\bibnamefont {Heine}}, \bibinfo
  {author} {\bibfnamefont {R.~L.}\ \bibnamefont {Garcea}},\ and\ \bibinfo
  {author} {\bibfnamefont {D.~T.}\ \bibnamefont {Wu}},\ }\href@noop {}
  {\bibfield  {journal} {\bibinfo  {journal} {Virology}\ }\textbf {\bibinfo
  {volume} {325}},\ \bibinfo {pages} {320} (\bibinfo {year}
  {2004})}\BibitemShut {NoStop}%
\bibitem [{\citenamefont {Medrano}\ \emph {et~al.}(2016)\citenamefont
  {Medrano}, \citenamefont {Fuertes}, \citenamefont {Valbuena}, \citenamefont
  {Carrillo}, \citenamefont {Rodr{\'\i}guez-Huete},\ and\ \citenamefont
  {Mateu}}]{medrano}%
  \BibitemOpen
  \bibfield  {author} {\bibinfo {author} {\bibfnamefont {M.}~\bibnamefont
  {Medrano}}, \bibinfo {author} {\bibfnamefont {M.~{\'A}.}\ \bibnamefont
  {Fuertes}}, \bibinfo {author} {\bibfnamefont {A.}~\bibnamefont {Valbuena}},
  \bibinfo {author} {\bibfnamefont {P.~J.}\ \bibnamefont {Carrillo}}, \bibinfo
  {author} {\bibfnamefont {A.}~\bibnamefont {Rodr{\'\i}guez-Huete}},\ and\
  \bibinfo {author} {\bibfnamefont {M.~G.}\ \bibnamefont {Mateu}},\ }\href@noop
  {} {\bibfield  {journal} {\bibinfo  {journal} {Journal of the American
  Chemical Society}\ }\textbf {\bibinfo {volume} {138}},\ \bibinfo {pages}
  {15385} (\bibinfo {year} {2016})}\BibitemShut {NoStop}%
\bibitem [{\citenamefont {Zandi}\ \emph {et~al.}(2006)\citenamefont {Zandi},
  \citenamefont {van~der Schoot}, \citenamefont {Reguera}, \citenamefont
  {Kegel},\ and\ \citenamefont {Reiss}}]{Zandi2006}%
  \BibitemOpen
  \bibfield  {author} {\bibinfo {author} {\bibfnamefont {R.}~\bibnamefont
  {Zandi}}, \bibinfo {author} {\bibfnamefont {P.}~\bibnamefont {van~der
  Schoot}}, \bibinfo {author} {\bibfnamefont {D.}~\bibnamefont {Reguera}},
  \bibinfo {author} {\bibfnamefont {W.}~\bibnamefont {Kegel}},\ and\ \bibinfo
  {author} {\bibfnamefont {H.}~\bibnamefont {Reiss}},\ }\href@noop {}
  {\bibfield  {journal} {\bibinfo  {journal} {Biophys. J.}\ }\textbf {\bibinfo
  {volume} {90}},\ \bibinfo {pages} {1939} (\bibinfo {year}
  {2006})}\BibitemShut {NoStop}%
\bibitem [{\citenamefont {Asor}\ \emph {et~al.}(2019)\citenamefont {Asor},
  \citenamefont {Selzer}, \citenamefont {Schlicksup}, \citenamefont {Zhao},
  \citenamefont {Zlotnick},\ and\ \citenamefont {Raviv}}]{asor2019}%
  \BibitemOpen
  \bibfield  {author} {\bibinfo {author} {\bibfnamefont {R.}~\bibnamefont
  {Asor}}, \bibinfo {author} {\bibfnamefont {L.}~\bibnamefont {Selzer}},
  \bibinfo {author} {\bibfnamefont {C.~J.}\ \bibnamefont {Schlicksup}},
  \bibinfo {author} {\bibfnamefont {Z.}~\bibnamefont {Zhao}}, \bibinfo {author}
  {\bibfnamefont {A.}~\bibnamefont {Zlotnick}},\ and\ \bibinfo {author}
  {\bibfnamefont {U.}~\bibnamefont {Raviv}},\ }\href@noop {} {\bibfield
  {journal} {\bibinfo  {journal} {ACS nano}\ }\textbf {\bibinfo {volume}
  {13}},\ \bibinfo {pages} {7610} (\bibinfo {year} {2019})}\BibitemShut
  {NoStop}%
\bibitem [{\citenamefont {Bryngelson}\ \emph {et~al.}(1995)\citenamefont
  {Bryngelson}, \citenamefont {Onuchic}, \citenamefont {Socci},\ and\
  \citenamefont {Wolynes}}]{bryngelson}%
  \BibitemOpen
  \bibfield  {author} {\bibinfo {author} {\bibfnamefont {J.~D.}\ \bibnamefont
  {Bryngelson}}, \bibinfo {author} {\bibfnamefont {J.~N.}\ \bibnamefont
  {Onuchic}}, \bibinfo {author} {\bibfnamefont {N.~D.}\ \bibnamefont {Socci}},\
  and\ \bibinfo {author} {\bibfnamefont {P.~G.}\ \bibnamefont {Wolynes}},\
  }\href@noop {} {\bibfield  {journal} {\bibinfo  {journal} {Proteins:
  Structure, Function, and Bioinformatics}\ }\textbf {\bibinfo {volume} {21}},\
  \bibinfo {pages} {167} (\bibinfo {year} {1995})}\BibitemShut {NoStop}%
\bibitem [{\citenamefont {Hu}\ and\ \citenamefont {Shklovskii}(2007)}]{hu2007}%
  \BibitemOpen
  \bibfield  {author} {\bibinfo {author} {\bibfnamefont {T.}~\bibnamefont
  {Hu}}\ and\ \bibinfo {author} {\bibfnamefont {B.~I.}\ \bibnamefont
  {Shklovskii}},\ }\href@noop {} {\bibfield  {journal} {\bibinfo  {journal}
  {Phys. Rev. E}\ }\textbf {\bibinfo {volume} {75}},\ \bibinfo {pages} {051901}
  (\bibinfo {year} {2007})}\BibitemShut {NoStop}%
\bibitem [{\citenamefont {Kler}\ \emph {et~al.}(2012)\citenamefont {Kler},
  \citenamefont {Asor}, \citenamefont {Li}, \citenamefont {Ginsburg},
  \citenamefont {Harries}, \citenamefont {Oppenheim}, \citenamefont
  {Zlotnick},\ and\ \citenamefont {Raviv}}]{kler2012}%
  \BibitemOpen
  \bibfield  {author} {\bibinfo {author} {\bibfnamefont {S.}~\bibnamefont
  {Kler}}, \bibinfo {author} {\bibfnamefont {R.}~\bibnamefont {Asor}}, \bibinfo
  {author} {\bibfnamefont {C.}~\bibnamefont {Li}}, \bibinfo {author}
  {\bibfnamefont {A.}~\bibnamefont {Ginsburg}}, \bibinfo {author}
  {\bibfnamefont {D.}~\bibnamefont {Harries}}, \bibinfo {author} {\bibfnamefont
  {A.}~\bibnamefont {Oppenheim}}, \bibinfo {author} {\bibfnamefont
  {A.}~\bibnamefont {Zlotnick}},\ and\ \bibinfo {author} {\bibfnamefont
  {U.}~\bibnamefont {Raviv}},\ }\href@noop {} {\bibfield  {journal} {\bibinfo
  {journal} {Journal of the American Chemical Society}\ }\textbf {\bibinfo
  {volume} {134}},\ \bibinfo {pages} {8823} (\bibinfo {year}
  {2012})}\BibitemShut {NoStop}%
\bibitem [{\citenamefont {Garmann}\ \emph {et~al.}(2019)\citenamefont
  {Garmann}, \citenamefont {Goldfain},\ and\ \citenamefont
  {Manoharan}}]{garmann2019}%
  \BibitemOpen
  \bibfield  {author} {\bibinfo {author} {\bibfnamefont {R.~F.}\ \bibnamefont
  {Garmann}}, \bibinfo {author} {\bibfnamefont {A.~M.}\ \bibnamefont
  {Goldfain}},\ and\ \bibinfo {author} {\bibfnamefont {V.~N.}\ \bibnamefont
  {Manoharan}},\ }\href@noop {} {\bibfield  {journal} {\bibinfo  {journal}
  {Proceedings of the National Academy of Sciences}\ }\textbf {\bibinfo
  {volume} {116}},\ \bibinfo {pages} {22485} (\bibinfo {year}
  {2019})}\BibitemShut {NoStop}%
\bibitem [{\citenamefont {Baker}\ \emph {et~al.}(1999)\citenamefont {Baker},
  \citenamefont {Olson},\ and\ \citenamefont {Fuller}}]{Baker}%
  \BibitemOpen
  \bibfield  {author} {\bibinfo {author} {\bibfnamefont {T.~S.}\ \bibnamefont
  {Baker}}, \bibinfo {author} {\bibfnamefont {N.~H.}\ \bibnamefont {Olson}},\
  and\ \bibinfo {author} {\bibfnamefont {S.~D.}\ \bibnamefont {Fuller}},\
  }\href@noop {} {\bibfield  {journal} {\bibinfo  {journal} {Microbiol. Mol.
  Biol. Rev.}\ }\textbf {\bibinfo {volume} {63}},\ \bibinfo {pages} {862}
  (\bibinfo {year} {1999})}\BibitemShut {NoStop}%
\bibitem [{\citenamefont {Tihova}\ \emph {et~al.}(2004)\citenamefont {Tihova},
  \citenamefont {Dryden}, \citenamefont {Le}, \citenamefont {Harvey},
  \citenamefont {Johnson}, \citenamefont {Yeager},\ and\ \citenamefont
  {Schneemann}}]{Tihova2004}%
  \BibitemOpen
  \bibfield  {author} {\bibinfo {author} {\bibfnamefont {M.}~\bibnamefont
  {Tihova}}, \bibinfo {author} {\bibfnamefont {K.~A.}\ \bibnamefont {Dryden}},
  \bibinfo {author} {\bibfnamefont {T.~V.~L.}\ \bibnamefont {Le}}, \bibinfo
  {author} {\bibfnamefont {S.~C.}\ \bibnamefont {Harvey}}, \bibinfo {author}
  {\bibfnamefont {J.~E.}\ \bibnamefont {Johnson}}, \bibinfo {author}
  {\bibfnamefont {M.}~\bibnamefont {Yeager}},\ and\ \bibinfo {author}
  {\bibfnamefont {A.}~\bibnamefont {Schneemann}},\ }\href@noop {} {\bibfield
  {journal} {\bibinfo  {journal} {J. Virol.}\ }\textbf {\bibinfo {volume}
  {78}},\ \bibinfo {pages} {2897} (\bibinfo {year} {2004})}\BibitemShut
  {NoStop}%
\bibitem [{\citenamefont {Johnson}\ \emph {et~al.}(2004)\citenamefont
  {Johnson}, \citenamefont {Willits}, \citenamefont {Young},\ and\
  \citenamefont {Zlotnick}}]{Johnson2004}%
  \BibitemOpen
  \bibfield  {author} {\bibinfo {author} {\bibfnamefont {J.~M.}\ \bibnamefont
  {Johnson}}, \bibinfo {author} {\bibfnamefont {D.~A.}\ \bibnamefont
  {Willits}}, \bibinfo {author} {\bibfnamefont {M.~J.}\ \bibnamefont {Young}},\
  and\ \bibinfo {author} {\bibfnamefont {A.}~\bibnamefont {Zlotnick}},\
  }\href@noop {} {\bibfield  {journal} {\bibinfo  {journal} {J. Mol. Biol.}\
  }\textbf {\bibinfo {volume} {335}},\ \bibinfo {pages} {455} (\bibinfo {year}
  {2004})}\BibitemShut {NoStop}%
\bibitem [{\citenamefont {Koning}\ \emph {et~al.}(2016)\citenamefont {Koning},
  \citenamefont {Gomez-Blanco}, \citenamefont {Akopjana}, \citenamefont
  {Vargas}, \citenamefont {Kazaks}, \citenamefont {Tars}, \citenamefont
  {Carazo},\ and\ \citenamefont {Koster}}]{koning}%
  \BibitemOpen
  \bibfield  {author} {\bibinfo {author} {\bibfnamefont {R.~I.}\ \bibnamefont
  {Koning}}, \bibinfo {author} {\bibfnamefont {J.}~\bibnamefont
  {Gomez-Blanco}}, \bibinfo {author} {\bibfnamefont {I.}~\bibnamefont
  {Akopjana}}, \bibinfo {author} {\bibfnamefont {J.}~\bibnamefont {Vargas}},
  \bibinfo {author} {\bibfnamefont {A.}~\bibnamefont {Kazaks}}, \bibinfo
  {author} {\bibfnamefont {K.}~\bibnamefont {Tars}}, \bibinfo {author}
  {\bibfnamefont {J.~M.}\ \bibnamefont {Carazo}},\ and\ \bibinfo {author}
  {\bibfnamefont {A.~J.}\ \bibnamefont {Koster}},\ }\href@noop {} {\bibfield
  {journal} {\bibinfo  {journal} {Nature communications}\ }\textbf {\bibinfo
  {volume} {7}},\ \bibinfo {pages} {1} (\bibinfo {year} {2016})}\BibitemShut
  {NoStop}%
\bibitem [{\citenamefont {Beren}\ \emph {et~al.}(2020)\citenamefont {Beren},
  \citenamefont {Cui}, \citenamefont {Chakravarty}, \citenamefont {Yang},
  \citenamefont {Rao}, \citenamefont {Knobler}, \citenamefont {Zhou},\ and\
  \citenamefont {Gelbart}}]{beren}%
  \BibitemOpen
  \bibfield  {author} {\bibinfo {author} {\bibfnamefont {C.}~\bibnamefont
  {Beren}}, \bibinfo {author} {\bibfnamefont {Y.}~\bibnamefont {Cui}}, \bibinfo
  {author} {\bibfnamefont {A.}~\bibnamefont {Chakravarty}}, \bibinfo {author}
  {\bibfnamefont {X.}~\bibnamefont {Yang}}, \bibinfo {author} {\bibfnamefont
  {A.}~\bibnamefont {Rao}}, \bibinfo {author} {\bibfnamefont {C.~M.}\
  \bibnamefont {Knobler}}, \bibinfo {author} {\bibfnamefont {Z.~H.}\
  \bibnamefont {Zhou}},\ and\ \bibinfo {author} {\bibfnamefont {W.~M.}\
  \bibnamefont {Gelbart}},\ }\href@noop {} {\bibfield  {journal} {\bibinfo
  {journal} {Proceedings of the National Academy of Sciences}\ }\textbf
  {\bibinfo {volume} {117}},\ \bibinfo {pages} {10673} (\bibinfo {year}
  {2020})}\BibitemShut {NoStop}%
\bibitem [{\citenamefont {Dykeman}\ \emph {et~al.}(2011)\citenamefont
  {Dykeman}, \citenamefont {Grayson}, \citenamefont {Toropova}, \citenamefont
  {Ranson}, \citenamefont {Stockley},\ and\ \citenamefont
  {Twarock}}]{Dykeman2011}%
  \BibitemOpen
  \bibfield  {author} {\bibinfo {author} {\bibfnamefont {E.~C.}\ \bibnamefont
  {Dykeman}}, \bibinfo {author} {\bibfnamefont {N.~E.}\ \bibnamefont
  {Grayson}}, \bibinfo {author} {\bibfnamefont {K.}~\bibnamefont {Toropova}},
  \bibinfo {author} {\bibfnamefont {N.~A.}\ \bibnamefont {Ranson}}, \bibinfo
  {author} {\bibfnamefont {P.~G.}\ \bibnamefont {Stockley}},\ and\ \bibinfo
  {author} {\bibfnamefont {R.}~\bibnamefont {Twarock}},\ }\href@noop {}
  {\bibfield  {journal} {\bibinfo  {journal} {J. Mol. Biol.}\ }\textbf
  {\bibinfo {volume} {408}},\ \bibinfo {pages} {399} (\bibinfo {year}
  {2011})}\BibitemShut {NoStop}%
\bibitem [{\citenamefont {Dai}\ \emph {et~al.}(2017)\citenamefont {Dai},
  \citenamefont {Li}, \citenamefont {Lai}, \citenamefont {Shu}, \citenamefont
  {Du}, \citenamefont {Zhou},\ and\ \citenamefont {Sun}}]{dai2017}%
  \BibitemOpen
  \bibfield  {author} {\bibinfo {author} {\bibfnamefont {X.}~\bibnamefont
  {Dai}}, \bibinfo {author} {\bibfnamefont {Z.}~\bibnamefont {Li}}, \bibinfo
  {author} {\bibfnamefont {M.}~\bibnamefont {Lai}}, \bibinfo {author}
  {\bibfnamefont {S.}~\bibnamefont {Shu}}, \bibinfo {author} {\bibfnamefont
  {Y.}~\bibnamefont {Du}}, \bibinfo {author} {\bibfnamefont {Z.~H.}\
  \bibnamefont {Zhou}},\ and\ \bibinfo {author} {\bibfnamefont
  {R.}~\bibnamefont {Sun}},\ }\href@noop {} {\bibfield  {journal} {\bibinfo
  {journal} {Nature}\ }\textbf {\bibinfo {volume} {541}},\ \bibinfo {pages}
  {112} (\bibinfo {year} {2017})}\BibitemShut {NoStop}%
\bibitem [{\citenamefont {Dimmock}\ \emph {et~al.}(2001)\citenamefont
  {Dimmock}, \citenamefont {Easton},\ and\ \citenamefont
  {Leppard}}]{Dimmock2001}%
  \BibitemOpen
  \bibfield  {author} {\bibinfo {author} {\bibfnamefont {N.}~\bibnamefont
  {Dimmock}}, \bibinfo {author} {\bibfnamefont {A.}~\bibnamefont {Easton}},\
  and\ \bibinfo {author} {\bibfnamefont {K.}~\bibnamefont {Leppard}},\
  }\href@noop {} {\emph {\bibinfo {title} {{Introduction to modern
  virology}}}}\ (\bibinfo  {publisher} {Blackwell Publishing},\ \bibinfo
  {address} {Malden, MA},\ \bibinfo {year} {2001})\BibitemShut {NoStop}%
\bibitem [{Note3()}]{Note3}%
  \BibitemOpen
  \bibinfo {note} {Selective nucleation was proposed by I. Rouzina in the
  context of the assembly of retroviruses}\BibitemShut {NoStop}%
\bibitem [{\citenamefont {Perlmutter}\ \emph {et~al.}(2014)\citenamefont
  {Perlmutter}, \citenamefont {Perkett},\ and\ \citenamefont
  {Hagan}}]{Perlmutter2014}%
  \BibitemOpen
  \bibfield  {author} {\bibinfo {author} {\bibfnamefont {J.~D.}\ \bibnamefont
  {Perlmutter}}, \bibinfo {author} {\bibfnamefont {M.~R.}\ \bibnamefont
  {Perkett}},\ and\ \bibinfo {author} {\bibfnamefont {M.~F.}\ \bibnamefont
  {Hagan}},\ }\bibfield  {journal} {\bibinfo  {journal} {J. Mol. Biol.}\ }\href
  {https://doi.org/10.1016/j.jmb.2014.07.004} {10.1016/j.jmb.2014.07.004}
  (\bibinfo {year} {2014})\BibitemShut {NoStop}%
\bibitem [{\citenamefont {Rudnick}\ and\ \citenamefont
  {Bruinsma}(2019)}]{rudnick2019}%
  \BibitemOpen
  \bibfield  {author} {\bibinfo {author} {\bibfnamefont {J.}~\bibnamefont
  {Rudnick}}\ and\ \bibinfo {author} {\bibfnamefont {R.}~\bibnamefont
  {Bruinsma}},\ }\href@noop {} {\bibfield  {journal} {\bibinfo  {journal}
  {Physical Review E}\ }\textbf {\bibinfo {volume} {100}},\ \bibinfo {pages}
  {012145} (\bibinfo {year} {2019})}\BibitemShut {NoStop}%
\bibitem [{\citenamefont {Garmann}\ \emph {et~al.}(2013)\citenamefont
  {Garmann}, \citenamefont {Comas-Garcia}, \citenamefont {Gopal}, \citenamefont
  {Knobler},\ and\ \citenamefont {Gelbart}}]{Garmann}%
  \BibitemOpen
  \bibfield  {author} {\bibinfo {author} {\bibfnamefont {R.~F.}\ \bibnamefont
  {Garmann}}, \bibinfo {author} {\bibfnamefont {M.}~\bibnamefont
  {Comas-Garcia}}, \bibinfo {author} {\bibfnamefont {A.}~\bibnamefont {Gopal}},
  \bibinfo {author} {\bibfnamefont {C.~M.}\ \bibnamefont {Knobler}},\ and\
  \bibinfo {author} {\bibfnamefont {W.~M.}\ \bibnamefont {Gelbart}},\ }\href
  {https://doi.org/10.1016/j.jmb.2013.10.017} {\bibfield  {journal} {\bibinfo
  {journal} {J. Mol. Biol.}\ }\textbf {\bibinfo {volume} {32}},\ \bibinfo
  {pages} {???} (\bibinfo {year} {2013})}\BibitemShut {NoStop}%
\bibitem [{\citenamefont {Van~Rosmalen}\ \emph {et~al.}(2020)\citenamefont
  {Van~Rosmalen}, \citenamefont {Kamsma}, \citenamefont {Biebricher},
  \citenamefont {Li}, \citenamefont {Zlotnick}, \citenamefont {Roos},\ and\
  \citenamefont {Wuite}}]{van2020}%
  \BibitemOpen
  \bibfield  {author} {\bibinfo {author} {\bibfnamefont {M.~G.}\ \bibnamefont
  {Van~Rosmalen}}, \bibinfo {author} {\bibfnamefont {D.}~\bibnamefont
  {Kamsma}}, \bibinfo {author} {\bibfnamefont {A.~S.}\ \bibnamefont
  {Biebricher}}, \bibinfo {author} {\bibfnamefont {C.}~\bibnamefont {Li}},
  \bibinfo {author} {\bibfnamefont {A.}~\bibnamefont {Zlotnick}}, \bibinfo
  {author} {\bibfnamefont {W.~H.}\ \bibnamefont {Roos}},\ and\ \bibinfo
  {author} {\bibfnamefont {G.~J.}\ \bibnamefont {Wuite}},\ }\href@noop {}
  {\bibfield  {journal} {\bibinfo  {journal} {Science advances}\ }\textbf
  {\bibinfo {volume} {6}},\ \bibinfo {pages} {eaaz1639} (\bibinfo {year}
  {2020})}\BibitemShut {NoStop}%
\bibitem [{\citenamefont {Zlotnick}(1994)}]{Zlotnick1994}%
  \BibitemOpen
  \bibfield  {author} {\bibinfo {author} {\bibfnamefont {A.}~\bibnamefont
  {Zlotnick}},\ }\href@noop {} {\bibfield  {journal} {\bibinfo  {journal} {J.
  Mol. Biol.}\ }\textbf {\bibinfo {volume} {241}},\ \bibinfo {pages} {59}
  (\bibinfo {year} {1994})}\BibitemShut {NoStop}%
\bibitem [{\citenamefont {Endres}\ and\ \citenamefont
  {Zlotnick}(2002)}]{Endres2002}%
  \BibitemOpen
  \bibfield  {author} {\bibinfo {author} {\bibfnamefont {D.}~\bibnamefont
  {Endres}}\ and\ \bibinfo {author} {\bibfnamefont {A.}~\bibnamefont
  {Zlotnick}},\ }\href@noop {} {\bibfield  {journal} {\bibinfo  {journal}
  {Biophys. J.}\ }\textbf {\bibinfo {volume} {83}},\ \bibinfo {pages} {1217}
  (\bibinfo {year} {2002})}\BibitemShut {NoStop}%
\bibitem [{\citenamefont {Zlotnick}(2007)}]{Zlotnick2007}%
  \BibitemOpen
  \bibfield  {author} {\bibinfo {author} {\bibfnamefont {A.}~\bibnamefont
  {Zlotnick}},\ }\href@noop {} {\bibfield  {journal} {\bibinfo  {journal} {J.
  Mol. Biol.}\ }\textbf {\bibinfo {volume} {366}},\ \bibinfo {pages} {14}
  (\bibinfo {year} {2007})}\BibitemShut {NoStop}%
\bibitem [{\citenamefont {Morozov}\ \emph {et~al.}(2009)\citenamefont
  {Morozov}, \citenamefont {Bruinsma},\ and\ \citenamefont
  {Rudnick}}]{Morozov2009}%
  \BibitemOpen
  \bibfield  {author} {\bibinfo {author} {\bibfnamefont {A.~Y.}\ \bibnamefont
  {Morozov}}, \bibinfo {author} {\bibfnamefont {R.~F.}\ \bibnamefont
  {Bruinsma}},\ and\ \bibinfo {author} {\bibfnamefont {J.}~\bibnamefont
  {Rudnick}},\ }\href@noop {} {\bibfield  {journal} {\bibinfo  {journal} {J.
  Chem. Phys.}\ }\textbf {\bibinfo {volume} {131}},\ \bibinfo {pages} {155101}
  (\bibinfo {year} {2009})}\BibitemShut {NoStop}%
\bibitem [{Note4()}]{Note4}%
  \BibitemOpen
  \bibinfo {note} {The assembly of larger icosahedral capsids can be
  represented by a version of the model that allows for pentameric and
  hexameric capsomers.}\BibitemShut {Stop}%
\bibitem [{\citenamefont {Bollob{\'a}s}(2013)}]{Bollobas}%
  \BibitemOpen
  \bibfield  {author} {\bibinfo {author} {\bibfnamefont {B.}~\bibnamefont
  {Bollob{\'a}s}},\ }\href@noop {} {\emph {\bibinfo {title} {Modern graph
  theory}}},\ Vol.\ \bibinfo {volume} {184}\ (\bibinfo  {publisher} {Springer
  Science \& Business Media},\ \bibinfo {year} {2013})\BibitemShut {NoStop}%
\bibitem [{\citenamefont {Graham}\ and\ \citenamefont {Hell}(1985)}]{Graham}%
  \BibitemOpen
  \bibfield  {author} {\bibinfo {author} {\bibfnamefont {R.~L.}\ \bibnamefont
  {Graham}}\ and\ \bibinfo {author} {\bibfnamefont {P.}~\bibnamefont {Hell}},\
  }\href@noop {} {\bibfield  {journal} {\bibinfo  {journal} {Annals of the
  History of Computing}\ }\textbf {\bibinfo {volume} {7}},\ \bibinfo {pages}
  {43} (\bibinfo {year} {1985})}\BibitemShut {NoStop}%
\bibitem [{\citenamefont {Fang}\ \emph {et~al.}(2011)\citenamefont {Fang},
  \citenamefont {Gelbart},\ and\ \citenamefont {Ben-Shaul}}]{fang}%
  \BibitemOpen
  \bibfield  {author} {\bibinfo {author} {\bibfnamefont {L.~T.}\ \bibnamefont
  {Fang}}, \bibinfo {author} {\bibfnamefont {W.~M.}\ \bibnamefont {Gelbart}},\
  and\ \bibinfo {author} {\bibfnamefont {A.}~\bibnamefont {Ben-Shaul}},\
  }\href@noop {} {\bibfield  {journal} {\bibinfo  {journal} {The Journal of
  Chemical Physics}\ }\textbf {\bibinfo {volume} {135}},\ \bibinfo {pages}
  {10B616} (\bibinfo {year} {2011})}\BibitemShut {NoStop}%
\bibitem [{\citenamefont {Rudnick}\ and\ \citenamefont
  {Bruinsma}(2005)}]{Rudnick2005}%
  \BibitemOpen
  \bibfield  {author} {\bibinfo {author} {\bibfnamefont {J.}~\bibnamefont
  {Rudnick}}\ and\ \bibinfo {author} {\bibfnamefont {R.}~\bibnamefont
  {Bruinsma}},\ }\href@noop {} {\bibfield  {journal} {\bibinfo  {journal}
  {Phys. Rev. Lett.}\ }\textbf {\bibinfo {volume} {94}},\ \bibinfo {pages}
  {038101} (\bibinfo {year} {2005})}\BibitemShut {NoStop}%
\bibitem [{\citenamefont {Dykeman}\ \emph {et~al.}(2013)\citenamefont
  {Dykeman}, \citenamefont {Stockley},\ and\ \citenamefont
  {Twarock}}]{Dykeman2013b}%
  \BibitemOpen
  \bibfield  {author} {\bibinfo {author} {\bibfnamefont {E.~C.}\ \bibnamefont
  {Dykeman}}, \bibinfo {author} {\bibfnamefont {P.~G.}\ \bibnamefont
  {Stockley}},\ and\ \bibinfo {author} {\bibfnamefont {R.}~\bibnamefont
  {Twarock}},\ }\href {https://doi.org/10.1016/j.jmb.2013.06.005} {\bibfield
  {journal} {\bibinfo  {journal} {J. Mol. Biol.}\ }\textbf {\bibinfo {volume}
  {425}},\ \bibinfo {pages} {3235} (\bibinfo {year} {2013})}\BibitemShut
  {NoStop}%
\bibitem [{\citenamefont {A.M.Gutin}\ \emph {et~al.}(1993)\citenamefont
  {A.M.Gutin}, \citenamefont {A.Y.Grosberg},\ and\ \citenamefont
  {E.I.Shakhnovich}}]{gutin}%
  \BibitemOpen
  \bibfield  {author} {\bibinfo {author} {\bibnamefont {A.M.Gutin}}, \bibinfo
  {author} {\bibnamefont {A.Y.Grosberg}},\ and\ \bibinfo {author} {\bibnamefont
  {E.I.Shakhnovich}},\ }\href@noop {} {\bibfield  {journal} {\bibinfo
  {journal} {Macromolecules}\ }\textbf {\bibinfo {volume} {26}},\ \bibinfo
  {pages} {1293} (\bibinfo {year} {1993})}\BibitemShut {NoStop}%
\bibitem [{Note5()}]{Note5}%
  \BibitemOpen
  \bibinfo {note} {In actuality, condensation of the RNA genome molecules takes
  place \protect \textit {during} encapsidation.}\BibitemShut {Stop}%
\bibitem [{\citenamefont {Parent}\ \emph {et~al.}(2006)\citenamefont {Parent},
  \citenamefont {Zlotnick},\ and\ \citenamefont {Teschke}}]{Parent2006}%
  \BibitemOpen
  \bibfield  {author} {\bibinfo {author} {\bibfnamefont {K.~N.}\ \bibnamefont
  {Parent}}, \bibinfo {author} {\bibfnamefont {A.}~\bibnamefont {Zlotnick}},\
  and\ \bibinfo {author} {\bibfnamefont {C.~M.}\ \bibnamefont {Teschke}},\
  }\href@noop {} {\bibfield  {journal} {\bibinfo  {journal} {J. Mol. Biol.}\
  }\textbf {\bibinfo {volume} {359}},\ \bibinfo {pages} {1097} (\bibinfo {year}
  {2006})}\BibitemShut {NoStop}%
\bibitem [{\citenamefont {Tuma}\ \emph {et~al.}(2008)\citenamefont {Tuma},
  \citenamefont {Tsuruta}, \citenamefont {French},\ and\ \citenamefont
  {Prevelige}}]{Tuma2008}%
  \BibitemOpen
  \bibfield  {author} {\bibinfo {author} {\bibfnamefont {R.}~\bibnamefont
  {Tuma}}, \bibinfo {author} {\bibfnamefont {H.}~\bibnamefont {Tsuruta}},
  \bibinfo {author} {\bibfnamefont {K.~H.}\ \bibnamefont {French}},\ and\
  \bibinfo {author} {\bibfnamefont {P.~E.}\ \bibnamefont {Prevelige}},\
  }\href@noop {} {\bibfield  {journal} {\bibinfo  {journal} {J. Mol. Biol.}\
  }\textbf {\bibinfo {volume} {381}},\ \bibinfo {pages} {1395} (\bibinfo {year}
  {2008})}\BibitemShut {NoStop}%
\bibitem [{\citenamefont {Basnak}\ \emph {et~al.}(2010)\citenamefont {Basnak},
  \citenamefont {Morton}, \citenamefont {Rolfsson}, \citenamefont {Stonehouse},
  \citenamefont {Ashcroft},\ and\ \citenamefont {Stockley}}]{Basnak2010}%
  \BibitemOpen
  \bibfield  {author} {\bibinfo {author} {\bibfnamefont {G.}~\bibnamefont
  {Basnak}}, \bibinfo {author} {\bibfnamefont {V.~L.}\ \bibnamefont {Morton}},
  \bibinfo {author} {\bibfnamefont {O.}~\bibnamefont {Rolfsson}}, \bibinfo
  {author} {\bibfnamefont {N.~J.}\ \bibnamefont {Stonehouse}}, \bibinfo
  {author} {\bibfnamefont {A.~E.}\ \bibnamefont {Ashcroft}},\ and\ \bibinfo
  {author} {\bibfnamefont {P.~G.}\ \bibnamefont {Stockley}},\ }\href@noop {}
  {\bibfield  {journal} {\bibinfo  {journal} {J. Mol. Biol.}\ }\textbf
  {\bibinfo {volume} {395}},\ \bibinfo {pages} {924} (\bibinfo {year}
  {2010})}\BibitemShut {NoStop}%
\bibitem [{\citenamefont {Johnston}\ \emph {et~al.}(2010)\citenamefont
  {Johnston}, \citenamefont {Louis},\ and\ \citenamefont
  {Doye}}]{Johnston2010}%
  \BibitemOpen
  \bibfield  {author} {\bibinfo {author} {\bibfnamefont {I.~G.}\ \bibnamefont
  {Johnston}}, \bibinfo {author} {\bibfnamefont {A.~A.}\ \bibnamefont
  {Louis}},\ and\ \bibinfo {author} {\bibfnamefont {J.~P.~K.}\ \bibnamefont
  {Doye}},\ }\href@noop {} {\bibfield  {journal} {\bibinfo  {journal} {J.
  Phys.: Condens. Matter}\ }\textbf {\bibinfo {volume} {22}},\ \bibinfo {pages}
  {104101} (\bibinfo {year} {2010})}\BibitemShut {NoStop}%
\bibitem [{\citenamefont {Hagan}\ \emph {et~al.}(2011)\citenamefont {Hagan},
  \citenamefont {Elrad},\ and\ \citenamefont {Jack}}]{Hagan2011}%
  \BibitemOpen
  \bibfield  {author} {\bibinfo {author} {\bibfnamefont {M.~F.}\ \bibnamefont
  {Hagan}}, \bibinfo {author} {\bibfnamefont {O.~M.}\ \bibnamefont {Elrad}},\
  and\ \bibinfo {author} {\bibfnamefont {R.~L.}\ \bibnamefont {Jack}},\
  }\href@noop {} {\bibfield  {journal} {\bibinfo  {journal} {J. Chem. Phys.}\
  }\textbf {\bibinfo {volume} {135}},\ \bibinfo {pages} {104115} (\bibinfo
  {year} {2011})}\BibitemShut {NoStop}%
\bibitem [{\citenamefont {Baschek}\ \emph {et~al.}(2012)\citenamefont
  {Baschek}, \citenamefont {Klein},\ and\ \citenamefont
  {Schwarz}}]{Baschek2012}%
  \BibitemOpen
  \bibfield  {author} {\bibinfo {author} {\bibfnamefont {J.~E.}\ \bibnamefont
  {Baschek}}, \bibinfo {author} {\bibfnamefont {H.~C.~R.}\ \bibnamefont
  {Klein}},\ and\ \bibinfo {author} {\bibfnamefont {U.~S.}\ \bibnamefont
  {Schwarz}},\ }\bibfield  {journal} {\bibinfo  {journal} {Bmc Biophysics}\
  }\textbf {\bibinfo {volume} {5}},\ \href
  {https://doi.org/10.1186/2046-1682-5-22} {10.1186/2046-1682-5-22} (\bibinfo
  {year} {2012}),\ \bibinfo {note} {baschek, Johanna E. Klein, Heinrich C. R.
  Schwarz, Ulrich S. Schwarz, Ulrich/K-4111-2014 Schwarz,
  Ulrich/0000-0003-1483-640X}\BibitemShut {NoStop}%
\bibitem [{\citenamefont {Perlmutter}\ and\ \citenamefont
  {Hagan}(2015)}]{Perlmutter2015b}%
  \BibitemOpen
  \bibfield  {author} {\bibinfo {author} {\bibfnamefont {J.~D.}\ \bibnamefont
  {Perlmutter}}\ and\ \bibinfo {author} {\bibfnamefont {M.~F.}\ \bibnamefont
  {Hagan}},\ }\bibfield  {journal} {\bibinfo  {journal} {{J. Mol. Biol.}}\
  }\href {https://doi.org/10.1016/j.jmb.2015.05.008}
  {10.1016/j.jmb.2015.05.008} (\bibinfo {year} {2015})\BibitemShut {NoStop}%
\bibitem [{\citenamefont {Safran}(2018)}]{safran}%
  \BibitemOpen
  \bibfield  {author} {\bibinfo {author} {\bibfnamefont {S.}~\bibnamefont
  {Safran}},\ }\href@noop {} {\emph {\bibinfo {title} {Statistical
  thermodynamics of surfaces, interfaces, and membranes}}}\ (\bibinfo
  {publisher} {CRC Press},\ \bibinfo {year} {2018})\BibitemShut {NoStop}%
\bibitem [{Note6()}]{Note6}%
  \BibitemOpen
  \bibinfo {note} {Since a virus is a system of limited size true phase
  transitions are not possible.}\BibitemShut {Stop}%
\bibitem [{Note7()}]{Note7}%
  \BibitemOpen
  \bibinfo {note} {The case of competing spanning trees is discussed in the
  paper, submitted to Plos Biocomputation.}\BibitemShut {Stop}%
\bibitem [{\citenamefont {Perkett}\ and\ \citenamefont
  {Hagan}(2014)}]{Perkett2014}%
  \BibitemOpen
  \bibfield  {author} {\bibinfo {author} {\bibfnamefont {M.~R.}\ \bibnamefont
  {Perkett}}\ and\ \bibinfo {author} {\bibfnamefont {M.~F.}\ \bibnamefont
  {Hagan}},\ }\href {https://doi.org/10.1063/1.4878494} {\bibfield  {journal}
  {\bibinfo  {journal} {J. Chem. Phys.}\ }\textbf {\bibinfo {volume} {140}},\
  \bibinfo {eid} {214101} (\bibinfo {year} {2014})}\BibitemShut {NoStop}%
\bibitem [{\citenamefont {Van~Kampen}(1992)}]{vanKampen}%
  \BibitemOpen
  \bibfield  {author} {\bibinfo {author} {\bibfnamefont {N.~G.}\ \bibnamefont
  {Van~Kampen}},\ }\href@noop {} {\emph {\bibinfo {title} {Stochastic processes
  in physics and chemistry}}},\ Vol.~\bibinfo {volume} {1}\ (\bibinfo
  {publisher} {Elsevier},\ \bibinfo {year} {1992})\BibitemShut {NoStop}%
\bibitem [{\citenamefont {Schulten}(1999)}]{schulten}%
  \BibitemOpen
  \bibfield  {author} {\bibinfo {author} {\bibfnamefont {K.}~\bibnamefont
  {Schulten}},\ }\href@noop {} {\emph {\bibinfo {title} {Non-equilibrium
  Statistical Mechanics}}}\ (\bibinfo  {publisher} {UIUC},\ \bibinfo {year}
  {1999})\BibitemShut {NoStop}%
\bibitem [{\citenamefont {Wu}(1992)}]{Wu1992}%
  \BibitemOpen
  \bibfield  {author} {\bibinfo {author} {\bibfnamefont {D.~T.}\ \bibnamefont
  {Wu}},\ }\href@noop {} {\bibfield  {journal} {\bibinfo  {journal} {J. Chem.
  Phys.}\ }\textbf {\bibinfo {volume} {97}},\ \bibinfo {pages} {2644} (\bibinfo
  {year} {1992})}\BibitemShut {NoStop}%
\bibitem [{Note8()}]{Note8}%
  \BibitemOpen
  \bibinfo {note} {Similar plots are obtained by time-dependent light
  scattering studies of the assembly of empty capsids with varying protein
  concentrations. \cite {Casini2004}.}\BibitemShut {Stop}%
\bibitem [{\citenamefont {Comas-Garcia}\ \emph {et~al.}(2012)\citenamefont
  {Comas-Garcia}, \citenamefont {Cadena-Nava}, \citenamefont {Rao},
  \citenamefont {Knobler},\ and\ \citenamefont {Gelbart}}]{Comas-Garcia}%
  \BibitemOpen
  \bibfield  {author} {\bibinfo {author} {\bibfnamefont {M.}~\bibnamefont
  {Comas-Garcia}}, \bibinfo {author} {\bibfnamefont {R.~D.}\ \bibnamefont
  {Cadena-Nava}}, \bibinfo {author} {\bibfnamefont {A.~L.~N.}\ \bibnamefont
  {Rao}}, \bibinfo {author} {\bibfnamefont {C.~M.}\ \bibnamefont {Knobler}},\
  and\ \bibinfo {author} {\bibfnamefont {W.~M.}\ \bibnamefont {Gelbart}},\
  }\href@noop {} {\bibfield  {journal} {\bibinfo  {journal} {Journal of
  Virology}\ }\textbf {\bibinfo {volume} {86}},\ \bibinfo {pages} {12271 }
  (\bibinfo {year} {2012})}\BibitemShut {NoStop}%
\bibitem [{\citenamefont {Hu}\ \emph {et~al.}(2006)\citenamefont {Hu},
  \citenamefont {Grosberg},\ and\ \citenamefont {Shklovskii}}]{Hu2006}%
  \BibitemOpen
  \bibfield  {author} {\bibinfo {author} {\bibfnamefont {T.}~\bibnamefont
  {Hu}}, \bibinfo {author} {\bibfnamefont {A.~Y.}\ \bibnamefont {Grosberg}},\
  and\ \bibinfo {author} {\bibfnamefont {B.~I.}\ \bibnamefont {Shklovskii}},\
  }\href@noop {} {\bibfield  {journal} {\bibinfo  {journal} {Biophys. J.}\
  }\textbf {\bibinfo {volume} {90}},\ \bibinfo {pages} {2731} (\bibinfo {year}
  {2006})}\BibitemShut {NoStop}%
\end{thebibliography}
\end{document}